\documentclass[12pt]{article}

\usepackage{placeins}
\usepackage{pifont}
\usepackage{graphicx}
\usepackage{booktabs}
\usepackage{multirow}
\usepackage{amsmath}
\usepackage{amssymb}
\usepackage{lmodern}
\usepackage{rotating}
\usepackage{float}
\usepackage{longtable}
\usepackage[T1]{fontenc}
\usepackage[a4paper,left=2.5cm,right=2.5cm,top=3cm,bottom=3cm]{geometry}
\usepackage[flushleft]{threeparttable}

\usepackage[style=authoryear-comp, backend=bibtex, maxcitenames=2, maxbibnames=99 ,dashed=false]{biblatex}
\bibliography{adj_curves}

\usepackage{hyperref}

\newcommand{\cmark}{\ding{51}}
\newcommand{\xmark}{\ding{55}}

\begin{document}

\title{A Comparison of Different Methods to Adjust Survival Curves for Confounders}
\date{}
\author{Robin Denz, Renate Klaaßen-Mielke, Nina Timmesfeld \\ \\ Ruhr-University of Bochum \\ Department of Medical Informatics, Biometry and Epidemiology}

\maketitle

\begin{abstract}
	Treatment specific survival curves are an important tool to illustrate the treatment effect in studies with time-to-event outcomes. In non-randomized studies, unadjusted estimates can lead to biased depictions due to confounding. Multiple methods to adjust survival curves for confounders exist. However, it is currently unclear which method is the most appropriate in which situation. Our goal is to compare forms of Inverse Probability of Treatment Weighting, the G-Formula, Propensity Score Matching, Empirical Likelihood Estimation and augmented estimators as well as their pseudo-values based counterparts in different scenarios with a focus on their bias and goodness-of-fit. We provide a short review of all methods and illustrate their usage by contrasting the survival of smokers and non-smokers, using data from the German Epidemiological Trial on Ankle-Brachial-Index. Subsequently, we compare the methods using a Monte-Carlo simulation. We consider scenarios in which correctly or incorrectly specified models for describing the treatment assignment and the time-to-event outcome are used with varying sample sizes. The bias and goodness-of-fit is determined by taking the entire survival curve into account. When used properly, all methods showed no systematic bias in medium to large samples. Cox regression based methods, however, showed systematic bias in small samples. The goodness-of-fit varied greatly between different methods and scenarios. Methods utilizing an outcome model were more efficient than other techniques, while augmented estimators using an additional treatment assignment model were unbiased when either model was correct with a goodness-of-fit comparable to other methods. These ``doubly-robust'' methods have important advantages in every considered scenario.
	\par
	\emph{Keywords:} adjusted survival curves, confounding, causal inference, time-to-event, observational data, simulation study
\end{abstract}

\section{Introduction} \label{chap::introduction}

In the analysis of clinical time-to-event data, treatment-specific survival curves are often used to graphically display the treatment effect in some population. The Kaplan-Meier estimator, stratified by treatment allocation, is usually used to calculate these curves. In sufficiently large randomized controlled trials with balanced groups this method yields unbiased results \parencite{Kaplan1958}. In reality, it is often impossible to conduct such studies because of ethical or administrative reasons, which is why observational study designs are very common. As is well known, the lack of randomization can lead to the occurrence of confounding \parencite{Rubin1974, Rubin1980}. Simple Kaplan-Meier estimates do not take confounders into account and therefore produce a systematically biased picture of the true treatment effect in such cases.
\par
The most popular way to adjust for confounders in medical time-to-event analysis is the use of the Cox proportional hazards model \parencite{Cox1972}. Communicating regression analysis results is, however, far more difficult than simply showing survival curves \parencite{Poole2010, Hernan2010, DeNeve2020}. It has been shown multiple times that graphics simplify the communication of statistical results and lead to a better understanding by the reader \parencite{Davis2010, Zipkin2014}. Many researchers report both adjusted hazard-ratios and unadjusted Kaplan-Meier estimates \parencite{Dey2020}. Since the latter did not correct for the presence of confounders, these results often differ and hence confuse the reader.
\par
Confounder-adjusted survival curves are a solution to this problem. Various methods for calculating these have been developed \parencite{Makuch1982, Cole2004, Austin2014, Wang2019, Zhang2012, Chatton2021a, Andersen2017}. Their properties have only been studied to a limited extent. Theoretical results can only be generalized to a certain degree because some assumptions are quite complex. Previous simulation studies deal exclusively with the properties of the survival curves at specific points in time \parencite{Wang2019, Wang2018, Cai2020, Chatton2020}. Our main goal is to fill this gap by investigating the properties of the available methods with respect to the entire survival curve using a Monte-Carlo simulation. We aim to determine which of these methods produce unbiased estimates and which methods show the least deviation from the true survival curve on average (goodness-of-fit).
\par
This paper is structured as follows. First, we give a formal description of confounder-adjusted survival curves and the background. Afterwards, a brief description of all included methods is given. Using real data from a large prospective cohort study, we illustrate the usage of these methods by comparing the survival of non-smokers and current or past smokers. Next, the design of the simulation study is described and the results are presented. Finally, we discuss the results and their implications for the practical applications of the adjustment methods.

\section{Background and Notation} \label{chap::notation}

Let $Z \in \{0, 1, ..., k\}$ denote the treatment group, where each value of $Z$ indicates one of $k$ possible treatments. For the sake of clarity, we use the term ``treatment groups'' throughout the paper, but it is important to note that any other grouping variable may be used as well. Let $T$ be the time to the occurrence of the event of interest. In reality, it is sometimes only known whether a person has suffered an event by time $C$ or not, which is known as right-censored data. In this case, only $T_{obs} = \min(T, C)$ would be observed with a corresponding event indicator $D = I(T < C)$. Although a crucial point for the estimation methods, it is unimportant for the definition of the target estimand.
\par
Under the Neyman-Rubin causal framework, every person has $k$ potential survival times $T^z \in \{T^0, T^1, ..., T^k\}$, one for each of the $k$ possible treatment strategies \parencite{Neyman1923, Rubin1974}. The goal is to estimate the counterfactual survival probability in the target population over time, where every person has received the same treatment. This population consists of $N$ individuals, indexed by $i$, $i = 1, 2, ..., N$, each with their own vector of baseline covariates $x_i$. The counterfactual survival probability of individual $i$ at time $t$ is defined by:

\begin{equation} \label{eq::individual_survival_prob}
	S(t | Z = z, X = x_i) = P(T^{z} > t | x_i),
\end{equation}

where $T^{z}$ denotes the failure time which would have been observed, if treatment $Z = z$ was actually administered \parencite{Cai2020}. Therefore, the target function is defined as:

\begin{equation} \label{eq::population_survival_prob}
	S_{z}(t) = E(I(T^z > t)).
\end{equation}

In the literature, this quantity is often called the \emph{causal survival curve}, the \emph{counterfactual survival curve} or the \emph{confounder-adjusted survival curve}. We use all of these terms interchangeably. It represents the survival probability that would be observed in the target-population, if every person in the population had received treatment $Z$. The difference or ratio between two treatment-specific counterfactual survival curves is sometimes used to define the \emph{average treatment effect} \parencite{Wang2019}. Since $S_{z}(t)$ refers to the entire target-population, irrespective of the observed treatment status, it may also be considered an average treatment effect by itself. Randomized controlled trials are the gold standard for their estimation. While it is fundamentally impossible to observe more than one potential survival time at once, randomization ensures that the distribution of all other variables does not differ between the treatment groups on average. Therefore, differences between the groups can only be attributed to the treatment itself \parencite{Rubin1974, Rubin1980}.
\par
In order to estimate such an effect without randomization, three assumptions have to be met: the \emph{stable unit treatment value assumption} (the potential survival time of one person is independent of the treatment assignment of other people in the study), the \emph{no unmeasured confounding assumption} (all relevant confounders have been measured) and the \emph{positivity assumption} (every person has a probability greater than 0 and smaller than 1 for receiving treatment $z$). Those are described in detail elsewhere \parencite{Rubin1980, Rosenbaum1983, Neugebauer2005, Guo2015}. Although the methods discussed in this paper are vastly different, they all share these fundamental assumptions. If any one of them is violated, $S_{z}(t)$ is not identifiable.

\section{Overview of Methods} \label{chap:methods}

In this article, we focus strictly on methods that can be used to adjust survival curves for measured baseline confounders, when random right-censoring is present. Methods which are concerned with covariate adjustment in order to increase statistical power only \parencite{Jiang2010}, corrections for covariate-dependent censoring \parencite{Zeng2004}, time-varying confounding \parencite{Clare2019a}, and unmeasured confounders \parencite{Martinez-Camblor2020} are disregarded. The \emph{Average Covariate} method, which entails fitting a Cox model to the data and plugging in the mean of all covariates in order to predict the survival probability for each treatment at a range of time points \parencite{Chang1982}, is also discarded. We choose to do this, because it has been shown repeatedly that this method produces biased estimates \parencite{Nieto1996, Ghali2001}. Multiple methods based on stratification \parencite{Nieto1996, Cupples1995, Gregory1988, Amato1988} are also excluded, because they are only defined for categorical confounders. Additionally, we choose to exclude \emph{Targeted Maximum Likelihood Estimation} based methods \parencite{Moore2009, Stitelman2010, Cai2020}, because they are currently only defined for discrete-time survival data. Artificial discretization of continuous-time survival data is problematic theoretically \parencite{Guerra2020} and has been shown to not work very well in practice \parencite{Sofrygin2019, Westling2021}. Current implementations of these methods are also very computationally complex, making it infeasible to include them in our simulation study.
\par
According to these constraints, the following adjustment methods are included in our simulation study: The \emph{G-Formula}, \emph{Inverse Probability of Treatment Weighting}, \emph{Propensity Score Matching}, \emph{Empirical Likelihood Estimation}, \emph{Augmented Inverse Probability of Treatment Weighting} and their \emph{Pseudo Values} based counterparts. A standard Kaplan-Meier estimator is also included to illustrate the impact of missing confounder-adjustment. An exhaustive description of each method is beyond the scope of this article, but a short review of each included method is given below. More details can be found in the cited literature. Table~\ref{tab::methods_summary} includes a brief summary of the most important aspects of each method and the online appendix includes a step-by-step guide to each method presented here.
\par
All methods mentioned above can be roughly divided into three categories: Methods utilizing the \emph{outcome mechanism}, methods that use the \emph{treatment assignment mechanism} and methods relying on \emph{both} types of mechanisms. In survival analysis, using the outcome mechanism refers to modelling the process which determines the time until the event of interest occurs. Formally, the goal is to use a statistical model to obtain a plug-in estimator of $S(t| Z, X)$ (see equation~\ref{eq::individual_survival_prob}). This model has to take right-censoring into account. The treatment assignment mechanism, on the other hand, describes the process by which an individual $i$ is assigned to one of the $k$ possible treatments. The goal is to estimate the probability of receiving treatment $z$ for each individual, denoted by $P(Z = z | X)$, which is formally known as the \emph{propensity score} \parencite{Rosenbaum1983, Rubin1978}. A statistical model is usually necessary to estimate this probability. How exactly the outcome and treatment models are used to calculate the causal survival curves is described below.

\subsection{G-Formula} \label{chap::methods_g-formula}

One well known method to adjust for confounders is the G-Formula, also known as G-computation, direct standardization, or corrected-group-prognosis method in epidemiology \parencite{Makuch1982, Chang1982, Robins1986}. Here, the confounders are adjusted for by correctly modelling the outcome mechanism. The Cox proportional hazards model \parencite{Cox1972} in conjunction with an estimate of the baseline-hazard function \parencite{Breslow1972} is usually used. However, any model allowing predictions for the conditional survival probability, given covariates and a point in time, may be used as well \parencite{Robins1986}. After the model has been estimated, it is used to make predictions for $S(t| Z=z, X=x_i)$ under each possible treatment for each individual. If the estimated conditional survival probabilities are unbiased, the resulting arithmetic mean of these predictions over all individuals is an unbiased estimate for the counterfactual survival probability at time $t$.
\par
For survival data, this method was first proposed independently by \textcite{Makuch1982} and \textcite{Chang1982}, using a simple Cox model. Other authors have used different models as well \parencite{Zhang2007, Chatton2021a}, but we will only consider the Cox model in this article. This method has been shown to outperform other methods in terms of efficiency, especially when strong predictors of the outcome are included in the model \parencite{Ozenne2020, Chatton2020}.

\subsection{Inverse Probability of Treatment Weighting (IPTW)}

Inverse Probability of Treatment Weighting (IPTW) is one of multiple methods utilizing the treatment assignment mechanism for confounder-adjustment \parencite{Cole2004, Xie2005}. First, the propensity scores are estimated for each individual and each treatment. Afterwards, the inverse of this probability is calculated. By using these values as weights in the analysis, the confounding is removed and unbiased estimates of the actual causal effect can be obtained. In practice, the propensity score is usually estimated using a logistic regression model, but other more sophisticated methods exist as well \parencite{McCaffrey2004, Imai2014}.
\par
For survival analysis, two very similar estimators of the causal survival curve based on this method have been proposed \parencite{Cole2004, Xie2005}. The estimator of \textcite{Cole2004} (IPTW HZ) is equivalent to fitting a weighted stratified Cox model, using the treatment indicator as stratification variable. The inverse probability weights are used as case weights. \textcite{Xie2005} on the other hand proposed a directly weighted Kaplan-Meier estimator (IPTW KM).
\par
Previous research on IPTW in the estimation of sample means has shown repeatedly that IPTW is less efficient than the G-Formula when both methods are used appropriately \parencite{Chatton2020a}. Efficiency is judged by the size of the standard errors in this context. This difference is particularly strong when the estimated weights are highly variable, close to 0 or extremely high, which often happens in small sample sizes \parencite{Raad2020}. Similar results have recently been reported for point estimates of the survival curve \parencite{Chatton2020}.

\subsection{Propensity Score Matching}

Another well known method is propensity score matching. Instead of using the propensity scores to construct weights for each observation, a new sample is constructed by matching patients with similar propensity scores. The new matched sample can subsequently be analysed using standard methods. Slightly different methods for creating adjusted survival curves from matched data have been proposed \parencite{Winnett2002, Galimberti2002, Austin2014, Austin2020}. We roughly follow the method described in Austin \parencite{Austin2014} and use the following three steps: (1) estimation of the propensity score, (2) matching patients with similar scores and (3) using a simple stratified Kaplan-Meier estimator on the matched sample. We used the algorithm of \textcite{Sekhon2011} with an euclidean distance and allowing both replacement and ties. If the propensity scores are correctly estimated and the algorithm used for matching the patients is appropriate, the resulting estimates of the survival curves are unbiased \parencite{Austin2014}. Propensity Score Matching has, however, been shown to be less efficient than IPTW and the G-Formula in other aspects of the analysis of time-to-event data \parencite{Borgne2016, Austin2016}.

\subsection{Augmented Inverse Probability of Treatment Weighting}

A different approach, called locally efficient augmented inverse probability weighting (AIPTW), was first proposed by \textcite{Robins1992} and \textcite{Hubbard2000} and further developed by different authors in the present context \parencite{Bai2013, Zhang2012, Ozenne2020}. Instead of using a single treatment assignment or outcome model, this method requires \emph{both} kinds of models. In simplified terms, the AIPTW methodology works by using the G-Formula estimate to augment the IPTW estimate, in order to make it more efficient. Essentially, it is just the IPTW estimator with the conditional survival predictions under each treatment added to it, after weighting them using the propensity score. The specific form of the estimating equation ensures that it is asymptotically unbiased if \emph{either} of the two models is correctly specified, making it \emph{doubly-robust}. This property is the main advantage of this method.
\par
It has been shown previously that this method is asymptotically at least as efficient as IPTW, when both models are correctly specified \parencite{Bai2013, Ozenne2020}. Research on using AIPTW for the estimation of biased sample means also indicates, however, that the method looses efficiency if both models are slightly incorrectly specified \parencite{Kang2007}. Because point estimates are made without any global constraints, there is no guarantee that the estimates will be monotonically decreasing, or that the estimates will be between 0 and 1. How often these problems occur in practice is currently not clear.

\subsection{G-Formula + IPTW}

Another possible way to combine the G-Formula estimator with the IPTW approach has recently been proposed by \textcite{Chatton2021a}, based on earlier work by \textcite{Vansteelandt2011}. This method works as follows. First, the inverse probability of treatment weights are estimated. Those weights are then used in the estimation of an outcome model. The outcome model should also include the relevant confounders as covariates. This model can then be used to calculate standard G-Formula estimates as described in section~\ref{chap::methods_g-formula} \parencite{Chatton2021a}. As in the usual G-Formula approach, any outcome model could theoretically be used, but we only consider a Cox model in this article.
\par
Although \textcite{Chatton2021a} performed a well designed Monte-Carlo simulation, where this method showed the doubly-robust property when estimating the Hazard-Ratio, they offer no mathematical proof. In addition, they did not study the estimation of the adjusted survival curve directly. Instead they focused on two other useful quantities, the Hazard-Ratio and the restricted mean survival time. It is unclear whether their results extend to the given context. We therefore use the name \emph{G-Formula IPTW} instead of \emph{Doubly-Robust Standardization}, which is the name given by the original authors.

\subsection{Empirical Likelihood Estimation (EL)}

Recently, \textcite{Wang2019} have proposed an estimator of the survival function, which is based on the Empirical Likelihood (EL) estimation methodology. Stated simply, EL is a likelihood based method, which does not require the assumption that the data was generated by any known family of distributions. It is a model free approach that works by forcing the moments of the covariates $X$ to be equal between treatment groups, through the maximization of a constrained likelihood function. The resulting equality of the distributions removes the bias created by the confounders. No specification of either outcome model or treatment allocation model is necessary. Due to the constraints in the optimization process, this method is guaranteed to produce non-increasing estimates lying in the probability bounds.
\par
Theoretical results and simulation studies of \textcite{Wang2019} and \textcite{Lee2022b} indicate that this method also shares the doubly-robust property. It has been demonstrated previously, that EL can outperform IPTW in terms of variance in some scenarios \parencite{Owen2001, Wang2019}. The method described in \textcite{Wang2019} itself is less efficient than the standard AIPTW estimator, but can be modified to increase its efficiency using the method described by \textcite{Lee2022b}. In this article we focus strictly on the version described in the original article by \textcite{Wang2019} because of code availability.

\subsection{Pseudo Values (PV)}

An alternative approach is the use of pseudo values (PV) \parencite{Andersen2017, Andersen2010}. The general idea is that one could use standard methods, such as generalized linear models, to model the outcome mechanism, if $T$ was known for each individual $i$. As described in section~\ref{chap::notation} this is usually not the case in practice. Generally, only $T_{obs}$ is observed because of right-censoring. Let $\hat{S}(t)$ be the standard Kaplan-Meier estimator. The PV for individual $i$ at time $t$ is then defined as:

\begin{equation}
	\hat{\theta}_i(t) = n \hat{S}(t) - (n - 1) \hat{S}^{-i}(t),
\end{equation}

where $\hat{S}^{-i}(t)$ is the Kaplan-Meier estimator applied to the data of size $n - 1$, where the $i$-th observation has been removed. The PV for individual $i$ can be interpreted as the contribution of the individual $i$ to the target estimate from a complete sample of size $n$ without censoring \parencite{Andersen2010, Zeng2021}.
\par 
After the PVs are estimated for each person at a fixed set of points in time, they can be used to construct G-Formula, IPTW or AIPTW estimates of the survival curve \parencite{Andersen2017, Wang2018, Zeng2021}. For the G-Formula estimate, a generalised estimating equation (GEE) model can be fit, using the PVs as response variable and all baseline covariates as independent variables for each time point. This model can then be used to obtain the required conditional survival probability predictions \parencite{Klein2008, Overgaard2017}. For the IPTW estimate, a simple propensity score weighted average of the PVs can be used \parencite{Andersen2017}. The AIPTW estimate based on PV can be obtained in a similar fashion \parencite{Wang2018, Zeng2021}. Little is known about the efficiency of these estimators compared to the non-PV based counterparts. Regardless which method based on PV is used, there is no guarantee that the survival probability will be monotonically decreasing over time. Isotonic regression can be used to correct these errors after estimating the curves \parencite{Westling2020}. The impact of using this correction method will also be assessed in the simulation study.

\begin{sidewaystable}[!htb]
	\caption{Summary of the qualitative properties of each method}
	\centering
	\label{tab::methods_summary}
	\begin{tabular}{lccccccc}
		\toprule
		\multirow{3}{*}{\textbf{Method}} & \textbf{Requires an} & \textbf{Requires a} & \textbf{Doubly} & \textbf{More than two} & \textbf{Monotone} & \textbf{Bounded} & \textbf{SE} \\
		& \textbf{Outcome} & \textbf{Treatment} & \textbf{Robust} & \textbf{Treatments} & \textbf{Estimates} & \textbf{Estimates} & \textbf{formula} \\
		& \textbf{Model} & \textbf{Model} & & \textbf{allowed} & & & \\
		\midrule
		G-Formula & \cmark & \xmark & \xmark & (\cmark)$^a$ & (\cmark)$^a$ & (\cmark)$^a$ & \cmark \\
		G-Formula PV & \cmark & \xmark & \xmark & \cmark & \xmark & \cmark & \xmark \\
		IPTW KM & \xmark & \cmark & \xmark & \cmark & \cmark & \cmark & \cmark \\
		IPTW HZ & \xmark & \cmark & \xmark & \cmark & \cmark & \cmark & \xmark \\
		IPTW PV & \xmark & \cmark & \xmark & \cmark & \xmark & \xmark & \cmark \\
		Matching & \xmark & \cmark & \xmark & \cmark & \cmark & \cmark & \xmark \\
		EL & \xmark & \xmark & \cmark & (\xmark)$^b$ & \cmark & \cmark & \xmark \\
		AIPTW & \cmark & \cmark & \cmark & \cmark & \xmark & \xmark & \cmark \\
		AIPTW PV & \cmark & \cmark & \cmark & \cmark & \xmark & \xmark & \cmark \\
		G-Formula IPTW & \cmark & \cmark & \textbf{?}$^c$ & (\cmark)$^a$ & (\cmark)$^a$ & (\cmark)$^a$ & \xmark \\
		\bottomrule
	\end{tabular}
	\begin{tablenotes}
		\item $^a$ This is true when using a Cox model, but might vary otherwise.
		\item $^b$ It has not been proposed in the literature, but can be extended to this case fairly easily.
		\item $^c$ This property is currently unclear.
	\end{tablenotes}
\end{sidewaystable}

\FloatBarrier

\section{Illustrative Example}

To illustrate the use of the methods described above we consider the estimation of counterfactual survival curves of smokers versus non-smokers. It is well known that smoking tobacco increases the probability of developing certain types of cancer and a multitude of other illnesses, thereby reducing the average life expectancy of smokers \parencite{Lee2012, Carter2015}. The total causal effect of smoking on the survival time is however difficult to assess, because it is impossible to conduct randomized trials due to ethical and administrative reasons. This has lead to a lot of discussion about confounding factors, especially in the late 1950s and early 1960s when the biological mechanisms that link smoking to a higher risk of cancer were not well known \parencite{Pearl2018}. By now it is known that age \parencite{Li2016}, gender \parencite{Gutterman2015} and the level of education \parencite{Schnohr2004} are some of the confounders that need to be accounted for when estimating the total causal effect of smoking on the survival time. These factors tend to increase the propensity to smoke and decrease the survival time. Crude Kaplan-Meier curves stratified by smoking status therefore usually show an \emph{overestimate} of the true causal effect, meaning that the curves are too far apart.
\par
We illustrate this using data from the German Epidemiological Trial on Ankle-Brachial-Index (getABI), which is a prospective observational cohort study including a total of 6880 primary care patients aged 65 or older. At the beginning of the trial in 2001, an examination was performed on every included patient. Data on both smoking status and the relevant confounders listed above was collected during this examination. In total, 3687 patients reported to have never smoked and 3134 patients had smoked in the past or are current smokers. Follow-Up information on these patients was collected for up to 7 years after the baseline examination. Detailed information about the design of the trial and the obtained results can be found elsewhere \parencite{StudyGroup2002, Diehm2004}.
\par
Figure~1 shows the crude and adjusted survival curves for people who have never smoked versus people who are current smokers or have smoked in the past. We used every method mentioned in section~\ref{chap:methods} to adjust the curves for age, gender and educational status. We used a Cox regression to model the survival time and a logistic regression to model the treatment assignment. Regardless of the method, it can clearly be seen that the adjusted survival curves are closer to each other than the crude Kaplan-Meier curves. Nonetheless, the survival curve of never-smokers is still above smokers in the entire interval, suggesting that the effect of smoking on the survival time is not entirely due to the considered confounders. The differences between the adjustment methods are rather small in this case. Additional information about the estimation of these curves, including checks of the proportional hazards assumption and positivity violations are given in the online appendix.
\par

\begin{figure}[!htb]
	\centering
	\includegraphics[width=1\linewidth]{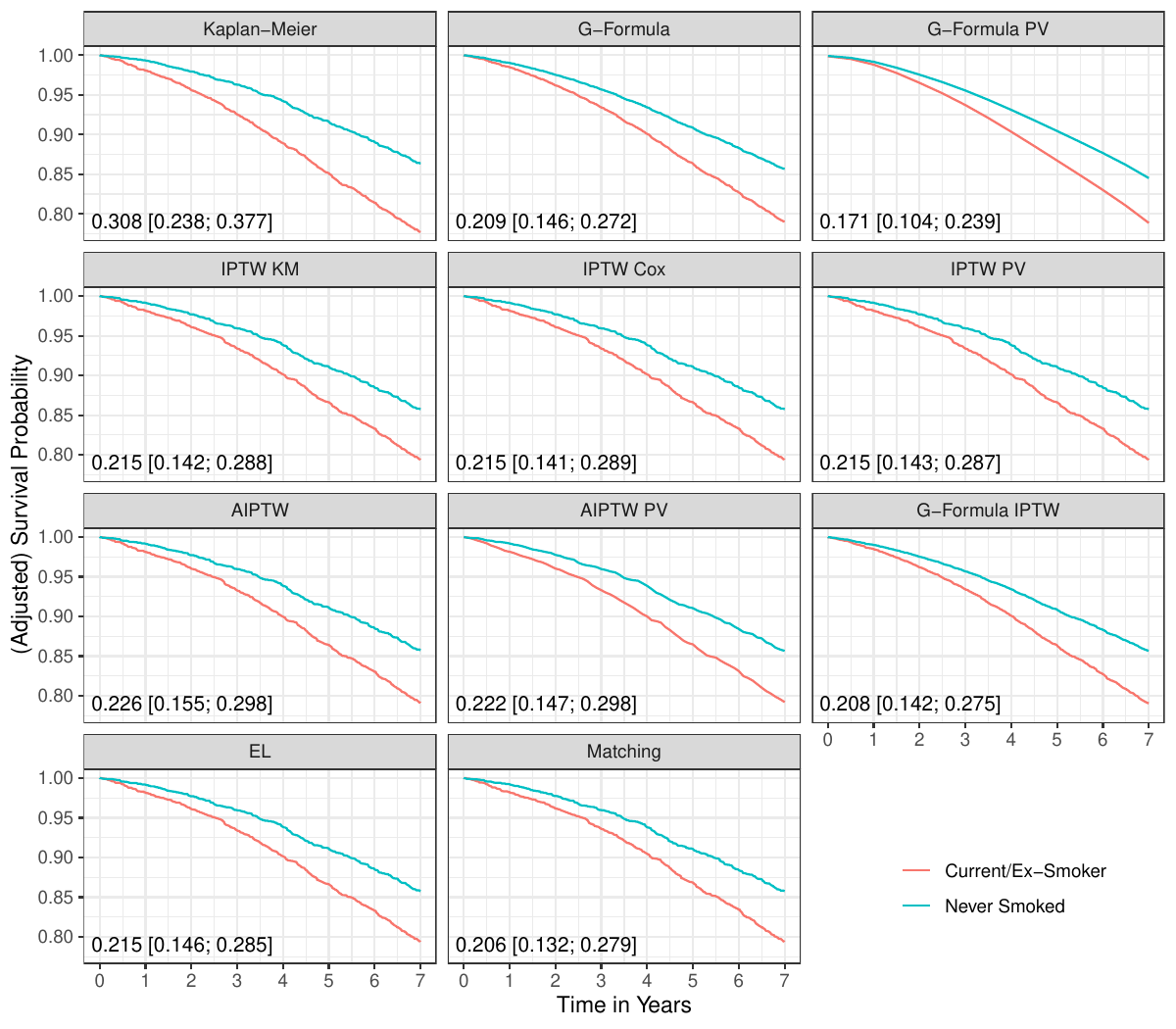}
	\caption{Crude and adjusted survival curves for people who smoked in the past or smoke currently versus people that never smoked. The numbers on the left of each facet are the area between the curves in the interval $[0, 7]$ years and its associated 95\% bootstrap confidence interval, calculated using $300$ bootstrap replications and numerical integration.}
	\label{fig::getABI_smoking}
\end{figure}

\FloatBarrier

\section{Simulation Study} \label{chap::simulation_design}

\subsection{Data Generation and Procedure}

The simulation is designed as a mixture of the approaches of \textcite{Austin2020} and \textcite{Chatton2020}. A causal diagram showing all causal pathways and associated coefficients is displayed in figure~2. The following five steps are executed:

\begin{description}
	\item[Step 1] First, a \emph{super-population} of one million people is generated. The distributions and causal coefficients used are listed below.
	\begin{description}
		\item[A] Four independent and identically distributed variables ($X_1, X_3, X_4, X_6$) are generated for each person.
		\item[B] Two further variables ($X_2$, $X_5$) are generated, where $X_2$ is partially caused by $X_3$ and $X_5$ is partially caused by $X_6$.
		\item[C] Based on those six variables and the causal coefficients listed below two survival times are simulated for each individual (using $Z$, $X_1$, $X_2$, $X_4$ with linear effects and $X_5$ with a quadratic effect). One for $Z = 0$ (control group) and one for $Z = 1$ (treatment group), using the same stream of random numbers. The result is a population of individuals in which both potential survival times are known. To generate the survival times we use the method of \textcite{Bender2005} with a Weibull distribution ($\lambda = 2, \gamma = 1.8$). 
		\item[D] Following \textcite{Chatton2020}, the probability of receiving the treatment $P(Z = 1|X)$ is determined by a logistic regression model, using the covariates $X_2$, $X_3$, $X_5$ and $X_6$ and the causal coefficients listed below, where $X_2$ has a quadratic effect on the probability. The intercept is set to $-1.2$, which results in approximately 50\% of all individuals receiving the treatment.
		\item[E] The true counterfactual survival curves $S_z(t)$ are calculated using simple proportions: \begin{equation}
			S_z(t) = \frac{1}{1000000} \sum_{i}^{1000000} I(t_{iz} > t),
		\end{equation}
		where $t_{iz}$ is the potential survival time of individual $i$ given treatment $z$. Those are displayed in the online appendix.
	\end{description}
	\item[Step 2] A simple random sample without replacement is drawn from the super-population.
	\item[Step 3] For each person in the sample, the treatment status $z_i$ is generated using a Bernoulli trial using the previously estimated probability of receiving treatment. Only the survival time corresponding to the drawn treatment status is kept in the sample.
	\item[Step 4] Random right-censoring is introduced using another Weibull distribution ($\lambda = 1, \gamma = 2$), resulting in $22.90$\% of right-censored individuals on average.
	\item[Step 5] All considered methods are used to estimate the treatment-specific causal survival probability for each treatment level at a fine grid of points in time (37 equally spaced points from 0.05 to 1.5).
\end{description}

Steps two to five are repeated numerous times in different simulation scenarios, which are described in more detail below. Since the probability of treatment allocation is dependent on the variables $X_2$ and $X_5$, which also have a direct causal effect on the outcome, the causal effect of the treatment on the survival time is confounded. Failing to adjust for this confounding would lead to biased estimates of the treatment specific survival curves. The causal structure of the samples created by carrying out the procedure described above is displayed in figure~2.
\par

\begin{figure}[!htb]
	\centering
	\includegraphics[width=1\linewidth]{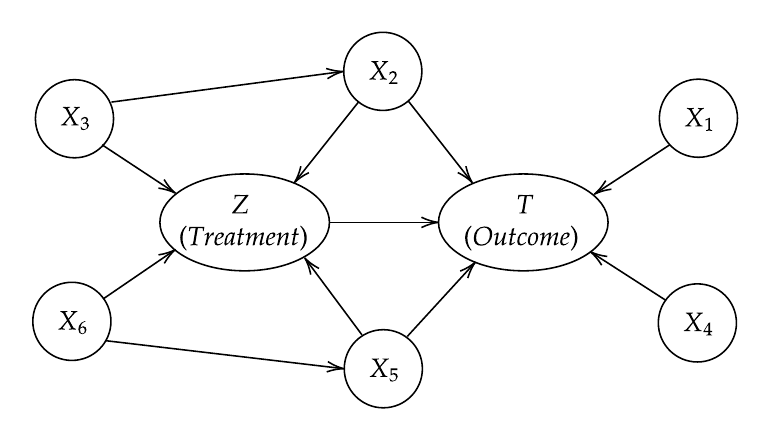}
	\caption{Causal diagram of the data generation mechanism used in the simulation. Adapted from \textcite{Chatton2020}.}
	\label{fig::causal_diagram}
\end{figure}

We used the following distributions and causal coefficients to generate the samples:

\begin{align*} 
	X_1 &\sim Bernoulli(0.5) \\ 
	X_2 &\sim Bernoulli(0.3 + X_3 \cdot 0.1) \\
	X_3 &\sim Bernoulli(0.5) \\
	X_4 &\sim N(0, 1) \\
	X_5 &\sim 0.3 + X_6 \cdot 0.1 + N(0, 1) \\
	X_6 &\sim N(0, 1) \\
	Z &\sim  Bernoulli\left(\frac{1}{1 + \exp(-(-1.2 + \log(3) \cdot X_2^2 + \log(1.5) \cdot X_3 + \log(1.5) \cdot X_5 + \log(2) \cdot X_6))}\right) \\
	T &\sim  \left(-\frac{\log\left(U(0, 1)\right)}{\exp(\log(1.8) \cdot X_1 + \log(1.8) \cdot X_2 + \log(1.8) \cdot X_4 + \log(2.3) \cdot X_5^2 - 1 \cdot Z)}\right)^{0.5} \\
\end{align*}

where $Bernoulli()$ corresponds to simple Bernoulli trials, $N(0, 1)$ is the standard normal distribution and $U(0, 1)$ is a uniform distribution of numbers between 0 and 1. All calculations were performed using the \textbf{R} programming language (version 4.2.1) with the \texttt{adjustedCurves} package available on CRAN \parencite{Denz2022a}, which was developed by the authors to facilitate the use of adjusted survival curves in practice. The code used is available in the online appendix.

\subsection{Scenarios}

We consider five main scenarios, in which different correctly and incorrectly specified models for the treatment assignment and outcome mechanism are used. In all of these scenarios the variables $X_1, X_2, X_4, X_5$ and $Z$ are used as independent variables to model the outcome mechanism and the variables $X_2$ and $X_5$ are used to model the treatment-assignment mechanism. We choose to include all outcome predictors in the outcome model to prevent the problem of unobserved heterogeneity. If, for example, $X_1$ was omitted from the model, this could lead to a violation of the proportional hazards assumption when using the Cox model and therefore lead to biased estimates, even if the confounders $X_2$ and $X_5$ were correctly adjusted for \parencite{Martinussen2013}. The only difference between the scenarios is whether the quadratic effects of $X_5$ on the outcome and of $X_2$ on the treatment-assignment was specified correctly:

\begin{enumerate}
	\item \textbf{Correct Outcome Mechanism \& Correct Treatment Assignment} (CO \& CT): $X_2$ is modelled as a quadratic effect on the treatment-assignment and as a linear effect on the outcome. $X_5$ is modelled as a linear effect on the treatment-assignment and as a quadratic effect on the outcome.
	\item \textbf{Correct Outcome Mechanism \& Incorrect Treatment Assignment} (CO \& ICT): $X_2$ is modelled as a linear effect on both the treatment-assignment and the outcome. $X_5$ is modelled as a quadratic effect on both the treatment-assignment and the outcome.
	\item \textbf{Incorrect Outcome Mechanism \& Correct Treatment Assignment} (ICO \& CT): $X_2$ is modelled as a quadratic effect on both the treatment-assignment and the outcome. $X_5$ is modelled as a linear effect on both the treatment-assignment and the outcome.
	\item \textbf{Incorrect Outcome Mechanism \& Incorrect Treatment Assignment} (ICO \& ICT): $X_2$ is modelled as a linear effect on the treatment-assignment and as a quadratic effect on the outcome. $X_5$ is modelled as a quadratic effect on the treatment-assignment and as a linear effect on the outcome.
	\item \textbf{Partially Correct Outcome Mechanism \& Partially Correct Treatment Assignment} (PCO \& PCT):  Both $X_2$ and $X_5$ are modelled as linear effects in both models.
\end{enumerate}

\par
Simple logistic regression models are used for methods relying on a treatment assignment model (IPTW KM, IPTW HZ, IPTW PV, PS Matching, AIPTW, AIPTW PV, G-Formula IPTW). Cox models are used for methods relying on an outcome model which takes right-censoring into account (G-Formula, AIPTW, G-Formula IPTW) and a GEE model is used in the corresponding PV based methods (G-Formula PV, AIPTW PV). The EL method does not utilize any models and therefore only received the corresponding raw design matrix as input. For example, in the CO \& CT scenario the EL method received a matrix containing the variables $X_1, X_2, X_2^2, X_4, X_5, X_5^2$ whereas it only received the variables $X_1, X_2, X_4, X_5, X_5^2$ in scenario CO \& ICT. For all five of the scenarios mentioned above the sample size is varied systematically. In doing so, a broad range of realistic scenarios is covered, allowing detailed judgments about the performance of each method. The online appendix additionally includes further simulation scenarios, in which the covariate sets used were varied.

\subsection{Performance Criteria} \label{chap::performance}

All estimates are compared to the true survival curves. In contrast to previous studies, our main interest is to estimate the performance of the estimators concerning the entire survival curves \parencite{Wang2019, Wang2018, Cai2020, Chatton2020}. Usually, the bias is defined as $E(S_{z}(t) - \hat{S}_{z}(t))$ for a particular point in time $t$. We remove this time dependency by defining the generalized bias in group $z$ as:

\begin{equation}
	G_{Bias}(z) = \int_{0}^{\infty} E\left(S_{z}(t) - \hat{S}_{z}(t)\right) dt = E\left(\int_{0}^{\infty} S_{z}(t) - \hat{S}_{z}(t) dt\right),
\end{equation}

where $\hat{S}_{z}(t)$ is the estimated survival function for treatment $Z = z$ and $S_{z}(t)$ is the true survival function for $Z = z$. To estimate the integral for a particular simulation run we use:

\begin{equation}
	\hat{\Delta}_{Bias}(z) = \int_{0}^{\tau} \left(S_{z}(t) - \hat{S}_{z}(t)\right) dt,
\end{equation}

with $\tau$ being defined as $\min\left(t_{max}, Q_{95}\left(S_{z}(t)\right)\right)$, where $t_{max}$ is the latest point in time at which an event occurred in group $z$ and $Q_{95}\left(S_{z}(t)\right)$ is the 95\% quantile of the true survival times. Using $t_{max}$ is necessary to ensure a fair comparison, because some estimators are only valid up to this point \parencite{Cole2004, Xie2005}.
\par
If $\hat{S}_{z}(t)$ is an unbiased estimator of $S_{z}(t)$, $G_{Bias}(z)$ is zero, and the average of $\hat{\Delta}_{Bias}(z)$ will tend to zero as the number of simulations increases. This quantity has previously been utilized to construct hypothesis tests to formally test if two survival functions are different in a given interval \parencite{Pepe1989, Pepe1991a, Zhao2012}. The arithmetic mean of $\hat{\Delta}_{Bias}(z)$ over all simulation repetitions is used as an estimate of the bias overall ($\hat{G}_{Bias}(z)$). The goodness-of-fit can be similarly estimated, by utilizing the generalized mean-squared-error of the estimated survival curve and the true survival curves, defined as \parencite{Klein1989}:

\begin{equation}
	G_{MSE}(z) = \int_{0}^{\infty} E\left(\left(S_{z}(t) - \hat{S}_{z}(t)\right)^2\right) dt = E\left(\int_{0}^{\infty} \left(S_{z}(t) - \hat{S}_{z}(t)\right)^2 dt\right).
\end{equation}

The integral for a particular simulation run is also similarly estimated using:

\begin{equation}
	\hat{\Delta}_{MSE}(z) = \int_{0}^{\tau} \left(S_{z}(t) - \hat{S}_{z}(t)\right)^2 dt.
\end{equation}

The arithmetic mean of $\hat{\Delta}_{MSE}(z)$ over all simulation repetitions is used as an estimator for the real $G_{MSE}(z)$. Both $\hat{G}_{Bias}(z)$ and $\hat{G}_{MSE}(z)$ are reported with associated Monte-Carlo simulation errors, estimated as the standard error of each quantity over all simulation repetitions. Additionally, the percentage of survival curves that are not monotonically decreasing as well as the percentage of estimated survival curves containing at least one point estimate falling outside of the probability bounds of 0 and 1 will be reported. For methods with existing approximate standard error equations, the coverage of point-wise confidence intervals and their average width are also given.

\subsection{Results}

Figure~3 shows the simulated distributions of $\hat{\Delta}_{Bias}(z=1)$ at different sample sizes in the scenarios described above. Results for the control group ($\hat{\Delta}_{Bias}(z=0)$) are very similar and can be seen alongside values for $\hat{G}_{Bias}(z)$ and Monte-Carlo errors in the online appendix. As expected, the simple Kaplan-Meier estimator is biased regardless of the sample size and scenario. When both the outcome mechanism and the treatment assignment are modelled correctly, all adjustment methods show negligible amounts of bias in large samples. Methods based on Cox models, however, show systematic bias in small samples, even when the model is correctly specified. The doubly-robust property of the two AIPTW based methods can also be seen clearly from the graph. Regardless of the sample size, when at least one of the utilized models is correctly specified, the estimators remain unbiased. Interestingly, this property also extends to the PCO \& PCT scenario, where the treatment-assignment model fails to include the quadratic effect of $X_2$ and the outcome model fails to include the quadratic effect of $X_5$. The current implementation of the EL method does not show this property. A small bias remains whenever the treatment-assignment model is incorrectly specified.
\par
$\hat{\Delta}_{Bias}(z=1)$ tends to zero for the G-Formula and the G-Formula IPTW methods when using a incorrectly specified outcome model and when both models are incorrectly specified. Therefore, these methods could be considered ``unbiased'' with respect to the definition of $\hat{\Delta}_{Bias}(z=1)$ in these scenarios. This counter-intuitive result can be explained by the fact that they happen to both overestimate and underestimate the counterfactual survival probability at different points in time, cancelling out their bias. This bias can be seen more clearly in figure~4, in which the simple average bias over time is displayed for each method and scenario for $Z = 1$ and $n = 1000$. In this figure we can also see that the G-Formula IPTW method is biased in the ICO \& CT scenario as well, indicating that it does not have the doubly-robust property. All methods show systematic bias when both the outcome model and the treatment model are misspecified. More graphics displaying the bias over time can be found in the appendix.
\par
\begin{figure}[!htb]
	\centering
	\includegraphics[width=1\linewidth]{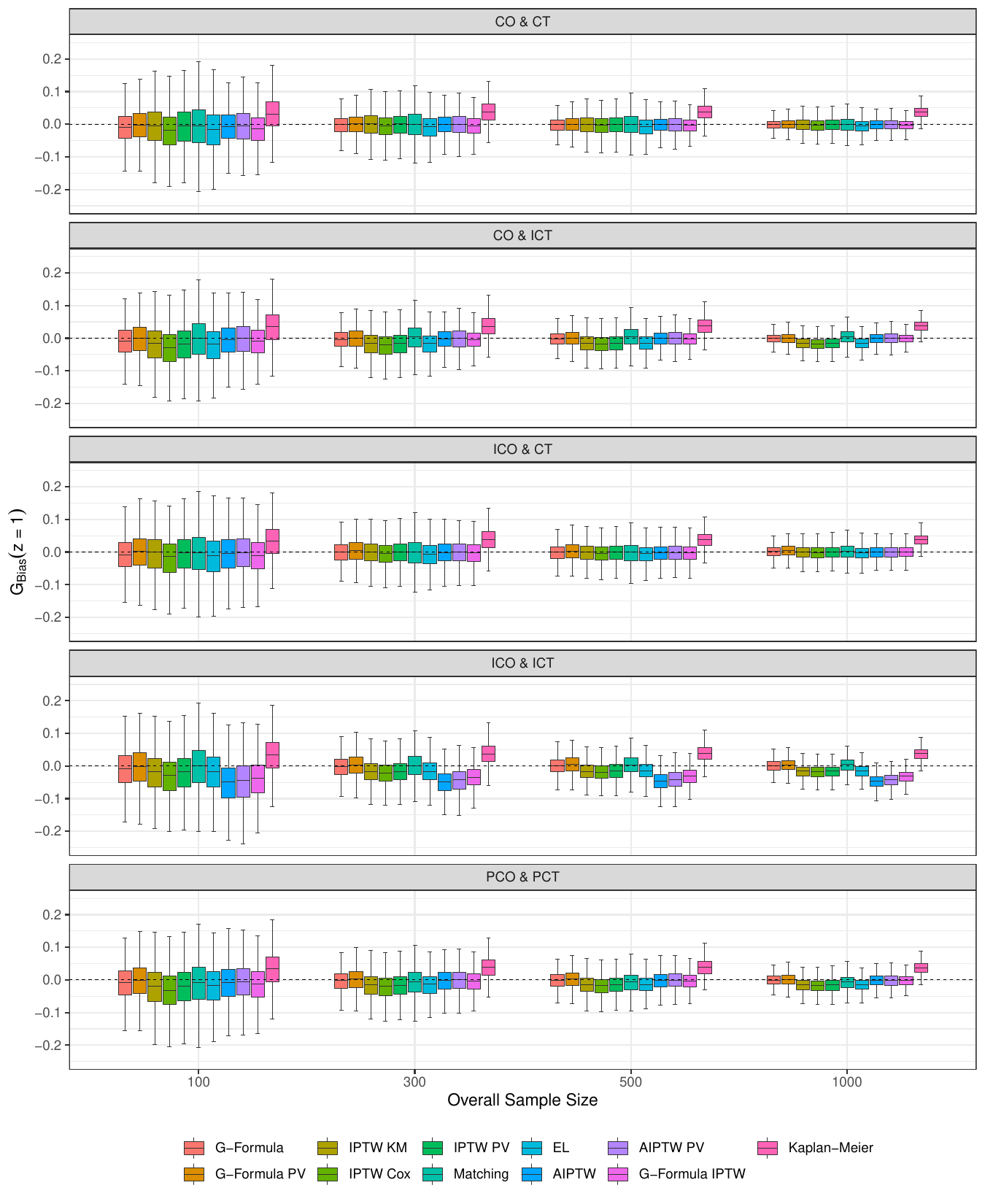}
	\caption{Distributions of $\Delta_{Bias}(z = 1)$ (treatment group) for all methods in each simulation scenario with varying sample sizes. Outliers are not shown. Estimates are based on 2000 simulation repetitions.}
	\label{fig::bias_boxplots}
\end{figure}

\begin{figure}[!htb]
	\centering
	\includegraphics[width=1\linewidth]{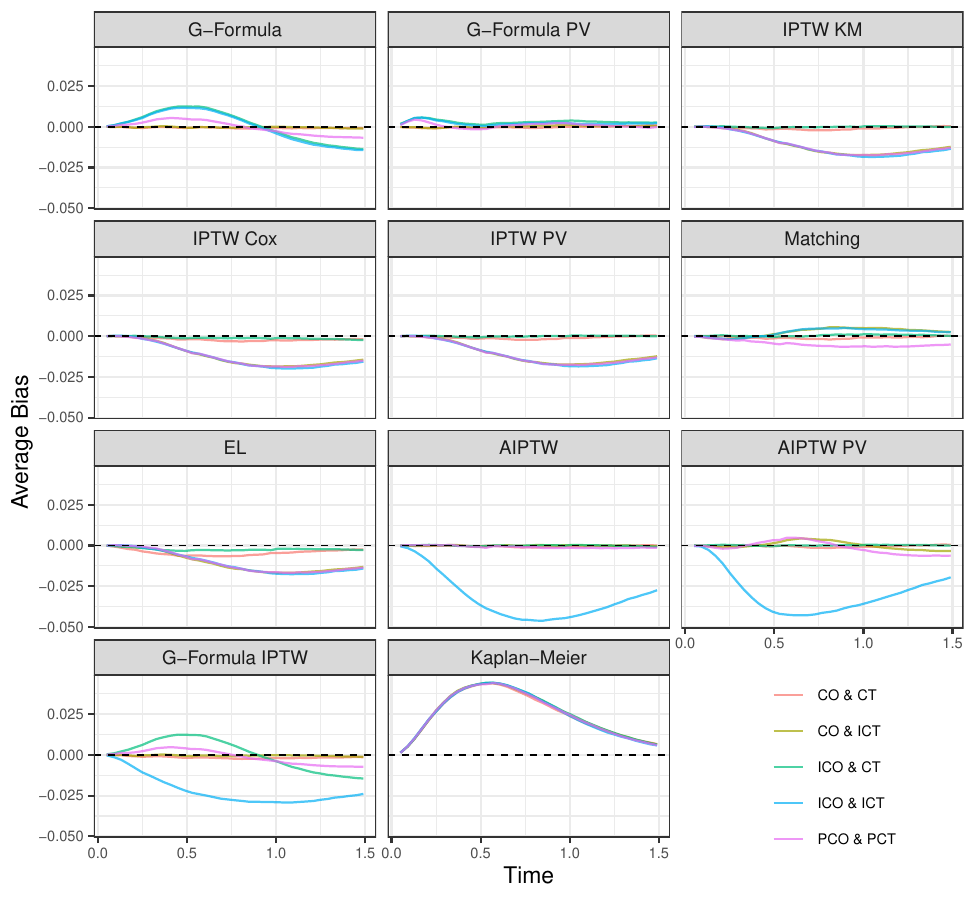}
	\caption{The average bias, defined as the arithmetic mean of the difference between the true survival probability and the estimates survival probability, over time for all simulation scenarios and all methods with $n = 1000$ and $Z = 1$. Estimates are based on $2000$ simulation repetitions.}
	\label{fig::bias_over_time}
\end{figure}

In figure~5 a graph similar to figure~3 for the goodness-of-fit in the treatment group is displayed. The same graph for the control group and values for $\hat{\Delta}_{MSE}(z)$ are shown in the appendix. Regardless of the simulation scenario and sample size, Matching shows the worst goodness-of-fit. On the opposite end of the spectrum, the G-Formula with a simple Cox model shows the least amount of variation, closely followed by the G-Formula using PV and a GEE model. This is true even when the Cox models are incorrectly specified. AIPTW based methods show similar amounts of variance as the IPTW based methods when only the treatment model is correctly specified, and outperform the IPTW based methods when the outcome model is also correctly specified. Empirical Likelihood Estimation has goodness-of-fit values that are very similar to the three IPTW methods. The sample size has no effect on the ranking of the methods with respect to $\hat{\Delta}_{MSE}(z)$ in any scenario.
\par
\begin{figure}[!htb]
	\centering
	\includegraphics[width=1\linewidth]{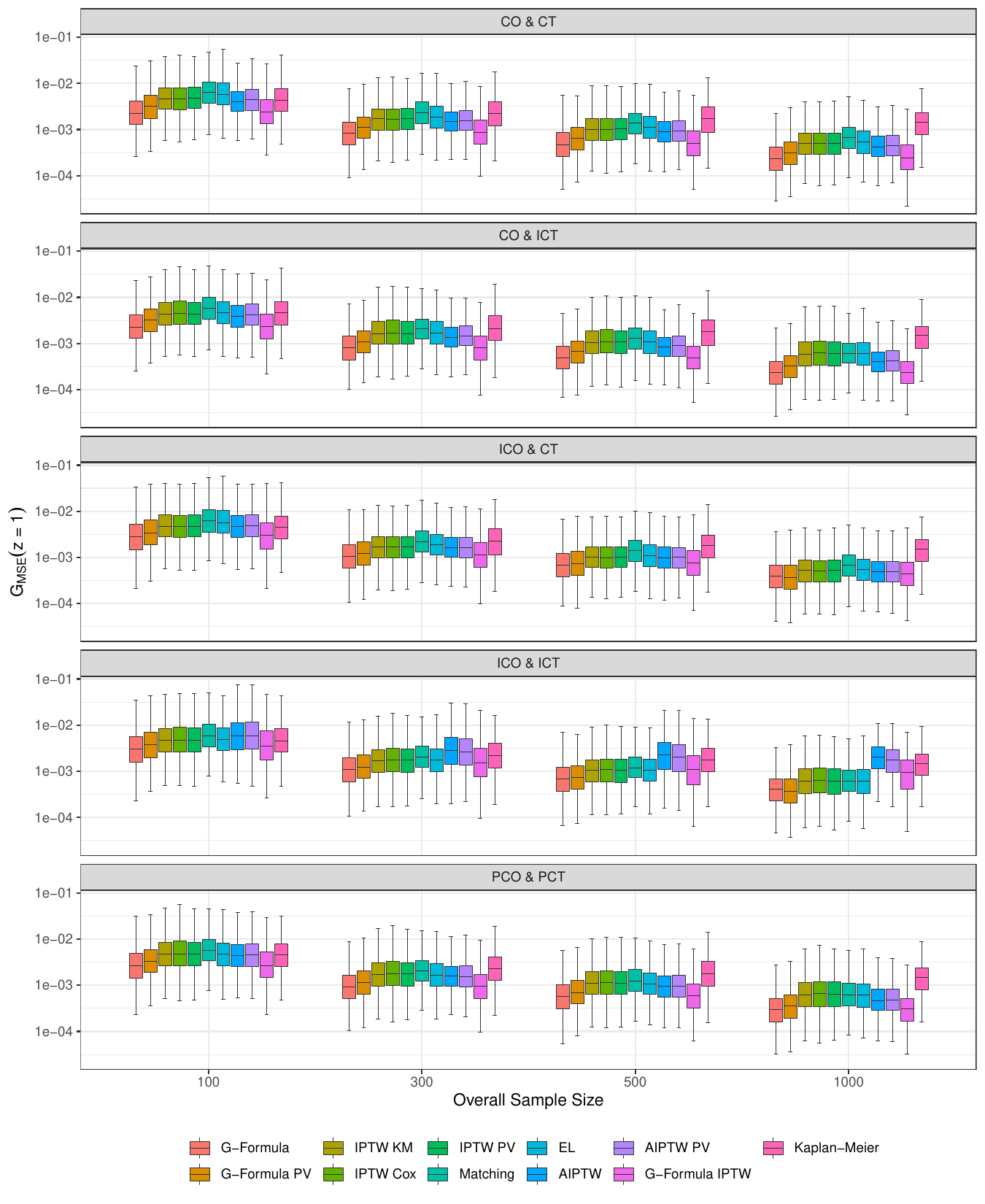}
	\caption{Distributions of $\Delta_{MSE}(z = 1)$ (treatment group) on the log-scale for all methods in each simulation scenario with varying sample sizes. Outliers are not shown. Estimates are based on 2000 simulation repetitions.}
	\label{fig::mse_boxplots}
\end{figure}

The percentages of estimated survival curves which are not monotonically decreasing over time (NM), or which contain at least one estimate outside of the 0 and 1 probability bounds (OOB), is displayed in table~\ref{tab::monotone}. The table only contains methods that are suspect to these problems (see table~\ref{tab::methods_summary}). OOB errors occur in roughly the same frequency in the three susceptible methods. While it seems to be a very frequent problem in small sample sizes, the effect diminishes as $n$ increases. A similar trend can be observed for the percentage of NM survival curves. The G-Formula method based on PV, however, is the least affected by this problem. Nonetheless, all four methods do exhibit some amount of these errors.
\par

\begin{table}[!htb]
	\centering
	\caption{Percentages of estimated survival curves in the simulation which are not strictly monotonically decreasing or which contain at least one estimate out of bounds}
	\label{tab::monotone}
	\begin{tabular}{rrcccccccccccc}
		\toprule
		\multirow{2}{*}{$n$} & \multirow{2}{*}{$Z$} & & \multicolumn{2}{l}{\textbf{G-Formula PV}} & & \multicolumn{2}{c}{\textbf{IPTW PV}} & & \multicolumn{2}{c}{\textbf{AIPTW}} & & \multicolumn{2}{c}{\textbf{AIPTW PV}} \\
		& & & OOB & NM & & OOB & NM & & OOB & NM & & OOB & NM \\
		\midrule
		100  & 0 & & 0.00 & 13.10 & & 14.00 & 75.75 & & 10.35 & 74.20 & & 21.85 & 77.00 \\
		100  & 1 & & 0.00 & 17.35 & & 20.90 & 81.35 & & 19.70 & 80.95 & & 41.10 & 85.05 \\
		300  & 0 & & 0.00 & 2.90  & & 11.85 & 69.40 & & 8.25 & 70.60 & & 14.45 & 70.55 \\
		300  & 1 & & 0.00 & 4.40  & & 16.60 & 58.40 & & 15.55 & 53.55 & & 21.65 & 58.40 \\
		500  & 0 & & 0.00 & 1.95  & & 6.50  & 66.30 & & 4.70  & 66.65 & & 7.65  & 66.75 \\
		500  & 1 & & 0.00 & 2.65  & & 10.90  & 41.20 & & 10.95  & 38.80 & & 13.15  & 41.90 \\
		1000 & 0 & & 0.00 & 0.45  & & 2.60  & 62.50 & & 2.85  & 63.55 & & 4.20  & 62.80 \\
		1000 & 1 & & 0.00 & 0.55  & & 2.15  & 24.65  & & 2.05  &  23.25 & & 2.75  & 23.70  \\
		\bottomrule
	\end{tabular}
	\begin{tablenotes}
		\item $n$: Sample Size; $Z$: Treatment indicator
		\item \textbf{OOB}: Percentage of survival curves containing at least one estimate which falls outside of the 0 and 1 probability bounds
		\item \textbf{NM}: Percentage of survival curves containing at least one instance of a not monotonically decreasing survival probability.
		\item Based on the assumption that all models are correctly specified (Scenario: CO \& CT). Using $2000$ simulation repetitions.
	\end{tablenotes}
\end{table}

To further study when these effects occur, we plotted the percentage of OOB estimates over the entire survival curve (see appendix). OOB errors occur much more frequently at the left and right end of the survival curve. They very rarely occur in the middle. To correct OOB estimates, the survival probability can simply be set to 1 if it is bigger than 1 and to 0 if it is negative. Isotonic regression can be used afterwards to correct NM survival curves \parencite{Westling2020}. This method uses the NM survival curve and augments it in such a way that it is changed as little as possible until it is non-decreasing, by using a weighted least squares fit subject to the monotonicity constraints. To study whether the use of these corrections has an impact on the asymptotic bias or goodness-of-fit, we applied it to all survival curves exhibiting these problems and recalculated $\hat{\Delta}_{Bias}(z)$ and $\hat{\Delta}_{MSE}(z)$. Applying these corrections resulted in a decreased bias in $53.94$\% of all cases and in a better goodness-of-fit in $85.88$\% of all cases. The changes in both bias and goodness-of-fit are, however, small on average. The average difference between the absolute value of the original $\hat{\Delta}_{Bias}(z)$ estimate and the absolute value of $\hat{\Delta}_{Bias}(z)$ after applying the corrections is $2.427934 \cdot 10^{-5}$. Similarly, the average difference between the original $\hat{\Delta}_{MSE}(z)$ and the $\hat{\Delta}_{MSE}(z)$ after applying the corrections is $8.262775 \cdot 10^{-5}$. Boxplots showing the distribution of these differences in each simulation scenario are displayed in the appendix.
\par
As described in table~\ref{tab::methods_summary}, for some methods approximate equations for the standard error of the causal survival probability have been derived. Where available, we used these standard errors in conjunction with the normal approximation to calculate pointwise 95\% confidence intervals over the whole range of the estimated survival curves. Figure~6 shows the percentage of these confidence intervals containing the true survival probability over time for different sample sizes and methods in the CO \& CT scenario. Regardless of method and sample size, the coverage is below 95\% for very early points in time and sometimes too high at the end of the curves. In most parts of the curves however, the confidence interval coverage is appropriate at every sample size and for every method. A similar graphic for the confidence interval width is included in the online appendix.
\par

\begin{figure}[!htb]
	\centering
	\includegraphics[width=1\linewidth]{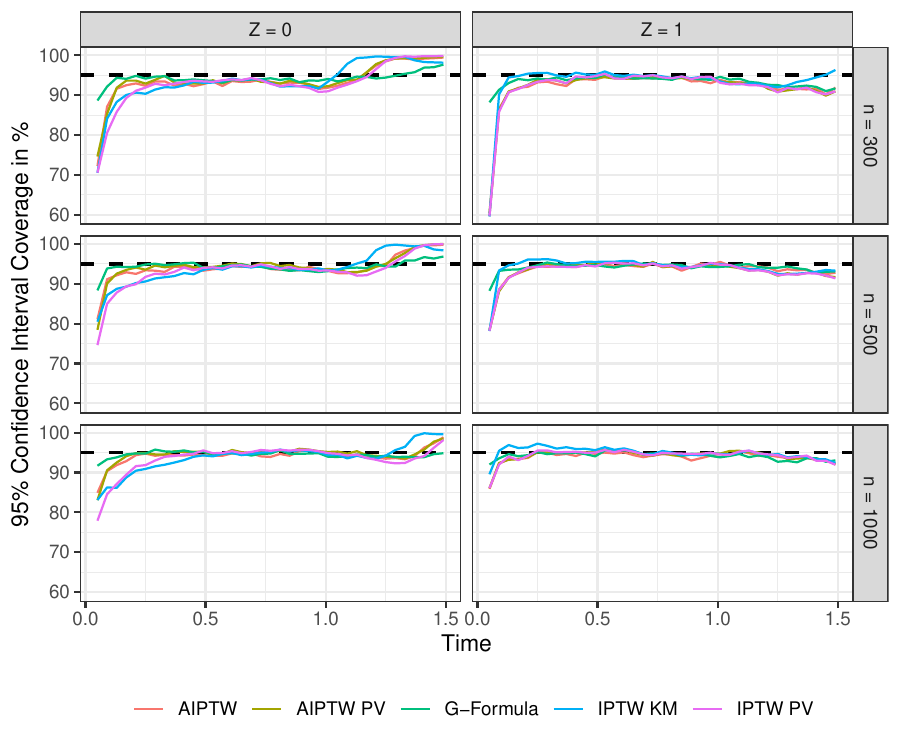}
	\caption{The percentage of approximate 95\% confidence intervals that contain the true survival probability for both treatment groups in the CO \& CT scenario for varying sample sizes. Estimates are based on $2000$ simulation repetitions.}
	\label{fig::ci_coverage}
\end{figure}

\FloatBarrier

\section{Discussion}

In this article we compared different methods for calculating confounder-adjusted survival curves with right-censored data under the assumption of proportional hazards. Our main focus was on the bias and the goodness-of-fit in different scenarios. The simulations showed that when used properly, all considered methods produce unbiased estimates for the whole survival curve in medium to large sample sizes. Methods utilizing the outcome mechanism were shown to be more efficient than methods relying on the treatment mechanism only. AIPTW based methods showed amounts of variation that are comparable to IPTW based methods when only one model was correctly specified and slightly outperformed the IPTW methods when both models were correct. These results are consistent with the previous literature on this topic \parencite{Cai2020, Chatton2020}. However, in contrast to the study by \textcite{Chatton2021a}, we found a small amount of bias when using the G-Formula IPTW method when a confounder was incorrectly modelled the outcome model. It follows that this method does not have the doubly-robust property when estimating the counterfactual survival curve. A possible reason for this may be that the addition of the weights results in a violation of the proportional hazards assumption. Using a different outcome model might therefore give better results.
\par
We were also able to show systematic bias in methods relying on a Cox model in small sample sizes ($n < 300$), even when it was correctly specified. An exception to this rule is the AIPTW method, which stays unbiased in small sample sizes when either the Cox model or the treatment assignment model are correctly specified. There were only small differences between the performance of PV based methods and their non-PV based counterparts. While AIPTW and PV based methods showed significant problems with monotonicity and estimates falling out of the 0 and 1 probability bounds, they should not be disregarded. Simple corrections, including isotonic regression and truncation, can be applied without introducing bias or loosing efficiency \parencite{Westling2020}.
\par
This study has several limitations. First, we only considered situations in which the true time-to-event process can be described using a Cox model, without time-varying confounders. This is an optimistic depiction of reality at best, but it reflects the assumptions made by most researchers in practice. Furthermore, we only considered a binary treatment variable, even though in reality, multi-arm studies are very common. We did this for the case of simplicity and because not all considered methods currently support categorical treatment variables \parencite{Wang2019}. Additionally, the generalized bias we used to judge the performance of the methods clearly has some shortcomings. If a method both overestimates and underestimates the true counterfactual survival probability in different points in time, the method can theoretically have a generalized bias of zero, despite it being biased at almost every point in time. We substituted the statistics with plots of the time-specific bias over time to eliminate this problem.
\par
Nevertheless, we were able to show that in the considered scenarios, all methods consistently outperformed the naive Kaplan-Meier estimates. We therefore think that the methods discussed here should be used instead of the standard Kaplan-Meier estimator when analysing observational data. Based on the results of this study and previous discussion on the topic \parencite{Robins1992, Bai2013} we recommend using AIPTW based methods, because they possess the doubly-robust property and showed goodness-of-fit similar to IPTW based methods. Although the G-Formula and G-Formula PV methods showed better goodness-of-fit overall, they do rely on one correctly specified model. When using the Cox model, this entails including all independent predictor variables in addition to the confounders, which can be difficult in practice. A drawback of the AIPTW and AIPTW PV methods is that they are more complex to use than other methods, because an implementation of the method itself and of isotonic regression is required. We believe, however, that this problem is mitigated by the user-friendly \textbf{R}-Code implementations in the \texttt{riskRegression}\parencite{Ozenne2020} and \texttt{adjustedCurves}\parencite{Denz2022a} packages.
\par
Although the EL method is also doubly-robust in theory, the only currently available implementation does not have this property, as demonstrated in our study. If such an implementation becomes available in the future, it would be a viable alternative to the AIPTW based methods because it does not rely on any kind of model. When there are known unmeasured confounders, the methods discussed here are insufficient to obtain unbiased estimates. Instrumental variable based methods might be preferable in this case \parencite{Martinez-Camblor2020}. 


\section*{Acknowledgments}

We would like to thank Xiaofei Wang and Fangfang Bai for supplying us with their R-Code used for empirical likelihood estimation. We also want to thank Jixian Wang for supplying us with the R-Code used for the calculation of the augmented inverse probability of treatment weighted estimator based on pseudo-values. Additionally we want to thank the anonymous reviewers for their extensive input, which greatly improved this manuscript. Last but not least, we want to thank Karl-Heinz Jöckel, Jale Basten, Marianne Tokic and Henrik Rudolf for their helpful comments and suggestions.

\newpage

\printbibliography

@Article{Makuch1982,
  author  = {Robert W. Makuch},
  title   = {Adjusted Survival Curve Estimation Using Covariates},
  journal = {Journal of Chronic Diseases},
  year    = {1982},
  volume  = {35},
  number  = {6},
  pages   = {437--443},
}

@Article{Nieto1996,
  author  = {F. Javier Nieto and Josef Coresh},
  title   = {Adjusting Survival Curves for Confounders: A Review and a New Method},
  journal = {American Journal of Epidemiology},
  year    = {1996},
  volume  = {143},
  number  = {10},
  pages   = {1059--1068},
}

@Article{Imai2014,
  author  = {Kosuke Imai and Marc Ratkovic},
  title   = {Covariate Balancing Propensity Score},
  journal = {Journal of the Royal Statistical Society: Series B},
  year    = {2014},
  volume  = {76},
  number  = {1},
  pages   = {243--263},
}

@Article{McCaffrey2004,
  author  = {Daniel F. McCaffrey and Greg Ridgeway and Andrew R. Morral},
  title   = {Propensity Score Estimation with Boosted Regression for Evaluating Causal Effects in Observational Studies},
  journal = {Psychological Methods},
  year    = {2004},
  volume  = {9},
  number  = {4},
  pages   = {403--425},
}

@Book{Guo2015,
  title     = {Propensity Score Analysis: Statistical Methods and Applications},
  publisher = {Sage},
  year      = {2015},
  author    = {Shengyang Guo and Mark W. Fraser},
  address   = {Los Angeles},
  edition   = {2},
}

@Article{Cole2004,
  author  = {Stephen R. Cole and Miguel A. Hernán},
  title   = {Adjusted Survival Curves with Inverse Probability Weights},
  journal = {Computer Methods and Programs in Biomedicine},
  year    = {2004},
  volume  = {2003},
  number  = {75},
  pages   = {45--49},
}

@Article{Xie2005,
  author  = {Jun Xie and Chaofeng Liu},
  title   = {Adjusted Kaplan-Meier Estimator and Log-Rank Test with Inverse Probability of Treatment Weighting for Survival Data},
  journal = {Statistics in Medicine},
  year    = {2005},
  volume  = {24},
  pages   = {3089--3110},
}

@Article{Austin2020,
  author  = {Peter C. Austin and Neal Thomas and Donald B. Rubin},
  title   = {Covariate-Adjusted Survival Analyses in Propensity-Score Matched Samples: Imputing Potential Time-To-Event Outcomes},
  journal = {Statistical Methods in Medical Research},
  year    = {2020},
  volume  = {29},
  number  = {3},
  pages   = {728--751},
}

@Article{Cupples1995,
  author  = {L. Adrienne Cupples and David R. Gragnon and Ratna Ramaswamy and Ralph D'Agostino},
  title   = {Age-Adjusted Survival Curves with Application in the Framingham Study},
  journal = {Statistics in Medicine},
  year    = {1995},
  volume  = {14},
  pages   = {1731--1744},
}

@Article{Ghali2001,
  author  = {William A. Ghali and Hude Quan and Rollin Brant and Guy {van Melle} and Colleen M. Norris and Peter D. Faris and P. Diane Gallbraith and Merril L. Knudtson},
  title   = {Comparison of 2 Methods for Calculating Adjusted Survival Curves from Proportional Hazards Models},
  journal = {JAMA},
  year    = {2001},
  volume  = {286},
  number  = {12},
  pages   = {1494--1497},
}

@Article{Gregory1988,
  author  = {W. M. Gregory},
  title   = {Adjusting Survival Curves for Imbalances in Prognostic Factors},
  journal = {British Journal of Cancer},
  year    = {1988},
  volume  = {58},
  pages   = {202--204},
}

@Article{Jiang2010,
  author  = {Honghua Jiang and James Symanowski and Yongming Qu and Xiao Ni and Yanping Wang},
  title   = {Covariate-Adjusted Non-Parametric Survival Curve Estimation},
  journal = {Statistics in Medicine},
  year    = {2010},
  volume  = {30},
  pages   = {1243--1253},
}

@Article{Borgne2016,
  author  = {Florent Le Borgne and Bruno Giraudeau and Anne Héléne Querard and Magali Giral and Yohann Foucher},
  title   = {Comparisons of the Performance of Different Statistical Tests for Time-To-Event Analysis with Confounding Factors: Practical Illustrations in Kidney Transplantation},
  journal = {Statistics in Medicine},
  year    = {2016},
  volume  = {35},
  pages   = {1103--1116},
}

@Article{Martinez-Camblor2020,
  author  = {Pablo Martínez-Camblor and Todd A. MacKenzie and Douglas O. Staiger and Phillip P. Goodney and A. James O'Malley},
  title   = {Summarizing Causal Differences in Survival Curves in the Presence of Unmeasured Confounding},
  journal = {The International Journal of Biostatistics},
  year    = {2020},
  volume  = {Inprint},
}

@Article{Wang2019,
  author  = {Xiaofei Wang and Fangfang Bai and Herbert Pang and Stephen L. George},
  title   = {Bias-Adjusted Kaplan-Meier Survival Curves for Marginal Treatment Effect in Observational Studies},
  journal = {Journal of Biopharmaceutical Statistics},
  year    = {2019},
  volume  = {29},
  number  = {4},
  pages   = {592--605},
}

@Article{Pepe1989,
  author  = {Margaret Sullivan Pepe and Thomas R. Fleming},
  title   = {Weighted Kaplan-Meier Statistics: A Class of Distance Tests for Censored Survival Data},
  journal = {Biometrics},
  year    = {1989},
  volume  = {45},
  number  = {2},
  pages   = {497--507},
}

@Article{Pepe1991a,
  author  = {Margaret Sullivan Pepe and Thomas R. Fleming},
  title   = {Weighted Kaplan-Meier Statistics: Large Sample and Optimality Considerations},
  journal = {Journal of the Royal Statistical Society: Series B},
  year    = {1991},
  volume  = {53},
  number  = {2},
  pages   = {341--352},
}

@Article{Zhao2012,
  author  = {Lihui Zhao and Lu Tian and Hajime Uno and Scott D. Solomon and Marc A. Pfeffer and Jerald S. Schindler and L. J. Wei},
  journal = {Clinical Trials},
  title   = {Utilizing the Integrated Difference of Two Survival Functions to Quantify the Treatment Contrast for Designing, Monitoring and Analyzing a Comparative Clinical Study},
  year    = {2012},
  number  = {5},
  pages   = {570--577},
  volume  = {9},
}

@Article{Austin2016,
  author  = {Peter C. Austin and Tibor Schuster},
  title   = {The Performance of Different Propensity Score Methods for Estimating Absolute Effects of Treatments on Survival Outcomes: A Simulation Study},
  journal = {Statistical Methods in Medical Research},
  year    = {2016},
  volume  = {25},
  number  = {5},
  pages   = {2214--2237},
}

@Article{Martinussen2013,
  author  = {Torben Martinussen and Stijn Vansteelandt},
  title   = {On Collapsibility and Confounding Bias in Cox and Aalen Regression Models},
  journal = {Lifetime Data Analysis},
  year    = {2013},
  volume  = {19},
  pages   = {279--296},
}

@Article{Rubin1978,
  author  = {Donald B. Rubin},
  title   = {Bayesian Inference for Causal Effects: The Role of Randomization},
  journal = {Annals of Statistics},
  year    = {1978},
  volume  = {6},
  number  = {1},
  pages   = {34--58},
}

@Article{Kaplan1958,
  author  = {E. L. Kaplan and Paul Meier},
  title   = {Nonparametric Estimation from Incomplete Observations},
  journal = {Journal of the American Statistical Association},
  year    = {1958},
  volume  = {53},
  number  = {282},
  pages   = {457--481},
}

@Article{Zhang2007,
  author  = {Xu Zhang and Fausto R. Loberiza and John P. Klein and Mei-Jie Zhang},
  title   = {A SAS Macro for Estimation of Direct Adjusted Survival Curves Based on a Stratified Cox Regression Model},
  journal = {Computer Methods and Programs in Biomedicine},
  year    = {2007},
  volume  = {88},
  pages   = {95--101},
}

@Article{Chang1982,
  author  = {I-Ming Chang and Rebecca Gelman and Marcello Pagano},
  title   = {Corrected Group Prognostic Curves and Summary Statistics},
  journal = {Journal of Chronic Diseases},
  year    = {1982},
  volume  = {35},
  pages   = {669--674},
}

@Article{Bai2013,
  author  = {Xiaofei Bai and Anastasios A. Tsiatis and Sean M. O'Brien},
  title   = {Doubly-Robust Estimators of Treatment-Specific Survival Distributions in Observational Studies with Stratified Sampling},
  journal = {Biometrics},
  year    = {2013},
  volume  = {69},
  pages   = {830--839},
}

@Article{Zhang2012,
  author  = {Min Zhang and Douglas E. Schaubel},
  title   = {Contrasting Treatment-Specific Survival Using Double-Robust Estimators},
  journal = {Statistics in Medicine},
  year    = {2012},
  volume  = {31},
  number  = {30},
  pages   = {4255--4268},
}

@Article{Wang2018,
  author  = {Jixian Wang},
  title   = {A Simple, Doubly Robust, Efficient Estimator for Survival Functions Using Pseudo Observations},
  journal = {Pharmaceutical Statistics},
  year    = {2018},
  volume  = {17},
  number  = {38--48},
}

@Misc{Chatton2020,
  author       = {Arthur Chatton and Florent Le Borgne and Clémence Leyrat and Yohann Foucher},
  howpublished = {arXiv:2006.16859v1},
  title        = {G-Computation and Inverse Probability Weighting for Time-To-Event Outcomes: A Comparative Study},
  year         = {2020},
}

@Article{Cai2020,
  author  = {Weixin Cai and Mark J. {van der Laan}},
  title   = {One-Step Targeted Maximum Likelihood Estimation for Time-To-Event Outcomes},
  journal = {Biometrics},
  year    = {2020},
  volume  = {76},
  pages   = {722--733},
}

@InCollection{Hubbard2000,
  author    = {Alan E. Hubbard and Mark J. {van der Laan} and James M. Robins},
  title     = {Nonparametric Locally Efficient Estimation of the Treatment Specific Survival Distribution with Right Censored Data and Covariates in Observational Studies},
  booktitle = {Statistical Models in Epidemiology, the Environment, and Clinical Trials},
  publisher = {Springer Science + Business Media},
  year      = {2000},
  editor    = {M. Elizabeth Halloran and Donald Berry},
  pages     = {135--177},
  address   = {New York},
}

@Article{Vansteelandt2011,
  author  = {Stijn Vansteelandt and Niels Keiding},
  title   = {Invited Commentary: G-Computation-Lost in Translation?},
  journal = {American Journal of Epidemiology},
  year    = {2011},
  volume  = {173},
  number  = {7},
  pages   = {739--742},
}

@Article{Ozenne2020,
  author  = {Brice Maxime Hugues Ozenne and Thomas Harder Scheike and Laila {St\ae rk}},
  title   = {On the Estimation of Average Treatment Effects with Right-Censored Time to Event Outcome and Competing Risks},
  journal = {Biometrical Journal},
  year    = {2020},
  volume  = {62},
  pages   = {751--763},
}

@Article{Andersen2017,
  author  = {Per Kragh Andersen and Elisavet Syriopoulou and Erik T. Parner},
  journal = {Statistics in Medicine},
  title   = {Causal Inference in Survival Analysis using Pseudo-Observations},
  year    = {2017},
  pages   = {2669--2681},
  volume  = {36},
}

@Article{Bender2005,
  author  = {Ralf Bender and Thomas Augustin and Maria Blettner},
  journal = {Statistics in Medicine},
  title   = {Generating Survival Times to Simulate Cox Proportional Hazards Models},
  year    = {2005},
  number  = {11},
  pages   = {1713--1723},
  volume  = {24},
}

@Article{Cox1972,
  author  = {D. R. Cox},
  journal = {Journal of the Royal Statistical Society: Series B (Methodological)},
  title   = {Regression Models and Life-Tables},
  year    = {1972},
  number  = {2},
  pages   = {187--220},
  volume  = {34},
}

@Article{Stitelman2010,
  author  = {Ori M. Stitelman and Mark J. {van der Laan}},
  journal = {The International Journal of Biostatistics},
  title   = {Collaborative Targeted Maximum Likelihood for Time to Event Data},
  year    = {2010},
  number  = {1},
  volume  = {6},
}

@Article{Chatton2020a,
  author  = {Arthur Chatton and Florent Le Borgne and Clémence Leyrat and Florence Gillaizeau and Chloé Rousseau and Laetitia Barbin and David Laplaud and Maxime Léger and Bruno Giraudeau and Yohann Foucher},
  journal = {Scientific Reports},
  title   = {G-Computation, Propensity Score-Based Methods, and Targeted Maximum Likelihood Estimator for Causal Inference with Different Covariates Sets: A Comparative Simulation Study},
  year    = {2020},
  number  = {9219},
  volume  = {10},
}

@Article{Rosenbaum1983,
  author  = {Paul R. Rosenbaum and Donald B. Rubin},
  journal = {Biometrika},
  title   = {The Central Role of the Propensity Score in Observational Studies for Causal Effects},
  year    = {1983},
  number  = {1},
  pages   = {41--55},
  volume  = {70},
}

@Article{Robins1986,
  author  = {James Robins},
  journal = {Mathematical Modelling},
  title   = {A New Approach to Causal Inference in Mortality Studies with a Sustained Exposure Period: Application to Control of the Healthy Worker Survivor Effect},
  year    = {1986},
  pages   = {1393--1512},
  volume  = {7},
}

@Article{Andersen2010,
  author  = {Per Kragh Andersen and Maja Pohar Perme},
  journal = {Statistical Methods in Medical Research},
  title   = {Pseudo-Observations in Survival Analysis},
  year    = {2010},
  pages   = {71--99},
  volume  = {19},
}

@Article{Overgaard2017,
  author  = {Norten Overgaard and Erik Thorlund Parner and Jan Pedersen},
  journal = {The Annals of Statistics},
  title   = {Asymptotic Theory of Generalized Estimating Equations Based on Jack-Knife Pseudo-Observations},
  year    = {2017},
  number  = {5},
  pages   = {1988--2015},
  volume  = {45},
}

@InCollection{Moore2009,
  author    = {Kelly L. Moore and Mark J. {van der Laan}},
  booktitle = {Design and Analysis of Clinical Trials with Time-To-Event Endpoints},
  publisher = {CRC Press},
  title     = {Application of Time-To-Event Methods in the Assessment of Safety in Clinical Trials},
  year      = {2009},
  editor    = {Karl E. Peace},
}

@Article{Zeng2004,
  author  = {Donglin Zeng},
  journal = {The Annals of Statistics},
  title   = {Estimating Marginal Survival Function by Adjusting for Dependent Censoring Using Many Covariates},
  year    = {2004},
  number  = {4},
  pages   = {1533--1555},
  volume  = {32},
}

@Article{Klein1989,
  author  = {John P. Klein and Melvin L. Moeschberger},
  journal = {Communications in Statistics: Simulation and Computation},
  title   = {The Robustness of Several Estimators of the Survivorship Function with Randomly Censored Data},
  year    = {1989},
  number  = {3},
  pages   = {1087--1112},
  volume  = {18},
}

@Article{Clare2019a,
  author  = {Philip J. Clare and Timothy A. Dobbins and Richard P. Mattick},
  journal = {International Journal of Epidemiology},
  title   = {Causal Models Adjusting for Time-Varying Confounding: A Systematic Review of the Literature},
  year    = {2019},
  number  = {1},
  pages   = {254--265},
  volume  = {48},
}

@Article{Raad2020,
  author  = {Hanaya Raad and Victoria Cornelius and Susan Chan and Elizabeth Williamson and Suzie Cro},
  journal = {BMC Medical Research Methodology},
  title   = {An Evaluation of Inverse Probability Weighting Using the Propensity Score for Baseline Covariate Adjustment in Smaller Population Randomised Controlled Trials with a Continuous Outcome},
  year    = {2020},
  number  = {70},
  volume  = {20},
}

@Article{Rubin1980,
  author  = {Donald B. Rubin},
  journal = {Journal of the American Statistical Association},
  title   = {Randomization Analysis of Experimental Data: The Fisher Randomization Test Comment},
  year    = {1980},
  number  = {371},
  pages   = {591--593},
  volume  = {75},
}

@Article{Poole2010,
  author  = {Charles Poole},
  journal = {Epidemiology},
  title   = {On the Origin of Risk Relativism},
  year    = {2010},
  number  = {1},
  pages   = {3--9},
  volume  = {21},
}

@Article{Hernan2010,
  author  = {Migueal A. Hernán},
  journal = {Epidemiology},
  title   = {The Hazards of Hazard Ratios},
  year    = {2010},
  number  = {1},
  pages   = {13--15},
  volume  = {21},
}

@Article{Davis2010,
  author  = {C. R. Davis and A. G. K. McNair and A. Brigic and M. G. Clarke and S. T. Brookes and M. G. Thomas and J. M. Blazeby},
  journal = {European Journal of Cancer},
  title   = {Optimising Methods for Communicating Survival Data to Patients Undergoing Cancer Surgery},
  year    = {2010},
  pages   = {3192--3199},
  volume  = {46},
}

@Article{Zipkin2014,
  author  = {Daniella A. Zipkin and Craig A. Umscheid and Nancy L. Keating and Elizabeth Allen and KoKo Aung and Rebecca Beyth and Scott Kaatz and Devin M. Mann and Jeremy B. Sussman and Deborah Korenstein and Connie Schardt and Avishek Nagi and Richard Sloane and David A. Feldstein},
  journal = {Annals of Internal Medicine},
  title   = {Evidence-Based Risk Communication: A Systematic Review},
  year    = {2014},
  number  = {4},
  pages   = {270--280},
  volume  = {161},
}

@Article{Rubin1974,
  author  = {Donald B. Rubin},
  journal = {Journal of Educational Psychology},
  title   = {Estimating Causal Effects of Treatments in Randomized and Nonrandomized Studies},
  year    = {1974},
  number  = {5},
  pages   = {688--701},
  volume  = {66},
}

@Book{Owen2001,
  author    = {Art B. Owen},
  publisher = {CRC Press},
  title     = {Empirical Likelihood},
  year      = {2001},
  address   = {Boca Raton},
}

@InCollection{Robins1992,
  author    = {James M. Robins and Andrea Rotnitzky},
  booktitle = {AIDS Epidemiology: Methodological Issues},
  publisher = {Springer Science $+$ Business Media},
  title     = {Recovery of Information and Adjustment for Dependent Censoring Using Surrogate Markers},
  year      = {1992},
  address   = {New York},
  editor    = {Nicholas P. Jewell and Klaus Dietz and Vernon T. Farewell},
  pages     = {297--331},
}

@Article{Kang2007,
  author  = {Joseph D. Y. Kang and Joseph L. Schafer},
  journal = {Statistical Science},
  title   = {Demystifying Double Robustness: A Comparison of Alternative Strategies for Estimating a Population Mean from Incomplete Data},
  year    = {2007},
  number  = {4},
  pages   = {523--539},
  volume  = {22},
}

@Article{Dey2020,
  author  = {Tanujit Dey and Anish Mukherjee and Sounak Chakraborty},
  journal = {CHEST},
  title   = {A Practical Overview and Reporting Strategies for Statistical Analysis of Survival Studies},
  year    = {2020},
  number  = {1, Supplement},
  pages   = {S39--S48},
  volume  = {158},
}

@Article{Winnett2002,
  author  = {Angela Winnett and Peter Sasieni},
  journal = {Journal of the American Statistical Association},
  title   = {Adjusted Nelson-Aalen Estimates with Retrospective Matching},
  year    = {2002},
  number  = {457},
  pages   = {245--256},
  volume  = {97},
}

@Article{Amato1988,
  author  = {David A. Amato},
  journal = {Communications in Statistics: Theory and Methods},
  title   = {A Generalized Kaplan-Meier Estimator for Heterogenous Populations},
  year    = {1988},
  number  = {1},
  pages   = {263--286},
  volume  = {17},
}

@Article{Westling2020,
  author  = {Ted Westling and Mark J. {van der Laan} and Marco Carone},
  journal = {Electronic Journal of Statistics},
  title   = {Correcting an Estimator of a Multivariate Monotone Function with Isotonic Regression},
  year    = {2020},
  pages   = {3032--3069},
  volume  = {14},
}

@Article{Austin2014,
  author  = {Peter C. Austin},
  journal = {Statistics in Medicine},
  title   = {The Use of Propensity Score Methods with Survival or Time-To-Event Outcomes: Reporting Measures of Effect Similar to those Used in Randomized Experiments},
  year    = {2014},
  pages   = {1242--1258},
  volume  = {33},
}

@Article{Klein2008,
  author  = {John P. Klein and Mette Gerster and Per Kragh Andersen and Sergey Tarima and Maja Pohar Perme},
  journal = {Computer Methods and Programs in Biomedicine},
  title   = {SAS and R Functions to Compute Pseudo-Values for Censored Data Regression},
  year    = {2008},
  number  = {3},
  pages   = {289--300},
  volume  = {89},
}

@Article{DeNeve2020,
  author  = {Jan {De Neve} and Thomas A. Gerds},
  journal = {Biometrical Journal},
  title   = {On the Intrepretation of the Hazard Ratio in Cox Regression},
  year    = {2020},
  pages   = {742--750},
  volume  = {62},
}

@Misc{Zeng2021,
  author       = {Shuxi Zeng and Fan Li and Liangyuan Hu and Fan Li},
  howpublished = {arXiv: 2103.00605v1},
  title        = {Propensity Score Weighting Analysis of Survival Outcomes Using Pseudo-Observations},
  year         = {2021},
}

@Article{Neugebauer2005,
  author  = {Romain Neugebauer and Mark J. {van der Laan}},
  journal = {Journal of Statistical Planning and Inference},
  title   = {Why Prefer Double Robust Estimators in Causal Inference},
  year    = {2005},
  pages   = {405--426},
  volume  = {129},
}

@Article{Sofrygin2019,
  author  = {Oleg Sofrygin and Zheng Zhu and Julie A. Schmittdiel and Alyce S. Adams and Richard W. Grant and Mark J. {van der Laan} and Romain Neugebauer},
  journal = {Statistics in Medicine},
  title   = {Targeted Learning with Daily EHR Data},
  year    = {2019},
  pages   = {3073--3090},
  volume  = {38},
}

@Article{Guerra2020,
  author  = {Steve Ferreira Guerra and Mireille E. Schnitzer and Amélie Forget and Lucie Blais},
  journal = {Statistics in Medicine},
  title   = {Impact of Discretization of the Timeline for Longitudinal Causal Inference Methods},
  year    = {2020},
  number  = {27},
  pages   = {4069--4085},
  volume  = {39},
}

@Misc{Westling2021,
  author       = {Ted Westling and Alex Luedtke and Peter Gilbert and Marco Carone},
  howpublished = {arXiv:2106.06602v1},
  title        = {Inference for Treatment-Specific Survival Curves using Machine Learning},
  year         = {2021},
}

@Article{Galimberti2002,
  author  = {Stefania Galimberti and Peter Sasieni and Maria Grazia Valsecchi},
  journal = {Statistics in Medicine},
  title   = {A Weighted Kaplan-Meier Estimator for Matched Data with Application to the Comparison of Chemotherapy and Bone-Marrow Transplant in Leukaemia},
  year    = {2002},
  pages   = {3847--3864},
  volume  = {21},
}

@Article{Neyman1923,
  author  = {J. S. Neyman},
  journal = {Journal of the Royal Statistical Society: Series B},
  title   = {Statistical Problems in Agricultural Experiments},
  year    = {1923},
  pages   = {107--180},
  volume  = {2},
}

@Article{Li2016,
  author  = {Kuibao Li and Chonghua Yao and Xuan Di and Xinchun Yang and Lei Dong and Li Xu and Meili Zheng},
  journal = {Medicine},
  title   = {Smoking and Risk of All-Cause Deaths in Younger and Older Adults},
  year    = {2016},
  number  = {3},
  volume  = {95},
}

@Article{Carter2015,
  author  = {Brian D. Carter and Christian C. Abnet and Diane Feskanich and Neal D. Freedman and Patricia Hartge and Cora E. Lewis and Judith K. Ockene and Ross L. Prentice and Frank E. Speizer and Michael J. Thun and Eric J. Jacobs},
  journal = {The New England Journal of Medicine},
  title   = {Smoking and Mortality: Beyond Established Causes},
  year    = {2015},
  number  = {7},
  pages   = {631--640},
  volume  = {372},
}

@Article{Schnohr2004,
  author  = {Christina Schnohr and Lise {H{\o}jbjerre} and Mette Riegels and Luise Ledet and Tine Larsen and Kirsten Schultz-Larsen and Liselotte Petersen and Eva Prescott and Morten {Gr{\o}nb{\ae}k}},
  journal = {Scandinavian Journal of Public Health},
  title   = {Does Educational Level Influence the Effects of Smoking, Alcohol, Physical Activity, and Obesity on Mortality? A Prospective Population Study},
  year    = {2004},
  number  = {4},
  pages   = {250--256},
  volume  = {32},
}

@Article{Gutterman2015,
  author  = {Sam Gutterman},
  journal = {North American Actuarial Journal},
  title   = {Mortality of Smoking by Gender},
  year    = {2015},
  number  = {3},
  pages   = {200--223},
  volume  = {19},
}

@Article{Breslow1972,
  author  = {N. Breslow},
  journal = {Journal of the Royal Statistical Society: Series B},
  title   = {Discussion of the Paper by D. R. Cox},
  year    = {1972},
  number  = {2},
  pages   = {216--217},
  volume  = {34},
}

@Book{Pearl2018,
  author    = {Judea Pearl and Dana Mackenzie},
  publisher = {Penguin Books},
  title     = {The Book of Why: The New Science of Cause and Effect},
  year      = {2018},
  address   = {London},
}

@Article{Lee2012,
  author  = {Peter N. Lee and Barbara A. Forey and Katharine J. Coombs},
  journal = {BMC Cancer},
  title   = {Systematic Review with Meta-Analysis of the Epidemiological Evidence in the 1900s Relating Smoking to Lung Cancer},
  year    = {2012},
  number  = {385},
  volume  = {12},
}

@Article{StudyGroup2002,
  author  = {{getABI Study Group}},
  journal = {VASA},
  title   = {getABI: German Epidemiological Trial on Ankle Brachial Index for Elderly Patients in Family Practice to Dedect Peripheral Arterial Disease, Significant Marker for High Mortality},
  year    = {2002},
  number  = {4},
  pages   = {241--248},
  volume  = {31},
}

@Article{Diehm2004,
  author  = {Curt Diehm and Alexander Schuster and Jens R. Allenberg and Harald Darius and Roman Haberl and Stefan Lange and David Pittrow and Berndt von Stritzky and Gerhart Tepohl and Hand-Joachim Trampisch},
  journal = {Atherosclerosis},
  title   = {High Prevalence of Peripheral Arterial Disease and Co-Morbidity in 6880 Primary Care Patients: Cross-Sectional Study},
  year    = {2004},
  number  = {1},
  pages   = {95--105},
  volume  = {172},
}

@Article{Chatton2021a,
  author  = {Arthur Chatton and Florent Le Borgne and Clémence Leyrat and Yohann Foucher},
  journal = {Statistical Methods in Medical Research},
  title   = {G-Computation and Doubly Robust Standardisation for Continuous-Time Data: A Comparison with Inverse Probability Weighting},
  year    = {2021},
  volume  = {OnlineFirst},
}

@Misc{Lee2022b,
  author       = {Dasom Lee and Shu Yang and Xiaofei Wang},
  howpublished = {arXiv:2201.06595v1},
  title        = {Generalizable Survival Analysis of Randomized Controlled Trials with Observational Data},
  year         = {2022},
}

@Article{Sekhon2011,
  author  = {Jasjeet S. Sekhon},
  journal = {Journal of Statistical Software},
  title   = {Multivariate and Propensity Score Matching Software with Automated Balance Optimization: The Matching Package for R},
  year    = {2011},
  number  = {7},
  volume  = {42},
}

@Misc{Denz2022a,
  author       = {Robin Denz},
  howpublished = {R package version 0.9.0},
  note         = {\url{https://cran.r-project.org/package=adjustedCurves}},
  title        = {adjustedCurves: Confounder-Adjusted Survival Curves and Cumulative Incidence Functions},
  year         = {2022},
}

\newpage

\appendix

\section{Additional Information about the Illustrative Example}

Table~\ref{tab::cox_reg} shows the hazard-ratios and associated 95\% confidence intervals of both the standard Cox model and the weighted Cox model used to create the counterfactual survival curves of people who never smoked versus ex-smokers or current smokers from the getABI data. Model (1) was used for the standard \emph{G-Formula} and the standard \emph{AIPTW} methods. Model (2) was used only for the \emph{G-Formula IPTW} method.
\par\medskip

\begin{table}[!htbp] \centering 
	\caption{Hazard-Ratios and associated 95\% confidence intervals for two Cox models for the survival time.} 
	\label{tab::cox_reg} 
	\begin{tabular}{@{\extracolsep{5pt}}lcc} 
		\\[-1.8ex]\hline 
		\hline \\[-1.8ex] 
		& \multicolumn{2}{c}{\textit{Dependent variable:}} \\ 
		\cline{2-3} 
		\\[-1.8ex] & \multicolumn{2}{c}{Death of all causes} \\ 
		\\[-1.8ex] & (1) & (2)\\ 
		\hline \\[-1.8ex] 
		Ever Smoker & 0.645 & 0.645 \\ 
		& (0.569, 0.731) & (0.570, 0.729) \\ 
		& & \\ 
		Age & 1.105 & 1.108 \\ 
		& (1.094, 1.115) & (1.097, 1.119) \\ 
		& & \\ 
		Sex & 1.760 & 1.754 \\ 
		& (1.542, 2.008) & (1.537, 2.001) \\ 
		& & \\ 
		Lehre (yes/no) & 0.774 & 0.790 \\ 
		& (0.673, 0.890) & (0.682, 0.916) \\ 
		& & \\ 
		University Degree (yes/no) & 0.552 & 0.561 \\ 
		& (0.416, 0.733) & (0.417, 0.754) \\ 
		& & \\ 
		Highschool Graduate (yes/no) & 0.931 & 0.926 \\ 
		& (0.806, 1.076) & (0.796, 1.077) \\ 
		& & \\ 
		A-Levels Graduate (yes/no) & 0.882 & 0.843 \\ 
		& (0.687, 1.133) & (0.646, 1.100) \\ 
		& & \\ 
		\hline \\[-1.8ex] 
		Observations & 6,752 & 6,752 \\ 
		R$^{2}$ & 0.079 & 0.233 \\ 
		Max. Possible R$^{2}$ & 0.962 & 1.000 \\ 
		Log Likelihood & $-$10,741.760 & $-$38,097.660 \\ 
		\hline 
		\hline \\[-1.8ex] 
	\end{tabular} 
\end{table} 

\begin{figure}[!htb]
	\centering
	\includegraphics[width=0.8\linewidth]{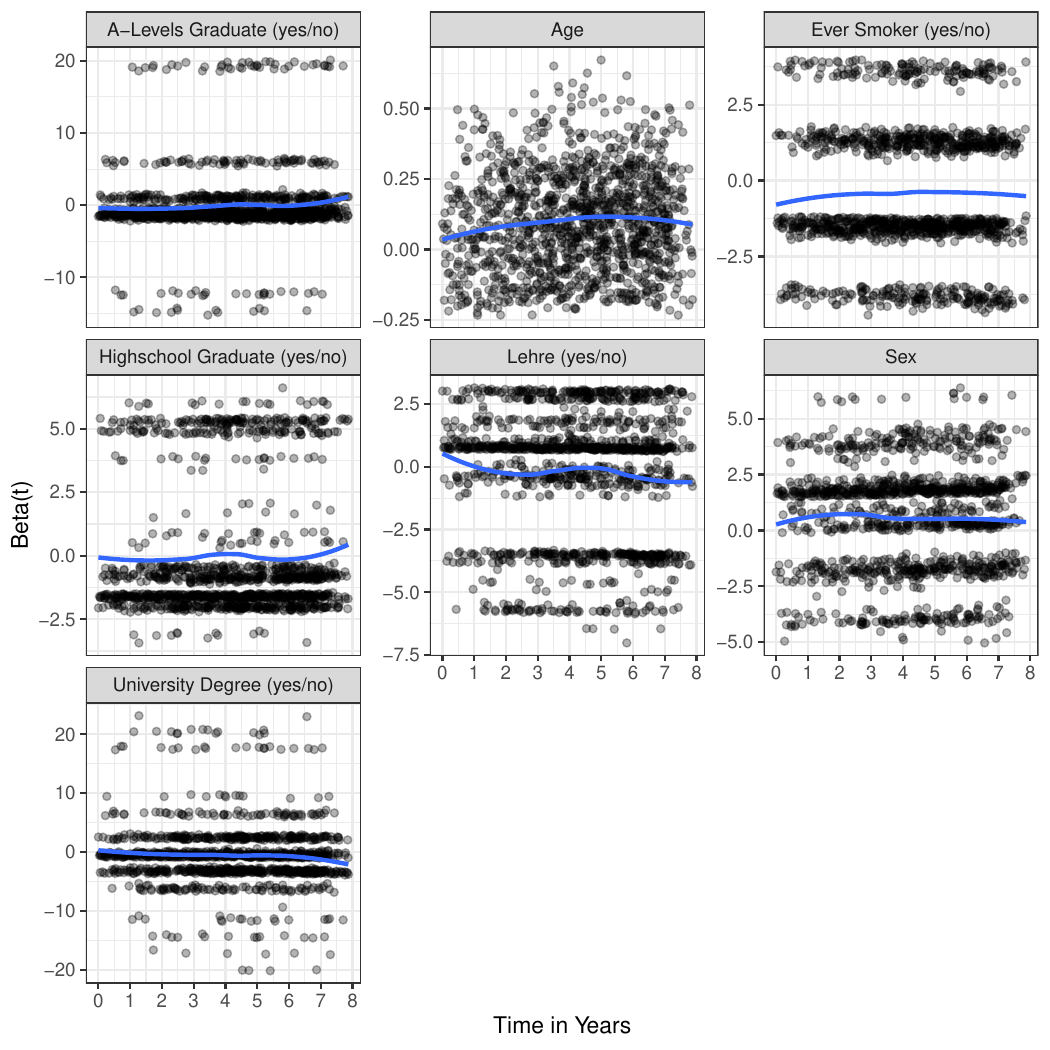}
	\caption{A graphical test of the proportional hazards assumption for the standard Cox model (model (1)). The blue lines are non-parametric locally weighted regressions.}
	\label{fig::getABI_cox_mod_ph}
\end{figure}

\begin{figure}[!htb]
	\centering
	\includegraphics[width=0.8\linewidth]{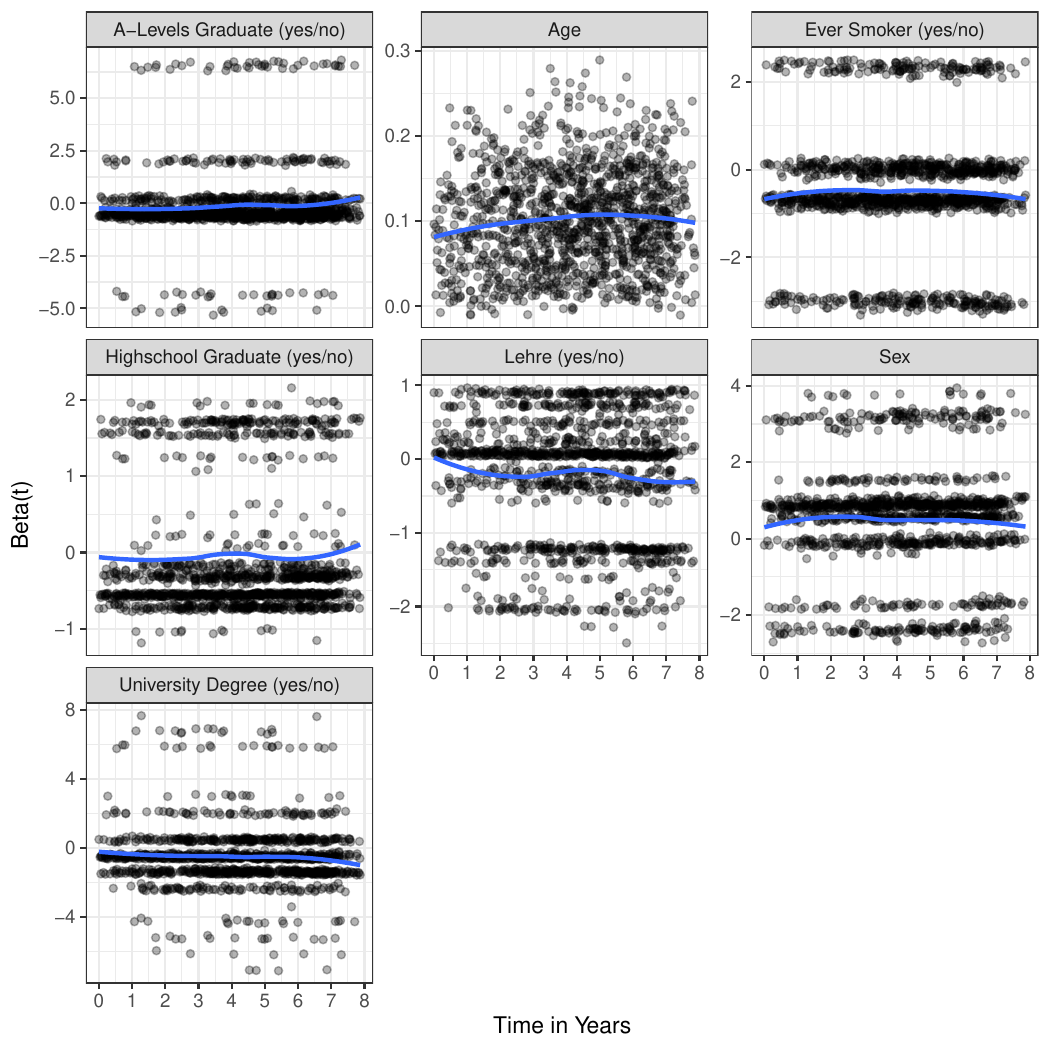}
	\caption{A graphical test of the proportional hazards assumption for the inverse probability of treatment weighted Cox model (model (2)). The blue lines are non-parametric locally weighted regressions.}
	\label{fig::getABI_cox_mod_iptw_ph}
\end{figure}

To graphically test if the proportional hazards assumption is appropriate for these models we plotted estimates of the time-dependent coefficients for each included covariate against time. If the proportional hazards assumption holds, the coefficients should be on a fairly straight line. Since there are a lot of points, we also added a non-parametric scatter-plot smoother to indicate the trend in each facet. Figure~\ref{fig::getABI_cox_mod_ph} displays the results of this procedure for model (1), while figure~\ref{fig::getABI_cox_mod_iptw_ph} shows the results for model (2).

Although the points do not fall on a perfect straight line for every included covariate, as indicated by the non-parametric regression line, it seems to be close enough to conclude that the proportional hazards assumption holds for these models.
\par
\begin{figure}[!htb]
	\centering
	\includegraphics[width=0.6\linewidth]{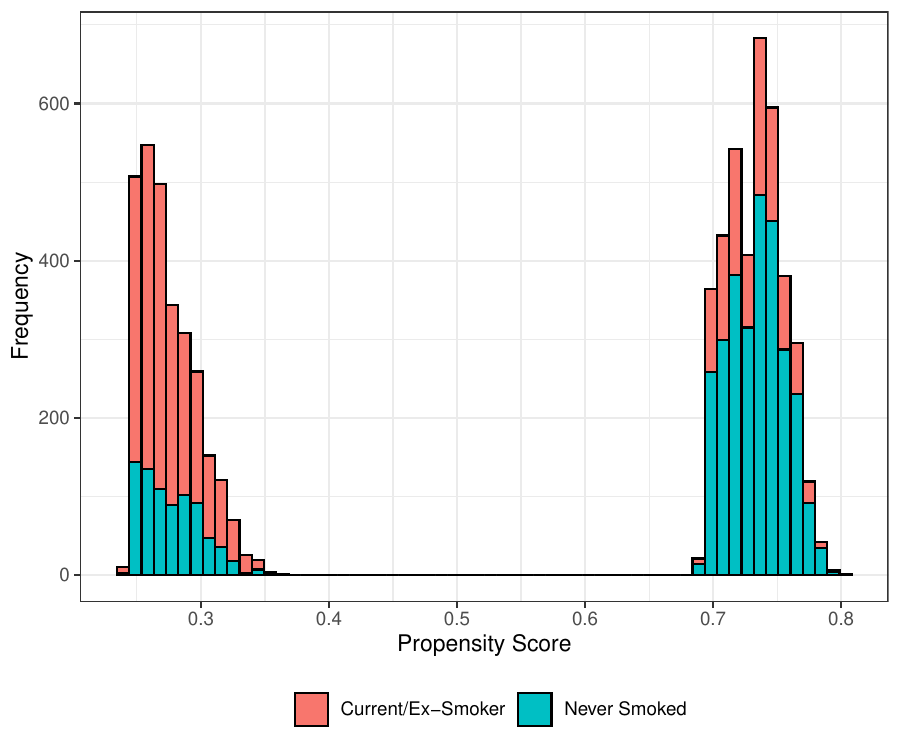}
	\caption{Histogram of the propensity score estimated by a logistic regression model for both people who never smoked and ex-smokers/current smokers.}
	\label{fig::getABI_ps_histogram}
\end{figure}

\begin{figure}[!htb]
	\centering
	\includegraphics[width=0.6\linewidth]{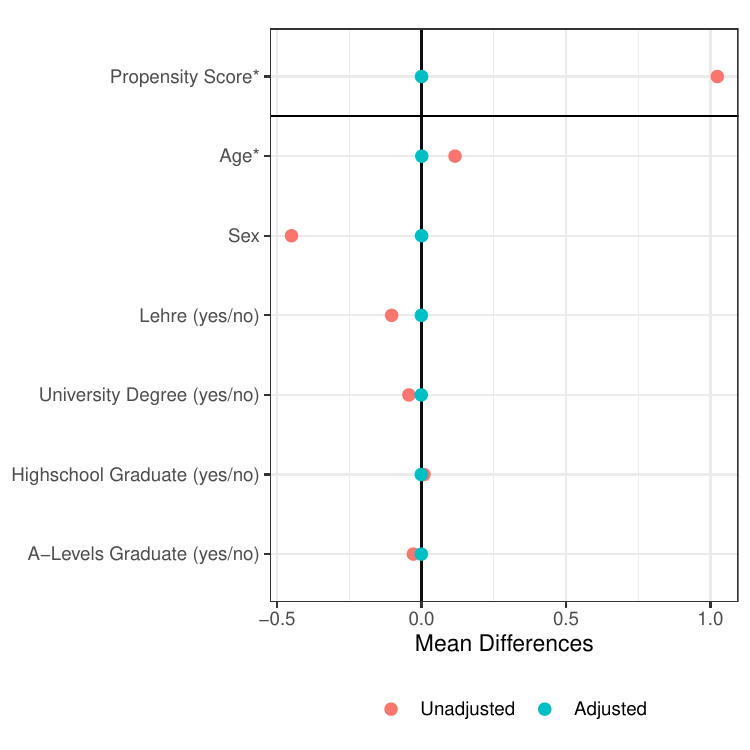}
	\caption{Mean differences between the two treatment groups (Never Smokers and Ex-Smokers/Current Smokers) before and after adjustment using the propensity score. The star indicates that standardized mean differences were used instead of normal mean differences.}
	\label{fig::getABI_ps_loveplot}
\end{figure}

\medskip
Table~\ref{tab::log_reg} shows the odds-ratios and associated 95\% confidence intervals of the logistic propensity score model used when calculating the survival curves using the \emph{IPTW KM}, \emph{IPTW HZ}, \emph{IPTW PV}, \emph{AIPTW}, \emph{AIPTW PV}, \emph{G-Formula IPTW} and \emph{Matching} methods. Figure~\ref{fig::getABI_ps_histogram} additionally shows the distribution of the estimated propensity scores for both groups and figure~\ref{fig::getABI_ps_loveplot} shows the (standardized) mean differences between the groups before and after adjustment using the propensity score. Overall, the model seems to be a good fit with adequate propensity score overlap. The mean differences after adjustment are also close to zero, indicating that the adjustment actually results in covariate balance between the groups.

\begin{table}[!htbp] \centering 
	\caption{Odds-Ratios and associated 95\% confidence intervals for the logistic regression model used to predict the propensity score.} 
	\label{tab::log_reg} 
	\begin{tabular}{@{\extracolsep{5pt}}lc} 
		\\[-1.8ex]\hline 
		\hline \\[-1.8ex] 
		& \multicolumn{1}{c}{\textit{Dependent variable:}} \\ 
		\cline{2-2} 
		\\[-1.8ex] & Ever Smoker \\ 
		\hline \\[-1.8ex] 
		Age & 1.012 \\ 
		& (1.002, 1.023) \\ 
		& \\ 
		Sex & 0.142 \\ 
		& (0.126, 0.159) \\ 
		& \\ 
		Lehre (yes/no) & 0.840 \\ 
		& (0.732, 0.965) \\ 
		& \\ 
		University Degree (yes/no) & 1.060 \\ 
		& (0.814, 1.380) \\ 
		& \\ 
		Highschool Graduate (yes/no) & 1.072 \\ 
		& (0.933, 1.233) \\ 
		& \\ 
		A-Levels Graduate (yes/no) & 0.956 \\ 
		& (0.755, 1.214) \\ 
		& \\ 
		Intercept & 1.242 \\ 
		& (0.582, 2.651) \\ 
		& \\ 
		\hline \\[-1.8ex] 
		Observations & 6,752 \\ 
		Log Likelihood & $-$3,931.601 \\ 
		Akaike Inf. Crit. & 7,877.202 \\ 
		\hline 
		\hline \\[-1.8ex] 
	\end{tabular} 
\end{table} 

\FloatBarrier
\newpage

\section{Step-By-Step Guide to the Methods}

In this section we will give a short semi-algorithmic description of each method for the reader who wants to know more details about the calculation of the adjusted survival probabilities without having to read the original articles cited in the main text.

\subsection{G-Formula}

\begin{enumerate}
	\item Fit a outcome model (for example a Cox model) with the time-to-event as response variable and the treatment $Z$ and the relevant confounders as independent variables.
	\item Set the treatment status of every person in the sample to $Z = z$.
	\item Use the outcome model from step 1 to predict the survival probability at some point in time $t$ for each individual in the sample, given their observed confounder realizations. If a Cox model was used in step 1, an estimate of the baseline hazard function is required for this step.
	\item Calculate the arithmetic mean of the predicted conditional survival probabilities at $t$ for all individuals in the sample. The result is the adjusted survival probability for $Z = z$.
\end{enumerate}

\subsection{G-Formula PV}

\begin{enumerate}
	\item Calculate pseudo values for every person in the sample at some point in time $t$.
	\item Add the pseudo values to the original dataset as a new column.
	\item Add the time to the original dataset as a new column.
	\item Repeat steps 1--3 for multiple $t$.
	\item Concatenate the resulting datasets into one dataset in the long format.
	\item Fit a generalized estimating equation model with the pseudo values as response variable and the treatment $Z$ and the relevant confounders as independent variables.
	\item Set the treatment status of every person in the longitudinal dataset to $Z = z$.
	\item Use the model from step 6 to predict the survival probability at the used points in time $t$ for each individual in the sample, given their observed confounder realizations.
	\item Calculate the arithmetic mean of the predicted survival probabilities for each $t$. The result is the adjusted survival probability for $Z = z$.
\end{enumerate}

\subsection{IPTW KM}

\begin{enumerate}
	\item Fit a propensity score model with the treatment $Z$ as response variable and the relevant confounders as independent variables.
	\item Use the model from step 1 to estimate the probability of receiving the treatment $Z$ for every person in the sample.
	\item Calculate inverse probability of treatment weights for every person in the sample.
	\item Calculate the weighted number of people at risk and the weighted number of events at some point in time $t$ as described in Xie \& Liu (2005) on page 3091.
	\item Use equation (1) in  Xie \& Liu (2005) to calculate the adjusted survival probability.
\end{enumerate}

\subsection{IPTW HZ}

\begin{enumerate}
	\item Fit a propensity score model with the treatment $Z$ as response variable and the relevant confounders as independent variables.
	\item Use the model from step 1 to estimate the probability of receiving the treatment $Z$ for every person in the sample.
	\item Calculate inverse probability of treatment weights for every person in the sample.
	\item Fit an empty weighted Cox model using the inverse probability of treatment weights from step 3 while including only the treatment $Z$ as \texttt{strata()} term with no further independent variables.
	\item Use the \texttt{survfit()} function form the \texttt{survival} \textbf{R}-package to get an estimate of the survival curve based on a weighted hazard function.
\end{enumerate}

\subsection{IPTW PV}

\begin{enumerate}
	\item Fit a propensity score model with the treatment $Z$ as response variable and the relevant confounders as independent variables.
	\item Use the model from step 1 to estimate the probability of receiving the treatment $Z$ for every person in the sample.
	\item Calculate inverse probability of treatment weights for every person in the sample.
	\item Calculate pseudo values for every person in the sample at some point in time $t$.
	\item Using the inverse probability of treatment weights from step 3, calculate a weighted mean of the pseudo values at $t$ for each group in $Z$. The result is the adjusted survival probability. 
\end{enumerate}

\subsection{Matching}

\begin{enumerate}
	\item Fit a propensity score model with the treatment $Z$ as response variable and the relevant confounders as independent variables.
	\item Use the model from step 1 to estimate the probability of receiving the treatment $Z$ for every person in the sample.
	\item Use the propensity scores from step 2 to perform propensity score matching.
	\item Use a standard Kaplan-Meier estimator to calculate the adjusted survival curves for each treatment group in $Z$ in the matched sample. 
\end{enumerate}

\subsection{Empirical Likelihood Estimation}

\begin{enumerate}
	\item Create a matrix containing the relevant confounders.
	\item Maximize the likelihood function defined in equation (2) in Wang et al. (2019), subject to the constraints listed in equations (2.1) and (2.2) in Wang et al. (2019) to obtain subject-specific weights. This can be done by using a newton-raphson type algorithm to numerically approximate the values for $\alpha$ and $\tau$ in equations (2.3) and (2.4) in Wang et al. (2019) which where obtained using the Lagrange multiplier algorithm.
	\item Use the weights (denoted as $p_i$ in the original article) to calculate a the weighted number of people at risk and the weighted number of events.
	\item Calculate the adjusted survival probability using equation (2.7) from Wang et al. (2019).
\end{enumerate}

\subsection{AIPTW}

\begin{enumerate}
	\item Fit a propensity score model with the treatment $Z$ as response variable and the relevant confounders as independent variables.
	\item Use the model from step 1 to estimate the probability of receiving the treatment $Z$ for every person in the sample.
	\item Fit a outcome model (such as a Cox model) with the time-to-event as response variable and the treatment $Z$ and the relevant confounders as independent variables.
	\item Set the treatment status of every person in the sample to $Z = z$.
	\item Use the outcome model from step 3 to predict the survival probability at some point in time $t$ for each individual in the sample, given their observed confounder realizations. If a Cox model was used in step 3, an estimate of the baseline hazard function is required for this step.
	\item Calculate the adjusted survival probability using equation (7) from Ozenne et al. (2020).
\end{enumerate}

\subsection{AIPTW PV}

\begin{enumerate}
	\item Calculate pseudo values for every person in the sample at some point in time $t$.
	\item Add the pseudo values to the original dataset as a new column.
	\item Add the time to the original dataset as a new column.
	\item Repeat steps 1--3 for multiple $t$.
	\item Concatenate the resulting datasets into one dataset in the long format.
	\item Fit a generalized estimating equation model with the pseudo values as response variable and the treatment $Z$ and the relevant confounders as independent variables.
	\item Set the treatment status of every person in the longitudinal dataset to $Z = z$.
	\item Use the model from step 6 to predict the survival probability at the used points in time $t$ for each individual in the sample, given their observed confounder realizations.
	\item Calculate the arithmetic mean of the predicted survival probabilities for each $t$. The result is the adjusted survival probability for $Z = z$.
	\item Fit a propensity score model with the treatment $Z$ as response variable and the relevant confounders as independent variables.
	\item Use the model from step 1 to estimate the probability of receiving the treatment $Z$ for every person in the sample.
	\item Use the estimated pseudo values, the estimated propensity scores and the predicted survival probabilities to calculate the adjusted survival probability using equation (16) from the article of Wang (2018).
\end{enumerate}

\subsection{G-Formula IPTW}

\begin{enumerate}
	\item Fit a propensity score model with the treatment $Z$ as response variable and the relevant confounders as independent variables.
	\item Use the model from step 1 to estimate the probability of receiving the treatment $Z$ for every person in the sample.
	\item Calculate inverse probability of treatment weights for every person in the sample.
	\item Fit a inverse probability of treatment weighted outcome model (such as the Cox model) with the time-to-event as response variable and the treatment $Z$ and the relevant confounders as independent variables. 
	\item Set the treatment status of every person in the sample to $Z = z$.
	\item Use the outcome model from step 4 to predict the survival probability at some point in time $t$ for each individual in the sample, given their observed confounder realizations. If a Cox model was used in step 4, an estimate of the baseline hazard function is required for this step.
	\item Calculate the arithmetic mean of the predicted conditional survival probabilities at $t$ for all individuals in the sample. The result is the adjusted survival probability for $Z = z$.
\end{enumerate}

\textbf{Literature}

\begin{itemize}
	\item[--] Ozenne BMH, Scheike TH, Stærk L. On the Estimation of Average Treatment Effects with Right-Censored Time to Event Outcome and Competing Risks. \emph{Biometrical Journal} 2020; 62: 751–763.
	\item[--] Wang J. A Simple, Doubly Robust, Efficient Estimator for Survival Functions Using Pseudo Observations. \emph{Pharmaceutical Statistics} 2018; 17(38–48).
	\item[--] Wang X, Bai F, Pang H, George SL. Bias-Adjusted Kaplan-Meier Survival Curves for Marginal Treatment Effect in Observational Studies. \emph{Journal of Biopharmaceutical Statistics} 2019; 29(4): 592–605.
	\item[--] Xie J, Liu C. Adjusted Kaplan-Meier Estimator and Log-Rank Test with Inverse Probability of Treatment Weighting for Survival Data. \emph{Statistics in Medicine} 2005; 24: 3089–3110.
\end{itemize}

\newpage

\section{Additional Results of the Main Simulation} \label{chap::appendix_add_results}

Below we show further results of the main simulation study. Tables \ref{tab::appendix_CO_CT}--\ref{tab::appendix_PCO_PCT} show the values for $\hat{G}_{Bias}(z)$ and $\hat{G}_{MSE}(z)$ and their associated Monte-Carlo standard errors, for each simulation scenario. The Monte-Carlo standard errors were calculated as:
\begin{equation}
	MCSE(\theta, m) = \frac{\sqrt{\frac{1}{m} \sum_{i = 1}^{m} \left(\bar{\theta} - \hat{\theta}\right)^2}}{\sqrt{m}},
\end{equation}
where $m$ is the number of simulation repetitions, $\hat{\theta}$ is the estimated quantity and $\bar{\theta}$ is the arithmetic mean of $\hat{\theta}$ over all simulation repetitions. In addition, the table includes the $\hat{G}_{Sup}(z)$ statistic, which we define as the average of the supremum over all simulation repetitions:
\begin{equation}
	\hat{G}_{sup}(z) = \frac{1}{m} \sum_{i = 1}^{m} \left( \sup \left|S_z(t) - \hat{S}_z(t)\right| \right),
\end{equation}
for the interval $t \in [0, \tau]$, using the same notation that was used in section~5.3 of the paper.
\par
\begin{figure}[!htb]
	\centering
	\includegraphics[width=0.8\linewidth]{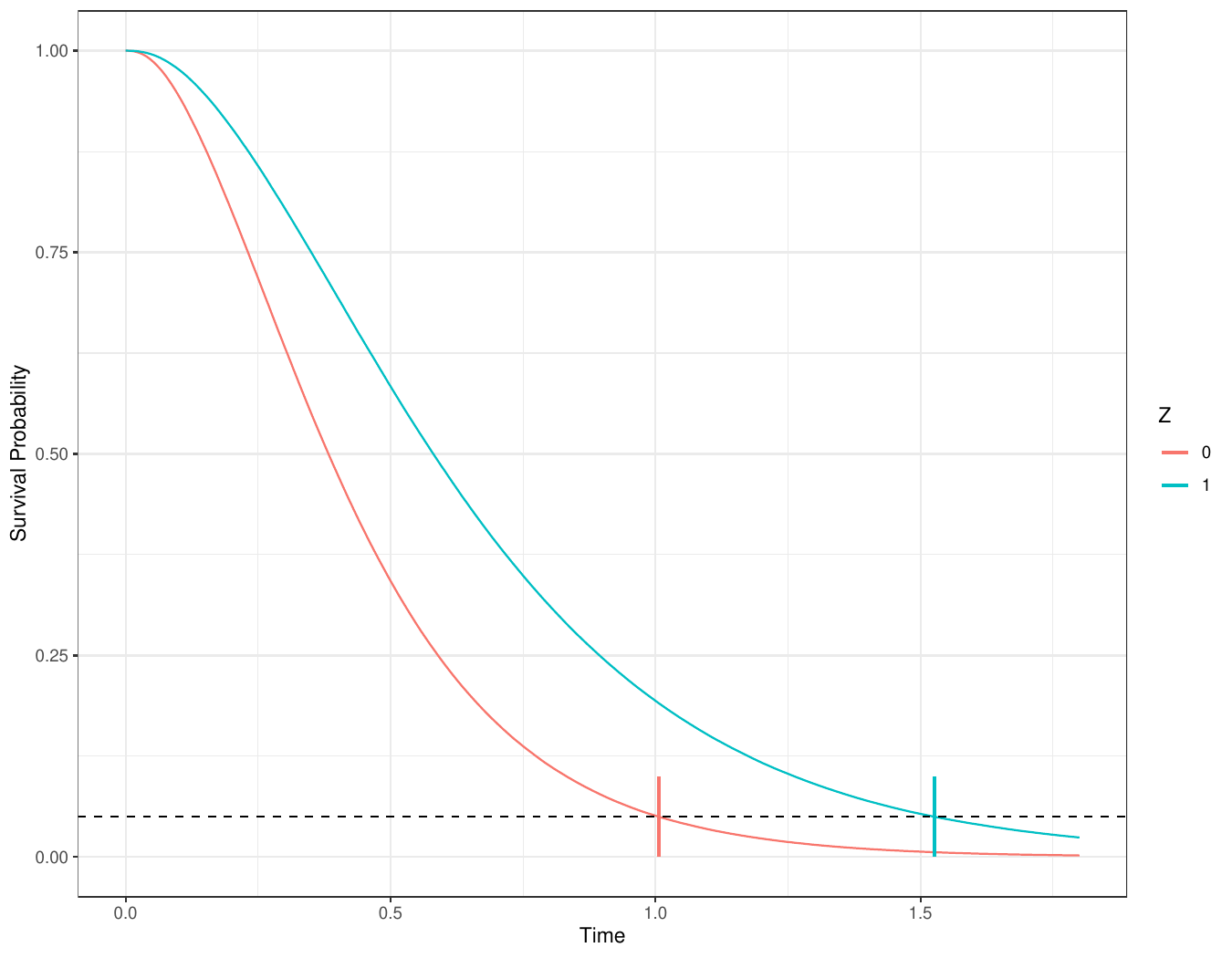}
	\caption{The true survival curves in the simulation study for both treatment groups ($Z$). The vertical lines indicate the 95\% quantile of the survival times.}
	\label{fig::true_surv}
\end{figure}
\begin{figure}[!htb]
	\centering
	\includegraphics[width=1\linewidth]{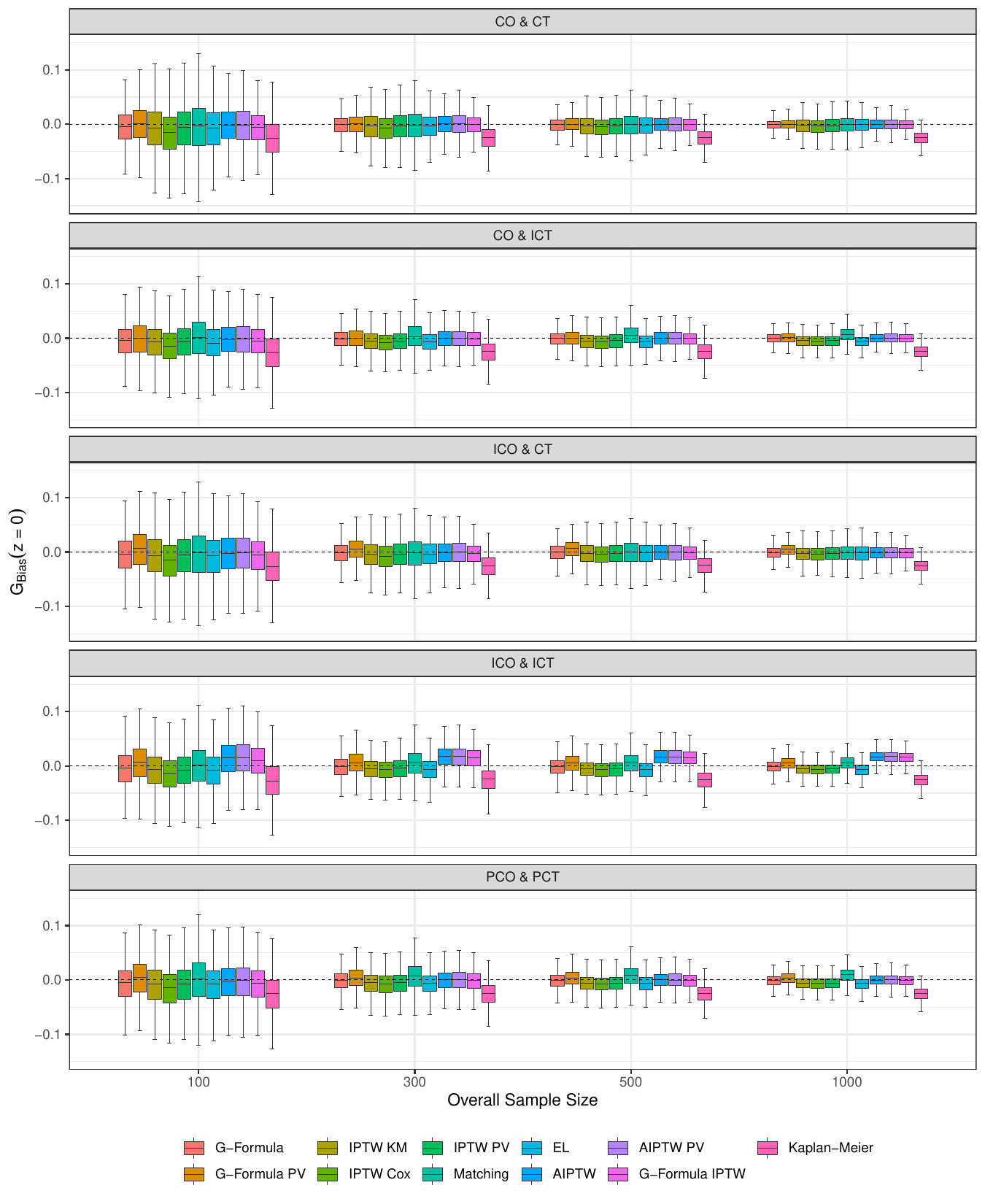}
	\caption{Distributions of $\hat{\Delta}_{Bias}(z = 0)$ (control group) for all methods in each simulation scenario with varying sample sizes. Outliers are not shown. Estimates are based on $2000$ simulation repetitions.}
	\label{fig::bias_over_scenarios_control}
\end{figure}
\begin{figure}[!htb]
	\centering
	\includegraphics[width=1\linewidth]{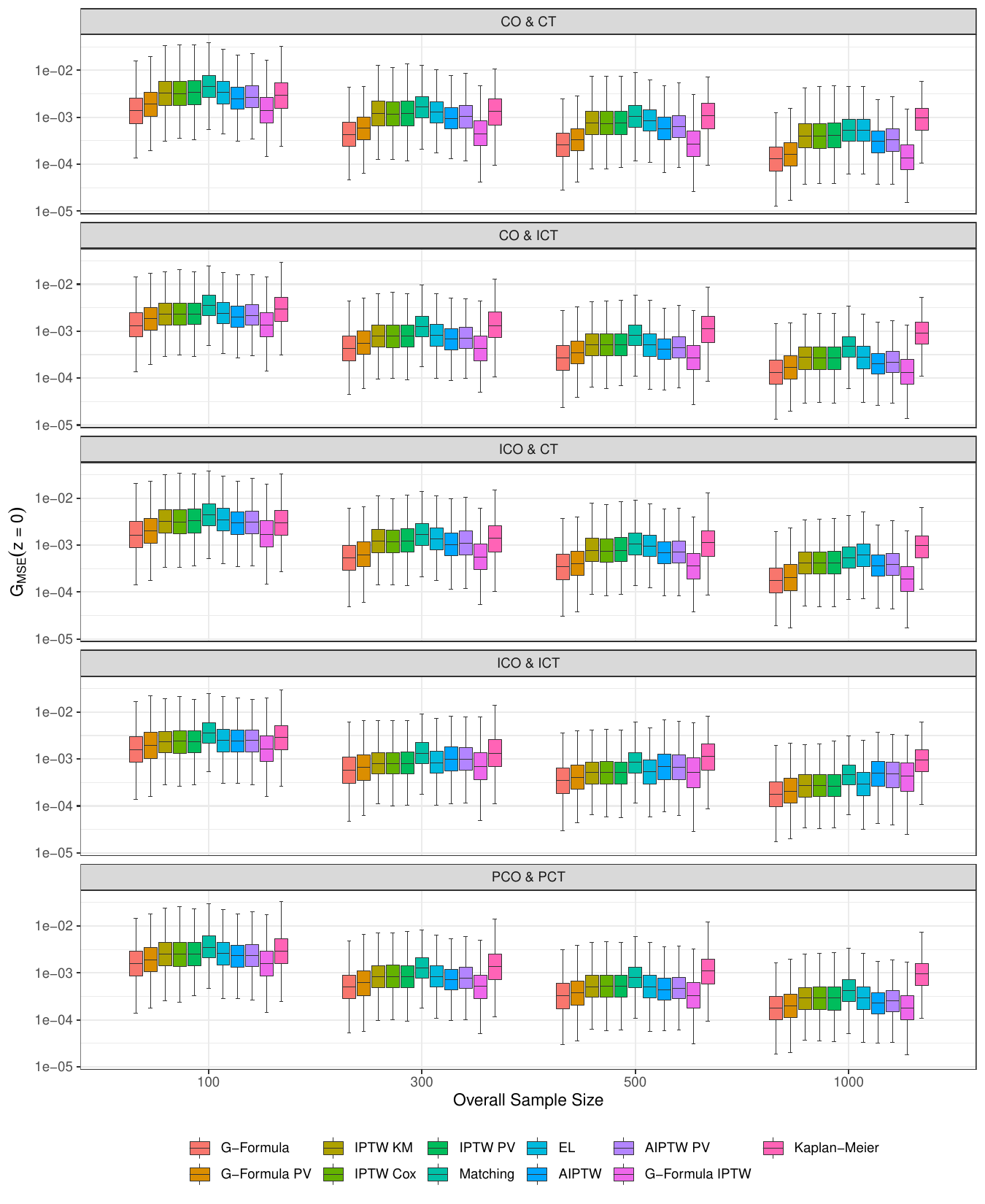}
	\caption{Distributions of $\hat{\Delta}_{MSE}(z = 0)$ (control group) on the log-scale for all methods in each simulation scenario with varying sample sizes. Outliers are not shown. Estimates are based on $2000$ simulation repetitions.}
	\label{fig::MSE_over_scenarios_control}
\end{figure}

These tables also include the percentage of survival curves including at least one value lying outside of the $[0, 1]$ probability bounds (\% OOB) and the percentage of survival curves that are not monotonically decreasing (\% NM). Figure~\ref{fig::true_surv} displays the true counterfactual survival curves of the super-population. Figure~\ref{fig::bias_over_scenarios_control} shows the distributions of $\hat{\Delta}_{Bias}(z = 0)$, and figure~\ref{fig::MSE_over_scenarios_control} shows the distributions of $\hat{\Delta}_{MSE}(z = 0)$. Additionally, figure~\ref{fig::bias_over_time_all} shows the average bias in relation to time for all simulation scenarios, sample sizes and both treatment groups. Figure~\ref{fig::mse_over_time} shows a similar graphic for the average mean-squared-error. We define the average bias as:
\begin{equation}
	\overline{Bias}\left(t, z\right) = \frac{1}{m} \sum_{i = 1}^{m} \left(S_z(t) - \hat{S}_z(t)\right),
\end{equation}
using the same notation as before. Similarly the average mean-squared-error is defined as:
\begin{equation}
	\overline{MSE}\left(t, z\right) = \frac{1}{m} \sum_{i = 1}^{m} \left(S_z(t) - \hat{S}_z(t)\right)^2,
\end{equation}
also using the same notation.
\par
Figure~\ref{fig::bias_diff} and figure~\ref{fig::mse_diff} display the impact of applying truncation and isotonic regression to survival curves exhibiting problems with out-of-bounds estimates non-monotonicity on the bias ($\hat{\Delta}_{Bias}(z)$) and goodness-of-fit ($\hat{\Delta}_{MSE}(z)$). Finally, figure~\ref{fig::ci_width} displays the width of approximate confidence intervals for methods allowing such estimations.

\newpage

\footnotesize

\begin{longtable}{rlrlllll}
	\caption{Estimates for all performance criteria in the \textbf{CO \& CT} scenario. The values in parentheses are the Monte-Carlo standard errors of the associated estimates. Estimates are based on $2000$ simulation repetitions. $\hat{G}_{Bias}(z)$, $\hat{G}_{MSE}(z)$, $\hat{G}_{Sup}(z)$ and their associated standard errors are rounded to the fourth decimal place, the other statistics are rounded to the second decimal place.}
	\label{tab::appendix_CO_CT}
	\endfirsthead
	\endhead
	\toprule
	$n$ & Method & $Z$ & $\hat{G}_{Bias}(z)$ & $\hat{G}_{MSE}(z)$ & $\hat{G}_{Sup}(z)$ & \% OOB & \% NM \\ 
	\midrule
	100 & G-Formula &   0 & -0.0048 [0.0007] & 0.0020 [0.0000] & 0.0866 [0.0007] &  0.00 &  0.00 \\ 
	100 & G-Formula &   1 & -0.0108 [0.0011] & 0.0034 [0.0001] & 0.0947 [0.0007] &  0.00 &  0.00 \\ 
	100 & G-Formula PV &   0 &  0.0002 [0.0008] & 0.0030 [0.0001] & 0.1153 [0.0025] &  0.00 & 13.10 \\ 
	100 & G-Formula PV &   1 & -0.0035 [0.0012] & 0.0048 [0.0001] & 0.1238 [0.0026] &  0.00 & 17.35 \\ 
	100 & IPTW KM &   0 & -0.0078 [0.0010] & 0.0049 [0.0001] & 0.1374 [0.0012] &  0.00 &  0.00 \\ 
	100 & IPTW KM &   1 & -0.0075 [0.0014] & 0.0065 [0.0001] & 0.1378 [0.0010] &  0.00 &  0.00 \\ 
	100 & IPTW Cox &   0 & -0.0172 [0.0010] & 0.0048 [0.0001] & 0.1351 [0.0011] &  0.00 &  0.00 \\ 
	100 & IPTW Cox &   1 & -0.0211 [0.0014] & 0.0066 [0.0001] & 0.1366 [0.0010] &  0.00 &  0.00 \\ 
	100 & IPTW PV &   0 & -0.0074 [0.0011] & 0.0050 [0.0001] & 0.1383 [0.0013] & 14.00 & 75.75 \\ 
	100 & IPTW PV &   1 & -0.0079 [0.0014] & 0.0067 [0.0001] & 0.1392 [0.0010] & 20.90 & 81.35 \\ 
	100 & Matching &   0 & -0.0081 [0.0012] & 0.0064 [0.0002] & 0.1589 [0.0012] &  0.00 &  0.00 \\ 
	100 & Matching &   1 & -0.0083 [0.0017] & 0.0088 [0.0002] & 0.1580 [0.0012] &  0.00 &  0.00 \\ 
	100 & EL &   0 & -0.0112 [0.0012] & 0.0060 [0.0005] & 0.1407 [0.0015] &  0.00 &  0.00 \\ 
	100 & EL &   1 & -0.0245 [0.0020] & 0.0118 [0.0009] & 0.1596 [0.0019] &  0.00 &  0.00 \\ 
	100 & AIPTW &   0 & -0.0017 [0.0009] & 0.0041 [0.0002] & 0.1279 [0.0017] & 10.35 & 74.20 \\ 
	100 & AIPTW &   1 & -0.0079 [0.0012] & 0.0055 [0.0001] & 0.1294 [0.0010] & 19.70 & 80.95 \\ 
	100 & AIPTW PV &   0 & -0.0015 [0.0009] & 0.0044 [0.0002] & 0.1316 [0.0017] & 21.85 & 77.00 \\ 
	100 & AIPTW PV &   1 & -0.0063 [0.0013] & 0.0060 [0.0001] & 0.1342 [0.0010] & 41.10 & 85.05 \\ 
	100 & G-Formula IPTW &   0 & -0.0059 [0.0007] & 0.0021 [0.0000] & 0.0880 [0.0007] &  0.00 &  0.00 \\ 
	100 & G-Formula IPTW &   1 & -0.0150 [0.0012] & 0.0037 [0.0001] & 0.0969 [0.0008] &  0.00 &  0.00 \\ 
	100 & Kaplan-Meier &   0 & -0.0266 [0.0009] & 0.0042 [0.0001] & 0.1248 [0.0010] &  0.00 &  0.00 \\ 
	100 & Kaplan-Meier &   1 &  0.0310 [0.0012] & 0.0060 [0.0001] & 0.1308 [0.0010] &  0.00 &  0.00 \\ 
	300 & G-Formula &   0 & -0.0012 [0.0004] & 0.0006 [0.0000] & 0.0490 [0.0004] &  0.00 &  0.00 \\ 
	300 & G-Formula &   1 & -0.0021 [0.0007] & 0.0012 [0.0000] & 0.0547 [0.0004] &  0.00 &  0.00 \\ 
	300 & G-Formula PV &   0 &  0.0004 [0.0004] & 0.0008 [0.0000] & 0.0575 [0.0009] &  0.00 &  2.90 \\ 
	300 & G-Formula PV &   1 &  0.0005 [0.0007] & 0.0016 [0.0000] & 0.0645 [0.0009] &  0.00 &  4.40 \\ 
	300 & IPTW KM &   0 & -0.0028 [0.0007] & 0.0019 [0.0001] & 0.0844 [0.0009] &  0.00 &  0.00 \\ 
	300 & IPTW KM &   1 & -0.0010 [0.0009] & 0.0023 [0.0000] & 0.0795 [0.0006] &  0.00 &  0.00 \\ 
	300 & IPTW Cox &   0 & -0.0068 [0.0006] & 0.0018 [0.0001] & 0.0828 [0.0008] &  0.00 &  0.00 \\ 
	300 & IPTW Cox &   1 & -0.0061 [0.0009] & 0.0023 [0.0000] & 0.0791 [0.0006] &  0.00 &  0.00 \\ 
	300 & IPTW PV &   0 & -0.0026 [0.0007] & 0.0020 [0.0001] & 0.0852 [0.0010] & 11.85 & 69.40 \\ 
	300 & IPTW PV &   1 & -0.0010 [0.0009] & 0.0023 [0.0000] & 0.0795 [0.0006] & 16.60 & 58.40 \\ 
	300 & Matching &   0 & -0.0031 [0.0007] & 0.0023 [0.0001] & 0.0962 [0.0008] &  0.00 &  0.00 \\ 
	300 & Matching &   1 & -0.0009 [0.0010] & 0.0031 [0.0001] & 0.0928 [0.0007] &  0.00 &  0.00 \\ 
	300 & EL &   0 & -0.0044 [0.0007] & 0.0020 [0.0002] & 0.0864 [0.0008] &  0.00 &  0.00 \\ 
	300 & EL &   1 & -0.0125 [0.0014] & 0.0048 [0.0008] & 0.0881 [0.0013] &  0.00 &  0.00 \\ 
	300 & AIPTW &   0 &  0.0001 [0.0005] & 0.0015 [0.0001] & 0.0797 [0.0011] &  8.25 & 70.60 \\ 
	300 & AIPTW &   1 & -0.0009 [0.0008] & 0.0019 [0.0000] & 0.0754 [0.0005] & 15.55 & 53.55 \\ 
	300 & AIPTW PV &   0 &  0.0003 [0.0006] & 0.0017 [0.0001] & 0.0820 [0.0012] & 14.45 & 70.55 \\ 
	300 & AIPTW PV &   1 & -0.0006 [0.0008] & 0.0021 [0.0000] & 0.0772 [0.0006] & 21.65 & 58.40 \\ 
	300 & G-Formula IPTW &   0 & -0.0014 [0.0004] & 0.0007 [0.0000] & 0.0498 [0.0004] &  0.00 &  0.00 \\ 
	300 & G-Formula IPTW &   1 & -0.0049 [0.0007] & 0.0013 [0.0000] & 0.0562 [0.0004] &  0.00 &  0.00 \\ 
	300 & Kaplan-Meier &   0 & -0.0249 [0.0005] & 0.0019 [0.0000] & 0.0806 [0.0007] &  0.00 &  0.00 \\ 
	300 & Kaplan-Meier &   1 &  0.0373 [0.0008] & 0.0030 [0.0001] & 0.0862 [0.0007] &  0.00 &  0.00 \\ 
	500 & G-Formula &   0 & -0.0011 [0.0003] & 0.0004 [0.0000] & 0.0375 [0.0003] &  0.00 &  0.00 \\ 
	500 & G-Formula &   1 & -0.0019 [0.0005] & 0.0007 [0.0000] & 0.0417 [0.0003] &  0.00 &  0.00 \\ 
	500 & G-Formula PV &   0 &  0.0000 [0.0003] & 0.0005 [0.0000] & 0.0435 [0.0007] &  0.00 &  1.95 \\ 
	500 & G-Formula PV &   1 & -0.0001 [0.0006] & 0.0009 [0.0000] & 0.0491 [0.0008] &  0.00 &  2.65 \\ 
	500 & IPTW KM &   0 & -0.0024 [0.0005] & 0.0012 [0.0000] & 0.0668 [0.0007] &  0.00 &  0.00 \\ 
	500 & IPTW KM &   1 & -0.0016 [0.0007] & 0.0014 [0.0000] & 0.0611 [0.0005] &  0.00 &  0.00 \\ 
	500 & IPTW Cox &   0 & -0.0048 [0.0005] & 0.0011 [0.0000] & 0.0659 [0.0006] &  0.00 &  0.00 \\ 
	500 & IPTW Cox &   1 & -0.0046 [0.0007] & 0.0014 [0.0000] & 0.0609 [0.0005] &  0.00 &  0.00 \\ 
	500 & IPTW PV &   0 & -0.0023 [0.0005] & 0.0012 [0.0000] & 0.0673 [0.0007] &  6.50 & 66.30 \\ 
	500 & IPTW PV &   1 & -0.0016 [0.0007] & 0.0014 [0.0000] & 0.0613 [0.0005] & 10.90 & 41.20 \\ 
	500 & Matching &   0 & -0.0023 [0.0006] & 0.0015 [0.0000] & 0.0776 [0.0006] &  0.00 &  0.00 \\ 
	500 & Matching &   1 & -0.0011 [0.0008] & 0.0018 [0.0000] & 0.0708 [0.0005] &  0.00 &  0.00 \\ 
	500 & EL &   0 & -0.0021 [0.0005] & 0.0012 [0.0000] & 0.0707 [0.0006] &  0.00 &  0.00 \\ 
	500 & EL &   1 & -0.0104 [0.0012] & 0.0031 [0.0007] & 0.0665 [0.0011] &  0.00 &  0.00 \\ 
	500 & AIPTW &   0 & -0.0001 [0.0004] & 0.0009 [0.0000] & 0.0619 [0.0008] &  4.70 & 66.65 \\ 
	500 & AIPTW &   1 & -0.0013 [0.0006] & 0.0011 [0.0000] & 0.0575 [0.0004] & 10.95 & 38.80 \\ 
	500 & AIPTW PV &   0 &  0.0000 [0.0005] & 0.0010 [0.0000] & 0.0633 [0.0008] &  7.65 & 66.75 \\ 
	500 & AIPTW PV &   1 & -0.0009 [0.0006] & 0.0012 [0.0000] & 0.0590 [0.0004] & 13.15 & 41.90 \\ 
	500 & G-Formula IPTW &   0 & -0.0012 [0.0003] & 0.0004 [0.0000] & 0.0383 [0.0003] &  0.00 &  0.00 \\ 
	500 & G-Formula IPTW &   1 & -0.0042 [0.0006] & 0.0008 [0.0000] & 0.0430 [0.0004] &  0.00 &  0.00 \\ 
	500 & Kaplan-Meier &   0 & -0.0251 [0.0004] & 0.0015 [0.0000] & 0.0696 [0.0005] &  0.00 &  0.00 \\ 
	500 & Kaplan-Meier &   1 &  0.0372 [0.0006] & 0.0023 [0.0000] & 0.0727 [0.0006] &  0.00 &  0.00 \\ 
	1000 & G-Formula &   0 & -0.0005 [0.0002] & 0.0002 [0.0000] & 0.0268 [0.0002] &  0.00 &  0.00 \\ 
	1000 & G-Formula &   1 & -0.0008 [0.0004] & 0.0003 [0.0000] & 0.0292 [0.0002] &  0.00 &  0.00 \\ 
	1000 & G-Formula PV &   0 & -0.0001 [0.0002] & 0.0002 [0.0000] & 0.0298 [0.0002] &  0.00 &  0.45 \\ 
	1000 & G-Formula PV &   1 & -0.0002 [0.0004] & 0.0004 [0.0000] & 0.0334 [0.0002] &  0.00 &  0.55 \\ 
	1000 & IPTW KM &   0 & -0.0016 [0.0004] & 0.0006 [0.0000] & 0.0492 [0.0005] &  0.00 &  0.00 \\ 
	1000 & IPTW KM &   1 & -0.0012 [0.0005] & 0.0007 [0.0000] & 0.0428 [0.0003] &  0.00 &  0.00 \\ 
	1000 & IPTW Cox &   0 & -0.0029 [0.0004] & 0.0006 [0.0000] & 0.0487 [0.0005] &  0.00 &  0.00 \\ 
	1000 & IPTW Cox &   1 & -0.0027 [0.0005] & 0.0007 [0.0000] & 0.0428 [0.0003] &  0.00 &  0.00 \\ 
	1000 & IPTW PV &   0 & -0.0016 [0.0004] & 0.0006 [0.0000] & 0.0496 [0.0005] &  2.60 & 62.50 \\ 
	1000 & IPTW PV &   1 & -0.0013 [0.0005] & 0.0007 [0.0000] & 0.0429 [0.0003] &  2.15 & 24.65 \\ 
	1000 & Matching &   0 & -0.0013 [0.0004] & 0.0007 [0.0000] & 0.0554 [0.0004] &  0.00 &  0.00 \\ 
	1000 & Matching &   1 & -0.0012 [0.0005] & 0.0009 [0.0000] & 0.0496 [0.0004] &  0.00 &  0.00 \\ 
	1000 & EL &   0 & -0.0010 [0.0004] & 0.0007 [0.0000] & 0.0569 [0.0005] &  0.00 &  0.00 \\ 
	1000 & EL &   1 & -0.0063 [0.0006] & 0.0011 [0.0003] & 0.0453 [0.0006] &  0.00 &  0.00 \\ 
	1000 & AIPTW &   0 &  0.0000 [0.0003] & 0.0004 [0.0000] & 0.0442 [0.0005] &  2.85 & 63.55 \\ 
	1000 & AIPTW &   1 & -0.0005 [0.0004] & 0.0006 [0.0000] & 0.0403 [0.0003] &  2.05 & 23.25 \\ 
	1000 & AIPTW PV &   0 & -0.0001 [0.0003] & 0.0005 [0.0000] & 0.0458 [0.0005] &  4.20 & 62.80 \\ 
	1000 & AIPTW PV &   1 & -0.0005 [0.0004] & 0.0006 [0.0000] & 0.0412 [0.0003] &  2.75 & 23.70 \\ 
	1000 & G-Formula IPTW &   0 & -0.0006 [0.0002] & 0.0002 [0.0000] & 0.0273 [0.0002] &  0.00 &  0.00 \\ 
	1000 & G-Formula IPTW &   1 & -0.0022 [0.0004] & 0.0004 [0.0000] & 0.0302 [0.0002] &  0.00 &  0.00 \\ 
	1000 & Kaplan-Meier &   0 & -0.0250 [0.0003] & 0.0012 [0.0000] & 0.0598 [0.0004] &  0.00 &  0.00 \\ 
	1000 & Kaplan-Meier &   1 &  0.0369 [0.0004] & 0.0017 [0.0000] & 0.0617 [0.0004] &  0.00 &  0.00 \\
	\bottomrule
\end{longtable}

\newpage

\begin{longtable}{rlrlllll}
	\caption{Estimates for all performance criteria in the \textbf{CO \& ICT} scenario. The values in parentheses are the Monte-Carlo standard errors of the associated estimates. Estimates are based on $2000$ simulation repetitions. $\hat{G}_{Bias}(z)$, $\hat{G}_{MSE}(z)$, $\hat{G}_{Sup}(z)$ and their associated standard errors are rounded to the fourth decimal place, the other statistics are rounded to the second decimal place.}
	\label{tab::appendix_CO_ICT}
	\endfirsthead
	\endhead
	\toprule
	$n$ & Method & $Z$ & $\hat{G}_{Bias}(z)$ & $\hat{G}_{MSE}(z)$ & $\hat{G}_{Sup}(z)$ & \% OOB & \% NM \\ 
	\midrule
	100 & G-Formula &   0 & -0.0055 [0.0007] & 0.0020 [0.0000] & 0.0863 [0.0007] &  0.00 &  0.00 \\ 
	100 & G-Formula &   1 & -0.0087 [0.0011] & 0.0034 [0.0001] & 0.0945 [0.0007] &  0.00 &  0.00 \\ 
	100 & G-Formula PV &   0 & -0.0010 [0.0008] & 0.0028 [0.0001] & 0.1092 [0.0021] &  0.00 & 12.55 \\ 
	100 & G-Formula PV &   1 & -0.0023 [0.0012] & 0.0047 [0.0001] & 0.1188 [0.0021] &  0.00 & 16.45 \\ 
	100 & IPTW KM &   0 & -0.0080 [0.0008] & 0.0032 [0.0001] & 0.1151 [0.0009] &  0.00 &  0.00 \\ 
	100 & IPTW KM &   1 & -0.0206 [0.0014] & 0.0063 [0.0001] & 0.1314 [0.0010] &  0.00 &  0.00 \\ 
	100 & IPTW Cox &   0 & -0.0153 [0.0008] & 0.0033 [0.0001] & 0.1152 [0.0009] &  0.00 &  0.00 \\ 
	100 & IPTW Cox &   1 & -0.0325 [0.0014] & 0.0068 [0.0002] & 0.1325 [0.0011] &  0.00 &  0.00 \\ 
	100 & IPTW PV &   0 & -0.0078 [0.0008] & 0.0032 [0.0001] & 0.1151 [0.0009] & 12.95 & 75.90 \\ 
	100 & IPTW PV &   1 & -0.0212 [0.0014] & 0.0065 [0.0001] & 0.1324 [0.0011] & 19.15 & 79.90 \\ 
	100 & Matching &   0 & -0.0008 [0.0010] & 0.0047 [0.0001] & 0.1423 [0.0011] &  0.00 &  0.00 \\ 
	100 & Matching &   1 & -0.0036 [0.0016] & 0.0078 [0.0002] & 0.1494 [0.0011] &  0.00 &  0.00 \\ 
	100 & EL &   0 & -0.0091 [0.0008] & 0.0034 [0.0001] & 0.1188 [0.0009] &  0.00 &  0.00 \\ 
	100 & EL &   1 & -0.0222 [0.0015] & 0.0072 [0.0004] & 0.1365 [0.0012] &  0.00 &  0.00 \\ 
	100 & AIPTW &   0 & -0.0029 [0.0007] & 0.0027 [0.0001] & 0.1081 [0.0008] &  4.20 & 78.05 \\ 
	100 & AIPTW &   1 & -0.0060 [0.0013] & 0.0052 [0.0001] & 0.1243 [0.0009] & 26.65 & 84.95 \\ 
	100 & AIPTW PV &   0 & -0.0024 [0.0008] & 0.0029 [0.0001] & 0.1108 [0.0008] & 17.15 & 77.60 \\ 
	100 & AIPTW PV &   1 & -0.0039 [0.0013] & 0.0056 [0.0001] & 0.1287 [0.0010] & 45.80 & 88.20 \\ 
	100 & G-Formula IPTW &   0 & -0.0058 [0.0007] & 0.0020 [0.0000] & 0.0866 [0.0007] &  0.00 &  0.00 \\ 
	100 & G-Formula IPTW &   1 & -0.0096 [0.0011] & 0.0034 [0.0001] & 0.0952 [0.0007] &  0.00 &  0.00 \\ 
	100 & Kaplan-Meier &   0 & -0.0275 [0.0008] & 0.0042 [0.0001] & 0.1244 [0.0010] &  0.00 &  0.00 \\ 
	100 & Kaplan-Meier &   1 &  0.0331 [0.0013] & 0.0062 [0.0001] & 0.1319 [0.0010] &  0.00 &  0.00 \\ 
	300 & G-Formula &   0 & -0.0012 [0.0004] & 0.0006 [0.0000] & 0.0488 [0.0004] &  0.00 &  0.00 \\ 
	300 & G-Formula &   1 & -0.0033 [0.0007] & 0.0011 [0.0000] & 0.0544 [0.0004] &  0.00 &  0.00 \\ 
	300 & G-Formula PV &   0 &  0.0005 [0.0004] & 0.0009 [0.0000] & 0.0586 [0.0012] &  0.00 &  2.60 \\ 
	300 & G-Formula PV &   1 & -0.0007 [0.0007] & 0.0016 [0.0000] & 0.0649 [0.0011] &  0.00 &  4.90 \\ 
	300 & IPTW KM &   0 & -0.0054 [0.0005] & 0.0011 [0.0000] & 0.0658 [0.0005] &  0.00 &  0.00 \\ 
	300 & IPTW KM &   1 & -0.0175 [0.0008] & 0.0024 [0.0000] & 0.0779 [0.0006] &  0.00 &  0.00 \\ 
	300 & IPTW Cox &   0 & -0.0079 [0.0005] & 0.0011 [0.0000] & 0.0661 [0.0005] &  0.00 &  0.00 \\ 
	300 & IPTW Cox &   1 & -0.0220 [0.0008] & 0.0025 [0.0001] & 0.0789 [0.0006] &  0.00 &  0.00 \\ 
	300 & IPTW PV &   0 & -0.0053 [0.0005] & 0.0011 [0.0000] & 0.0659 [0.0005] &  8.85 & 68.65 \\ 
	300 & IPTW PV &   1 & -0.0178 [0.0008] & 0.0024 [0.0000] & 0.0783 [0.0006] & 17.00 & 58.85 \\ 
	300 & Matching &   0 &  0.0033 [0.0006] & 0.0017 [0.0000] & 0.0841 [0.0006] &  0.00 &  0.00 \\ 
	300 & Matching &   1 &  0.0022 [0.0010] & 0.0028 [0.0001] & 0.0880 [0.0006] &  0.00 &  0.00 \\ 
	300 & EL &   0 & -0.0067 [0.0005] & 0.0011 [0.0000] & 0.0667 [0.0005] &  0.00 &  0.00 \\ 
	300 & EL &   1 & -0.0172 [0.0008] & 0.0024 [0.0000] & 0.0784 [0.0006] &  0.00 &  0.00 \\ 
	300 & AIPTW &   0 & -0.0002 [0.0004] & 0.0009 [0.0000] & 0.0616 [0.0005] &  1.85 & 67.50 \\ 
	300 & AIPTW &   1 & -0.0022 [0.0007] & 0.0018 [0.0000] & 0.0716 [0.0005] & 22.15 & 65.75 \\ 
	300 & AIPTW PV &   0 & -0.0002 [0.0004] & 0.0010 [0.0000] & 0.0629 [0.0005] &  7.70 & 63.35 \\ 
	300 & AIPTW PV &   1 & -0.0014 [0.0008] & 0.0019 [0.0000] & 0.0736 [0.0005] & 24.25 & 74.50 \\ 
	300 & G-Formula IPTW &   0 & -0.0013 [0.0004] & 0.0006 [0.0000] & 0.0489 [0.0004] &  0.00 &  0.00 \\ 
	300 & G-Formula IPTW &   1 & -0.0037 [0.0007] & 0.0011 [0.0000] & 0.0546 [0.0004] &  0.00 &  0.00 \\ 
	300 & Kaplan-Meier &   0 & -0.0250 [0.0005] & 0.0019 [0.0000] & 0.0813 [0.0007] &  0.00 &  0.00 \\ 
	300 & Kaplan-Meier &   1 &  0.0367 [0.0008] & 0.0030 [0.0001] & 0.0858 [0.0007] &  0.00 &  0.00 \\ 
	500 & G-Formula &   0 & -0.0007 [0.0003] & 0.0004 [0.0000] & 0.0383 [0.0003] &  0.00 &  0.00 \\ 
	500 & G-Formula &   1 & -0.0012 [0.0005] & 0.0007 [0.0000] & 0.0424 [0.0003] &  0.00 &  0.00 \\ 
	500 & G-Formula PV &   0 &  0.0003 [0.0004] & 0.0005 [0.0000] & 0.0431 [0.0003] &  0.00 &  1.30 \\ 
	500 & G-Formula PV &   1 &  0.0009 [0.0006] & 0.0010 [0.0000] & 0.0493 [0.0006] &  0.00 &  2.45 \\ 
	500 & IPTW KM &   0 & -0.0051 [0.0004] & 0.0007 [0.0000] & 0.0524 [0.0004] &  0.00 &  0.00 \\ 
	500 & IPTW KM &   1 & -0.0155 [0.0007] & 0.0015 [0.0000] & 0.0614 [0.0005] &  0.00 &  0.00 \\ 
	500 & IPTW Cox &   0 & -0.0066 [0.0004] & 0.0007 [0.0000] & 0.0525 [0.0004] &  0.00 &  0.00 \\ 
	500 & IPTW Cox &   1 & -0.0182 [0.0006] & 0.0016 [0.0000] & 0.0622 [0.0005] &  0.00 &  0.00 \\ 
	500 & IPTW PV &   0 & -0.0050 [0.0004] & 0.0007 [0.0000] & 0.0525 [0.0004] &  5.45 & 66.60 \\ 
	500 & IPTW PV &   1 & -0.0156 [0.0007] & 0.0015 [0.0000] & 0.0617 [0.0005] & 10.70 & 43.25 \\ 
	500 & Matching &   0 &  0.0055 [0.0005] & 0.0011 [0.0000] & 0.0673 [0.0005] &  0.00 &  0.00 \\ 
	500 & Matching &   1 &  0.0040 [0.0007] & 0.0017 [0.0000] & 0.0693 [0.0005] &  0.00 &  0.00 \\ 
	500 & EL &   0 & -0.0064 [0.0004] & 0.0007 [0.0000] & 0.0525 [0.0004] &  0.00 &  0.00 \\ 
	500 & EL &   1 & -0.0151 [0.0006] & 0.0015 [0.0000] & 0.0619 [0.0005] &  0.00 &  0.00 \\ 
	500 & AIPTW &   0 &  0.0000 [0.0003] & 0.0006 [0.0000] & 0.0484 [0.0004] &  0.60 & 58.50 \\ 
	500 & AIPTW &   1 & -0.0001 [0.0006] & 0.0011 [0.0000] & 0.0560 [0.0004] & 14.15 & 55.65 \\ 
	500 & AIPTW PV &   0 & -0.0001 [0.0004] & 0.0006 [0.0000] & 0.0495 [0.0004] &  3.85 & 57.45 \\ 
	500 & AIPTW PV &   1 &  0.0008 [0.0006] & 0.0012 [0.0000] & 0.0575 [0.0004] & 13.85 & 63.65 \\ 
	500 & G-Formula IPTW &   0 & -0.0007 [0.0003] & 0.0004 [0.0000] & 0.0384 [0.0003] &  0.00 &  0.00 \\ 
	500 & G-Formula IPTW &   1 & -0.0014 [0.0005] & 0.0007 [0.0000] & 0.0424 [0.0003] &  0.00 &  0.00 \\ 
	500 & Kaplan-Meier &   0 & -0.0248 [0.0004] & 0.0015 [0.0000] & 0.0699 [0.0005] &  0.00 &  0.00 \\ 
	500 & Kaplan-Meier &   1 &  0.0379 [0.0006] & 0.0024 [0.0000] & 0.0738 [0.0006] &  0.00 &  0.00 \\ 
	1000 & G-Formula &   0 &  0.0000 [0.0002] & 0.0002 [0.0000] & 0.0271 [0.0002] &  0.00 &  0.00 \\ 
	1000 & G-Formula &   1 & -0.0004 [0.0004] & 0.0003 [0.0000] & 0.0294 [0.0002] &  0.00 &  0.00 \\ 
	1000 & G-Formula PV &   0 &  0.0005 [0.0002] & 0.0002 [0.0000] & 0.0303 [0.0002] &  0.00 &  0.75 \\ 
	1000 & G-Formula PV &   1 &  0.0003 [0.0004] & 0.0004 [0.0000] & 0.0336 [0.0002] &  0.00 &  0.90 \\ 
	1000 & IPTW KM &   0 & -0.0048 [0.0003] & 0.0004 [0.0000] & 0.0380 [0.0003] &  0.00 &  0.00 \\ 
	1000 & IPTW KM &   1 & -0.0158 [0.0004] & 0.0008 [0.0000] & 0.0448 [0.0003] &  0.00 &  0.00 \\ 
	1000 & IPTW Cox &   0 & -0.0055 [0.0003] & 0.0004 [0.0000] & 0.0381 [0.0003] &  0.00 &  0.00 \\ 
	1000 & IPTW Cox &   1 & -0.0171 [0.0004] & 0.0009 [0.0000] & 0.0454 [0.0003] &  0.00 &  0.00 \\ 
	1000 & IPTW PV &   0 & -0.0048 [0.0003] & 0.0004 [0.0000] & 0.0380 [0.0003] &  1.95 & 64.25 \\ 
	1000 & IPTW PV &   1 & -0.0158 [0.0005] & 0.0008 [0.0000] & 0.0450 [0.0003] &  2.65 & 22.75 \\ 
	1000 & Matching &   0 &  0.0068 [0.0003] & 0.0006 [0.0000] & 0.0499 [0.0004] &  0.00 &  0.00 \\ 
	1000 & Matching &   1 &  0.0038 [0.0005] & 0.0008 [0.0000] & 0.0474 [0.0003] &  0.00 &  0.00 \\ 
	1000 & EL &   0 & -0.0061 [0.0003] & 0.0004 [0.0000] & 0.0383 [0.0003] &  0.00 &  0.00 \\ 
	1000 & EL &   1 & -0.0151 [0.0004] & 0.0008 [0.0000] & 0.0448 [0.0003] &  0.00 &  0.00 \\ 
	1000 & AIPTW &   0 &  0.0004 [0.0002] & 0.0003 [0.0000] & 0.0338 [0.0003] &  0.05 & 50.05 \\ 
	1000 & AIPTW &   1 &  0.0000 [0.0004] & 0.0005 [0.0000] & 0.0388 [0.0003] &  3.70 & 37.15 \\ 
	1000 & AIPTW PV &   0 &  0.0003 [0.0003] & 0.0003 [0.0000] & 0.0348 [0.0003] &  0.95 & 53.40 \\ 
	1000 & AIPTW PV &   1 &  0.0003 [0.0004] & 0.0006 [0.0000] & 0.0398 [0.0003] &  3.60 & 39.40 \\ 
	1000 & G-Formula IPTW &   0 &  0.0000 [0.0002] & 0.0002 [0.0000] & 0.0271 [0.0002] &  0.00 &  0.00 \\ 
	1000 & G-Formula IPTW &   1 & -0.0005 [0.0004] & 0.0003 [0.0000] & 0.0295 [0.0002] &  0.00 &  0.00 \\ 
	1000 & Kaplan-Meier &   0 & -0.0245 [0.0003] & 0.0011 [0.0000] & 0.0597 [0.0004] &  0.00 &  0.00 \\ 
	1000 & Kaplan-Meier &   1 &  0.0377 [0.0004] & 0.0017 [0.0000] & 0.0618 [0.0004] &  0.00 &  0.00 \\
	\bottomrule
\end{longtable}

\newpage

\begin{longtable}{rlrlllll}
	\caption{Estimates for all performance criteria in the \textbf{ICO \& CT} scenario. The values in parentheses are the Monte-Carlo standard errors of the associated estimates. Estimates are based on $2000$ simulation repetitions. $\hat{G}_{Bias}(z)$, $\hat{G}_{MSE}(z)$, $\hat{G}_{Sup}(z)$ and their associated standard errors are rounded to the fourth decimal place, the other statistics are rounded to the second decimal place.}
	\label{tab::appendix_ICO_CT}
	\endfirsthead
	\endhead
	\toprule
	$n$ & Method & $Z$ & $\hat{G}_{Bias}(z)$ & $\hat{G}_{MSE}(z)$ & $\hat{G}_{Sup}(z)$ & \% OOB & \% NM \\ 
	\midrule
	100 & G-Formula &   0 & -0.0051 [0.0008] & 0.0025 [0.0001] & 0.0930 [0.0008] &  0.00 &  0.00 \\ 
	100 & G-Formula &   1 & -0.0088 [0.0013] & 0.0042 [0.0001] & 0.1018 [0.0008] &  0.00 &  0.00 \\ 
	100 & G-Formula PV &   0 &  0.0050 [0.0009] & 0.0033 [0.0001] & 0.1119 [0.0023] &  0.00 & 13.00 \\ 
	100 & G-Formula PV &   1 &  0.0006 [0.0014] & 0.0056 [0.0002] & 0.1248 [0.0024] &  0.00 & 17.05 \\ 
	100 & IPTW KM &   0 & -0.0069 [0.0011] & 0.0049 [0.0001] & 0.1383 [0.0013] &  0.00 &  0.00 \\ 
	100 & IPTW KM &   1 & -0.0048 [0.0015] & 0.0066 [0.0001] & 0.1387 [0.0011] &  0.00 &  0.00 \\ 
	100 & IPTW Cox &   0 & -0.0164 [0.0010] & 0.0048 [0.0001] & 0.1353 [0.0012] &  0.00 &  0.00 \\ 
	100 & IPTW Cox &   1 & -0.0183 [0.0014] & 0.0067 [0.0001] & 0.1368 [0.0011] &  0.00 &  0.00 \\ 
	100 & IPTW PV &   0 & -0.0067 [0.0011] & 0.0051 [0.0001] & 0.1394 [0.0013] & 13.50 & 75.05 \\ 
	100 & IPTW PV &   1 & -0.0051 [0.0015] & 0.0068 [0.0001] & 0.1396 [0.0011] & 22.45 & 80.80 \\ 
	100 & Matching &   0 & -0.0073 [0.0012] & 0.0065 [0.0002] & 0.1590 [0.0013] &  0.00 &  0.00 \\ 
	100 & Matching &   1 & -0.0069 [0.0017] & 0.0088 [0.0002] & 0.1574 [0.0012] &  0.00 &  0.00 \\ 
	100 & EL &   0 & -0.0113 [0.0013] & 0.0066 [0.0006] & 0.1455 [0.0016] &  0.00 &  0.00 \\ 
	100 & EL &   1 & -0.0183 [0.0021] & 0.0120 [0.0010] & 0.1590 [0.0019] &  0.00 &  0.00 \\ 
	100 & AIPTW &   0 & -0.0034 [0.0010] & 0.0050 [0.0003] & 0.1363 [0.0019] &  7.60 & 71.70 \\ 
	100 & AIPTW &   1 & -0.0062 [0.0014] & 0.0064 [0.0001] & 0.1366 [0.0010] & 23.55 & 79.60 \\ 
	100 & AIPTW PV &   0 & -0.0027 [0.0010] & 0.0055 [0.0004] & 0.1386 [0.0018] & 18.40 & 72.00 \\ 
	100 & AIPTW PV &   1 & -0.0042 [0.0014] & 0.0066 [0.0001] & 0.1387 [0.0011] & 40.40 & 84.50 \\ 
	100 & G-Formula IPTW &   0 & -0.0072 [0.0009] & 0.0026 [0.0001] & 0.0945 [0.0008] &  0.00 &  0.00 \\ 
	100 & G-Formula IPTW &   1 & -0.0134 [0.0013] & 0.0046 [0.0001] & 0.1052 [0.0009] &  0.00 &  0.00 \\ 
	100 & Kaplan-Meier &   0 & -0.0269 [0.0009] & 0.0043 [0.0001] & 0.1250 [0.0010] &  0.00 &  0.00 \\ 
	100 & Kaplan-Meier &   1 &  0.0327 [0.0013] & 0.0062 [0.0001] & 0.1324 [0.0010] &  0.00 &  0.00 \\ 
	300 & G-Formula &   0 & -0.0015 [0.0005] & 0.0008 [0.0000] & 0.0532 [0.0004] &  0.00 &  0.00 \\ 
	300 & G-Formula &   1 & -0.0004 [0.0008] & 0.0015 [0.0000] & 0.0602 [0.0005] &  0.00 &  0.00 \\ 
	300 & G-Formula PV &   0 &  0.0057 [0.0005] & 0.0010 [0.0000] & 0.0596 [0.0011] &  0.00 &  5.05 \\ 
	300 & G-Formula PV &   1 &  0.0042 [0.0008] & 0.0018 [0.0001] & 0.0684 [0.0013] &  0.00 &  9.70 \\ 
	300 & IPTW KM &   0 & -0.0033 [0.0007] & 0.0019 [0.0001] & 0.0841 [0.0008] &  0.00 &  0.00 \\ 
	300 & IPTW KM &   1 & -0.0010 [0.0009] & 0.0023 [0.0000] & 0.0789 [0.0006] &  0.00 &  0.00 \\ 
	300 & IPTW Cox &   0 & -0.0070 [0.0006] & 0.0018 [0.0000] & 0.0828 [0.0008] &  0.00 &  0.00 \\ 
	300 & IPTW Cox &   1 & -0.0061 [0.0009] & 0.0023 [0.0000] & 0.0785 [0.0006] &  0.00 &  0.00 \\ 
	300 & IPTW PV &   0 & -0.0032 [0.0007] & 0.0019 [0.0001] & 0.0847 [0.0009] & 10.60 & 69.20 \\ 
	300 & IPTW PV &   1 & -0.0010 [0.0009] & 0.0023 [0.0000] & 0.0790 [0.0006] & 17.30 & 54.20 \\ 
	300 & Matching &   0 & -0.0032 [0.0007] & 0.0023 [0.0001] & 0.0972 [0.0008] &  0.00 &  0.00 \\ 
	300 & Matching &   1 & -0.0012 [0.0010] & 0.0030 [0.0001] & 0.0914 [0.0007] &  0.00 &  0.00 \\ 
	300 & EL &   0 & -0.0043 [0.0006] & 0.0019 [0.0000] & 0.0883 [0.0007] &  0.00 &  0.00 \\ 
	300 & EL &   1 & -0.0105 [0.0014] & 0.0049 [0.0008] & 0.0879 [0.0013] &  0.00 &  0.00 \\ 
	300 & AIPTW &   0 & -0.0011 [0.0006] & 0.0018 [0.0001] & 0.0812 [0.0011] &  6.15 & 71.25 \\ 
	300 & AIPTW &   1 & -0.0014 [0.0009] & 0.0022 [0.0000] & 0.0782 [0.0006] & 15.25 & 50.85 \\ 
	300 & AIPTW PV &   0 & -0.0008 [0.0006] & 0.0019 [0.0002] & 0.0821 [0.0011] & 12.00 & 67.10 \\ 
	300 & AIPTW PV &   1 & -0.0006 [0.0009] & 0.0022 [0.0000] & 0.0788 [0.0006] & 21.35 & 55.20 \\ 
	300 & G-Formula IPTW &   0 & -0.0027 [0.0005] & 0.0008 [0.0000] & 0.0542 [0.0005] &  0.00 &  0.00 \\ 
	300 & G-Formula IPTW &   1 & -0.0034 [0.0008] & 0.0017 [0.0000] & 0.0621 [0.0005] &  0.00 &  0.00 \\ 
	300 & Kaplan-Meier &   0 & -0.0254 [0.0005] & 0.0020 [0.0000] & 0.0816 [0.0007] &  0.00 &  0.00 \\ 
	300 & Kaplan-Meier &   1 &  0.0382 [0.0008] & 0.0031 [0.0001] & 0.0864 [0.0007] &  0.00 &  0.00 \\ 
	500 & G-Formula &   0 & -0.0003 [0.0004] & 0.0005 [0.0000] & 0.0422 [0.0004] &  0.00 &  0.00 \\ 
	500 & G-Formula &   1 & -0.0007 [0.0006] & 0.0010 [0.0000] & 0.0484 [0.0004] &  0.00 &  0.00 \\ 
	500 & G-Formula PV &   0 &  0.0058 [0.0004] & 0.0006 [0.0000] & 0.0459 [0.0006] &  0.00 &  3.45 \\ 
	500 & G-Formula PV &   1 &  0.0027 [0.0006] & 0.0011 [0.0001] & 0.0522 [0.0009] &  0.00 &  6.40 \\ 
	500 & IPTW KM &   0 & -0.0021 [0.0005] & 0.0012 [0.0000] & 0.0676 [0.0007] &  0.00 &  0.00 \\ 
	500 & IPTW KM &   1 & -0.0014 [0.0007] & 0.0014 [0.0000] & 0.0614 [0.0005] &  0.00 &  0.00 \\ 
	500 & IPTW Cox &   0 & -0.0045 [0.0005] & 0.0012 [0.0000] & 0.0667 [0.0006] &  0.00 &  0.00 \\ 
	500 & IPTW Cox &   1 & -0.0044 [0.0007] & 0.0014 [0.0000] & 0.0612 [0.0005] &  0.00 &  0.00 \\ 
	500 & IPTW PV &   0 & -0.0020 [0.0005] & 0.0012 [0.0000] & 0.0680 [0.0007] &  7.15 & 65.90 \\ 
	500 & IPTW PV &   1 & -0.0015 [0.0007] & 0.0014 [0.0000] & 0.0615 [0.0005] & 10.15 & 44.80 \\ 
	500 & Matching &   0 & -0.0016 [0.0006] & 0.0015 [0.0000] & 0.0773 [0.0006] &  0.00 &  0.00 \\ 
	500 & Matching &   1 & -0.0018 [0.0008] & 0.0018 [0.0000] & 0.0711 [0.0005] &  0.00 &  0.00 \\ 
	500 & EL &   0 & -0.0029 [0.0006] & 0.0015 [0.0002] & 0.0754 [0.0007] &  0.00 &  0.00 \\ 
	500 & EL &   1 & -0.0070 [0.0011] & 0.0028 [0.0006] & 0.0665 [0.0010] &  0.00 &  0.00 \\ 
	500 & AIPTW &   0 & -0.0006 [0.0004] & 0.0010 [0.0000] & 0.0646 [0.0007] &  3.15 & 69.50 \\ 
	500 & AIPTW &   1 & -0.0021 [0.0007] & 0.0013 [0.0000] & 0.0607 [0.0004] & 10.00 & 42.75 \\ 
	500 & AIPTW PV &   0 & -0.0003 [0.0005] & 0.0011 [0.0000] & 0.0656 [0.0007] &  9.00 & 65.60 \\ 
	500 & AIPTW PV &   1 & -0.0017 [0.0007] & 0.0013 [0.0000] & 0.0613 [0.0004] & 12.25 & 44.65 \\ 
	500 & G-Formula IPTW &   0 & -0.0013 [0.0004] & 0.0005 [0.0000] & 0.0428 [0.0004] &  0.00 &  0.00 \\ 
	500 & G-Formula IPTW &   1 & -0.0028 [0.0007] & 0.0011 [0.0000] & 0.0502 [0.0004] &  0.00 &  0.00 \\ 
	500 & Kaplan-Meier &   0 & -0.0250 [0.0004] & 0.0015 [0.0000] & 0.0703 [0.0006] &  0.00 &  0.00 \\ 
	500 & Kaplan-Meier &   1 &  0.0371 [0.0006] & 0.0023 [0.0000] & 0.0742 [0.0006] &  0.00 &  0.00 \\ 
	1000 & G-Formula &   0 & -0.0005 [0.0003] & 0.0003 [0.0000] & 0.0301 [0.0002] &  0.00 &  0.00 \\ 
	1000 & G-Formula &   1 &  0.0016 [0.0004] & 0.0005 [0.0000] & 0.0358 [0.0003] &  0.00 &  0.00 \\ 
	1000 & G-Formula PV &   0 &  0.0049 [0.0003] & 0.0003 [0.0000] & 0.0324 [0.0005] &  0.00 &  2.15 \\ 
	1000 & G-Formula PV &   1 &  0.0045 [0.0005] & 0.0006 [0.0000] & 0.0360 [0.0005] &  0.00 &  2.60 \\ 
	1000 & IPTW KM &   0 & -0.0024 [0.0004] & 0.0006 [0.0000] & 0.0490 [0.0005] &  0.00 &  0.00 \\ 
	1000 & IPTW KM &   1 &  0.0000 [0.0005] & 0.0007 [0.0000] & 0.0434 [0.0003] &  0.00 &  0.00 \\ 
	1000 & IPTW Cox &   0 & -0.0037 [0.0004] & 0.0006 [0.0000] & 0.0485 [0.0005] &  0.00 &  0.00 \\ 
	1000 & IPTW Cox &   1 & -0.0015 [0.0005] & 0.0007 [0.0000] & 0.0433 [0.0003] &  0.00 &  0.00 \\ 
	1000 & IPTW PV &   0 & -0.0024 [0.0004] & 0.0006 [0.0000] & 0.0492 [0.0005] &  2.60 & 64.40 \\ 
	1000 & IPTW PV &   1 &  0.0000 [0.0005] & 0.0007 [0.0000] & 0.0435 [0.0003] &  2.05 & 22.60 \\ 
	1000 & Matching &   0 & -0.0016 [0.0004] & 0.0008 [0.0000] & 0.0563 [0.0004] &  0.00 &  0.00 \\ 
	1000 & Matching &   1 &  0.0007 [0.0005] & 0.0009 [0.0000] & 0.0497 [0.0004] &  0.00 &  0.00 \\ 
	1000 & EL &   0 & -0.0023 [0.0004] & 0.0009 [0.0000] & 0.0601 [0.0005] &  0.00 &  0.00 \\ 
	1000 & EL &   1 & -0.0032 [0.0007] & 0.0011 [0.0003] & 0.0453 [0.0006] &  0.00 &  0.00 \\ 
	1000 & AIPTW &   0 & -0.0012 [0.0003] & 0.0005 [0.0000] & 0.0469 [0.0005] &  1.05 & 68.10 \\ 
	1000 & AIPTW &   1 &  0.0000 [0.0005] & 0.0007 [0.0000] & 0.0426 [0.0003] &  2.50 & 22.10 \\ 
	1000 & AIPTW PV &   0 & -0.0011 [0.0003] & 0.0006 [0.0000] & 0.0475 [0.0005] &  4.35 & 63.40 \\ 
	1000 & AIPTW PV &   1 &  0.0003 [0.0005] & 0.0007 [0.0000] & 0.0431 [0.0003] &  2.10 & 23.40 \\ 
	1000 & G-Formula IPTW &   0 & -0.0016 [0.0003] & 0.0003 [0.0000] & 0.0307 [0.0003] &  0.00 &  0.00 \\ 
	1000 & G-Formula IPTW &   1 &  0.0009 [0.0005] & 0.0006 [0.0000] & 0.0375 [0.0003] &  0.00 &  0.00 \\ 
	1000 & Kaplan-Meier &   0 & -0.0251 [0.0003] & 0.0012 [0.0000] & 0.0599 [0.0004] &  0.00 &  0.00 \\ 
	1000 & Kaplan-Meier &   1 &  0.0379 [0.0004] & 0.0018 [0.0000] & 0.0623 [0.0004] &  0.00 &  0.00 \\
	\bottomrule
\end{longtable}

\newpage

\begin{longtable}{rlrlllll}
	\caption{Estimates for all performance criteria in the \textbf{ICO \& ICT} scenario. The values in parentheses are the Monte-Carlo standard errors of the associated estimates. Estimates are based on $2000$ simulation repetitions. $\hat{G}_{Bias}(z)$, $\hat{G}_{MSE}(z)$, $\hat{G}_{Sup}(z)$ and their associated standard errors are rounded to the fourth decimal place, the other statistics are rounded to the second decimal place.}
	\label{tab::appendix_ICO_ICT}
	\endfirsthead
	\endhead
	\toprule
	$n$ & Method & $Z$ & $\hat{G}_{Bias}(z)$ & $\hat{G}_{MSE}(z)$ & $\hat{G}_{Sup}(z)$ & \% OOB & \% NM \\ 
	\midrule
	100 & G-Formula &   0 & -0.0049 [0.0008] & 0.0025 [0.0001] & 0.0927 [0.0008] &  0.00 &  0.00 \\ 
	100 & G-Formula &   1 & -0.0101 [0.0013] & 0.0045 [0.0001] & 0.1039 [0.0008] &  0.00 &  0.00 \\ 
	100 & G-Formula PV &   0 &  0.0059 [0.0009] & 0.0032 [0.0001] & 0.1109 [0.0023] &  0.00 & 12.60 \\ 
	100 & G-Formula PV &   1 & -0.0026 [0.0014] & 0.0059 [0.0002] & 0.1278 [0.0024] &  0.00 & 16.45 \\ 
	100 & IPTW KM &   0 & -0.0082 [0.0008] & 0.0032 [0.0001] & 0.1147 [0.0009] &  0.00 &  0.00 \\ 
	100 & IPTW KM &   1 & -0.0207 [0.0014] & 0.0067 [0.0001] & 0.1352 [0.0010] &  0.00 &  0.00 \\ 
	100 & IPTW Cox &   0 & -0.0155 [0.0008] & 0.0033 [0.0001] & 0.1149 [0.0009] &  0.00 &  0.00 \\ 
	100 & IPTW Cox &   1 & -0.0326 [0.0014] & 0.0071 [0.0002] & 0.1358 [0.0011] &  0.00 &  0.00 \\ 
	100 & IPTW PV &   0 & -0.0082 [0.0008] & 0.0032 [0.0001] & 0.1146 [0.0009] & 11.55 & 74.25 \\ 
	100 & IPTW PV &   1 & -0.0218 [0.0015] & 0.0069 [0.0002] & 0.1366 [0.0011] & 22.60 & 81.10 \\ 
	100 & Matching &   0 & -0.0005 [0.0010] & 0.0048 [0.0001] & 0.1421 [0.0011] &  0.00 &  0.00 \\ 
	100 & Matching &   1 & -0.0038 [0.0016] & 0.0082 [0.0002] & 0.1531 [0.0011] &  0.00 &  0.00 \\ 
	100 & EL &   0 & -0.0087 [0.0008] & 0.0034 [0.0001] & 0.1188 [0.0009] &  0.00 &  0.00 \\ 
	100 & EL &   1 & -0.0199 [0.0015] & 0.0070 [0.0001] & 0.1388 [0.0010] &  0.00 &  0.00 \\ 
	100 & AIPTW &   0 &  0.0139 [0.0008] & 0.0033 [0.0001] & 0.1155 [0.0009] &  6.00 & 84.40 \\ 
	100 & AIPTW &   1 & -0.0526 [0.0015] & 0.0090 [0.0002] & 0.1476 [0.0012] & 37.55 & 92.35 \\ 
	100 & AIPTW PV &   0 &  0.0141 [0.0008] & 0.0033 [0.0001] & 0.1162 [0.0009] & 17.60 & 84.50 \\ 
	100 & AIPTW PV &   1 & -0.0486 [0.0016] & 0.0091 [0.0002] & 0.1498 [0.0013] & 46.10 & 90.85 \\ 
	100 & G-Formula IPTW &   0 &  0.0089 [0.0008] & 0.0025 [0.0001] & 0.0941 [0.0008] &  0.00 &  0.00 \\ 
	100 & G-Formula IPTW &   1 & -0.0404 [0.0014] & 0.0058 [0.0001] & 0.1117 [0.0010] &  0.00 &  0.00 \\ 
	100 & Kaplan-Meier &   0 & -0.0275 [0.0009] & 0.0042 [0.0001] & 0.1248 [0.0010] &  0.00 &  0.00 \\ 
	100 & Kaplan-Meier &   1 &  0.0324 [0.0013] & 0.0064 [0.0001] & 0.1332 [0.0011] &  0.00 &  0.00 \\ 
	300 & G-Formula &   0 & -0.0010 [0.0005] & 0.0008 [0.0000] & 0.0547 [0.0005] &  0.00 &  0.00 \\ 
	300 & G-Formula &   1 & -0.0019 [0.0008] & 0.0016 [0.0000] & 0.0610 [0.0005] &  0.00 &  0.00 \\ 
	300 & G-Formula PV &   0 &  0.0064 [0.0005] & 0.0011 [0.0000] & 0.0626 [0.0015] &  0.00 &  5.20 \\ 
	300 & G-Formula PV &   1 &  0.0031 [0.0009] & 0.0019 [0.0001] & 0.0697 [0.0015] &  0.00 &  7.80 \\ 
	300 & IPTW KM &   0 & -0.0047 [0.0005] & 0.0011 [0.0000] & 0.0662 [0.0005] &  0.00 &  0.00 \\ 
	300 & IPTW KM &   1 & -0.0171 [0.0008] & 0.0024 [0.0000] & 0.0779 [0.0006] &  0.00 &  0.00 \\ 
	300 & IPTW Cox &   0 & -0.0072 [0.0005] & 0.0011 [0.0000] & 0.0664 [0.0005] &  0.00 &  0.00 \\ 
	300 & IPTW Cox &   1 & -0.0216 [0.0008] & 0.0025 [0.0001] & 0.0789 [0.0006] &  0.00 &  0.00 \\ 
	300 & IPTW PV &   0 & -0.0046 [0.0005] & 0.0011 [0.0000] & 0.0663 [0.0005] &  8.95 & 67.35 \\ 
	300 & IPTW PV &   1 & -0.0172 [0.0009] & 0.0024 [0.0000] & 0.0782 [0.0006] & 16.50 & 57.80 \\ 
	300 & Matching &   0 &  0.0051 [0.0006] & 0.0017 [0.0000] & 0.0858 [0.0006] &  0.00 &  0.00 \\ 
	300 & Matching &   1 &  0.0018 [0.0009] & 0.0027 [0.0000] & 0.0876 [0.0006] &  0.00 &  0.00 \\ 
	300 & EL &   0 & -0.0064 [0.0005] & 0.0012 [0.0000] & 0.0680 [0.0005] &  0.00 &  0.00 \\ 
	300 & EL &   1 & -0.0161 [0.0009] & 0.0024 [0.0000] & 0.0793 [0.0006] &  0.00 &  0.00 \\ 
	300 & AIPTW &   0 &  0.0174 [0.0005] & 0.0014 [0.0000] & 0.0714 [0.0006] & 10.85 & 91.60 \\ 
	300 & AIPTW &   1 & -0.0481 [0.0009] & 0.0040 [0.0001] & 0.0936 [0.0007] & 32.65 & 47.05 \\ 
	300 & AIPTW PV &   0 &  0.0176 [0.0005] & 0.0014 [0.0000] & 0.0722 [0.0006] & 12.40 & 89.70 \\ 
	300 & AIPTW PV &   1 & -0.0433 [0.0009] & 0.0039 [0.0001] & 0.0927 [0.0007] & 30.15 & 48.10 \\ 
	300 & G-Formula IPTW &   0 &  0.0148 [0.0005] & 0.0011 [0.0000] & 0.0606 [0.0005] &  0.00 &  0.00 \\ 
	300 & G-Formula IPTW &   1 & -0.0339 [0.0008] & 0.0024 [0.0001] & 0.0690 [0.0006] &  0.00 &  0.00 \\ 
	300 & Kaplan-Meier &   0 & -0.0243 [0.0005] & 0.0019 [0.0000] & 0.0804 [0.0007] &  0.00 &  0.00 \\ 
	300 & Kaplan-Meier &   1 &  0.0364 [0.0008] & 0.0030 [0.0001] & 0.0861 [0.0007] &  0.00 &  0.00 \\ 
	500 & G-Formula &   0 & -0.0012 [0.0004] & 0.0005 [0.0000] & 0.0422 [0.0003] &  0.00 &  0.00 \\ 
	500 & G-Formula &   1 &  0.0010 [0.0006] & 0.0010 [0.0000] & 0.0480 [0.0004] &  0.00 &  0.00 \\ 
	500 & G-Formula PV &   0 &  0.0049 [0.0004] & 0.0006 [0.0000] & 0.0462 [0.0008] &  0.00 &  4.85 \\ 
	500 & G-Formula PV &   1 &  0.0049 [0.0006] & 0.0011 [0.0000] & 0.0515 [0.0008] &  0.00 &  6.50 \\ 
	500 & IPTW KM &   0 & -0.0055 [0.0004] & 0.0007 [0.0000] & 0.0520 [0.0004] &  0.00 &  0.00 \\ 
	500 & IPTW KM &   1 & -0.0157 [0.0006] & 0.0015 [0.0000] & 0.0607 [0.0005] &  0.00 &  0.00 \\ 
	500 & IPTW Cox &   0 & -0.0070 [0.0004] & 0.0007 [0.0000] & 0.0522 [0.0004] &  0.00 &  0.00 \\ 
	500 & IPTW Cox &   1 & -0.0184 [0.0006] & 0.0015 [0.0000] & 0.0615 [0.0005] &  0.00 &  0.00 \\ 
	500 & IPTW PV &   0 & -0.0055 [0.0004] & 0.0007 [0.0000] & 0.0521 [0.0004] &  4.35 & 67.85 \\ 
	500 & IPTW PV &   1 & -0.0158 [0.0006] & 0.0015 [0.0000] & 0.0610 [0.0005] & 10.50 & 42.00 \\ 
	500 & Matching &   0 &  0.0052 [0.0005] & 0.0011 [0.0000] & 0.0675 [0.0005] &  0.00 &  0.00 \\ 
	500 & Matching &   1 &  0.0031 [0.0007] & 0.0016 [0.0000] & 0.0673 [0.0005] &  0.00 &  0.00 \\ 
	500 & EL &   0 & -0.0074 [0.0004] & 0.0007 [0.0000] & 0.0537 [0.0004] &  0.00 &  0.00 \\ 
	500 & EL &   1 & -0.0145 [0.0007] & 0.0015 [0.0000] & 0.0614 [0.0005] &  0.00 &  0.00 \\ 
	500 & AIPTW &   0 &  0.0169 [0.0004] & 0.0009 [0.0000] & 0.0572 [0.0004] & 15.25 & 94.45 \\ 
	500 & AIPTW &   1 & -0.0463 [0.0007] & 0.0031 [0.0001] & 0.0791 [0.0006] & 20.60 & 23.75 \\ 
	500 & AIPTW PV &   0 &  0.0167 [0.0004] & 0.0009 [0.0000] & 0.0575 [0.0004] & 12.75 & 92.30 \\ 
	500 & AIPTW PV &   1 & -0.0417 [0.0007] & 0.0028 [0.0001] & 0.0774 [0.0006] & 16.95 & 26.90 \\ 
	500 & G-Formula IPTW &   0 &  0.0153 [0.0004] & 0.0008 [0.0000] & 0.0500 [0.0004] &  0.00 &  0.00 \\ 
	500 & G-Formula IPTW &   1 & -0.0316 [0.0006] & 0.0017 [0.0000] & 0.0561 [0.0005] &  0.00 &  0.00 \\ 
	500 & Kaplan-Meier &   0 & -0.0253 [0.0004] & 0.0015 [0.0000] & 0.0705 [0.0006] &  0.00 &  0.00 \\ 
	500 & Kaplan-Meier &   1 &  0.0382 [0.0006] & 0.0023 [0.0000] & 0.0733 [0.0006] &  0.00 &  0.00 \\ 
	1000 & G-Formula &   0 & -0.0003 [0.0003] & 0.0003 [0.0000] & 0.0301 [0.0003] &  0.00 &  0.00 \\ 
	1000 & G-Formula &   1 &  0.0005 [0.0004] & 0.0005 [0.0000] & 0.0359 [0.0003] &  0.00 &  0.00 \\ 
	1000 & G-Formula PV &   0 &  0.0051 [0.0003] & 0.0003 [0.0000] & 0.0323 [0.0003] &  0.00 &  2.90 \\ 
	1000 & G-Formula PV &   1 &  0.0033 [0.0005] & 0.0005 [0.0000] & 0.0355 [0.0003] &  0.00 &  3.50 \\ 
	1000 & IPTW KM &   0 & -0.0054 [0.0003] & 0.0004 [0.0000] & 0.0377 [0.0003] &  0.00 &  0.00 \\ 
	1000 & IPTW KM &   1 & -0.0165 [0.0005] & 0.0009 [0.0000] & 0.0452 [0.0004] &  0.00 &  0.00 \\ 
	1000 & IPTW Cox &   0 & -0.0062 [0.0003] & 0.0004 [0.0000] & 0.0379 [0.0003] &  0.00 &  0.00 \\ 
	1000 & IPTW Cox &   1 & -0.0178 [0.0005] & 0.0009 [0.0000] & 0.0458 [0.0004] &  0.00 &  0.00 \\ 
	1000 & IPTW PV &   0 & -0.0054 [0.0003] & 0.0004 [0.0000] & 0.0378 [0.0003] &  1.75 & 63.30 \\ 
	1000 & IPTW PV &   1 & -0.0166 [0.0005] & 0.0009 [0.0000] & 0.0454 [0.0004] &  2.35 & 23.50 \\ 
	1000 & Matching &   0 &  0.0060 [0.0003] & 0.0006 [0.0000] & 0.0501 [0.0004] &  0.00 &  0.00 \\ 
	1000 & Matching &   1 &  0.0034 [0.0005] & 0.0008 [0.0000] & 0.0483 [0.0004] &  0.00 &  0.00 \\ 
	1000 & EL &   0 & -0.0071 [0.0003] & 0.0004 [0.0000] & 0.0392 [0.0003] &  0.00 &  0.00 \\ 
	1000 & EL &   1 & -0.0153 [0.0005] & 0.0009 [0.0000] & 0.0454 [0.0003] &  0.00 &  0.00 \\ 
	1000 & AIPTW &   0 &  0.0173 [0.0003] & 0.0007 [0.0000] & 0.0456 [0.0004] & 21.05 & 96.45 \\ 
	1000 & AIPTW &   1 & -0.0472 [0.0005] & 0.0025 [0.0000] & 0.0674 [0.0005] &  5.00 &  6.90 \\ 
	1000 & AIPTW PV &   0 &  0.0171 [0.0003] & 0.0006 [0.0000] & 0.0456 [0.0004] & 15.20 & 93.65 \\ 
	1000 & AIPTW PV &   1 & -0.0427 [0.0005] & 0.0022 [0.0000] & 0.0650 [0.0005] &  4.95 & 11.70 \\ 
	1000 & G-Formula IPTW &   0 &  0.0163 [0.0003] & 0.0006 [0.0000] & 0.0419 [0.0004] &  0.00 &  0.00 \\ 
	1000 & G-Formula IPTW &   1 & -0.0319 [0.0004] & 0.0013 [0.0000] & 0.0466 [0.0004] &  0.00 &  0.00 \\ 
	1000 & Kaplan-Meier &   0 & -0.0252 [0.0003] & 0.0012 [0.0000] & 0.0602 [0.0004] &  0.00 &  0.00 \\ 
	1000 & Kaplan-Meier &   1 &  0.0373 [0.0004] & 0.0017 [0.0000] & 0.0621 [0.0004] &  0.00 &  0.00 \\
	\bottomrule
\end{longtable}

\newpage

\begin{longtable}{rlrlllll}
	\caption{Estimates for all performance criteria in the \textbf{PCO \& PCT} scenario. The values in parentheses are the Monte-Carlo standard errors of the associated estimates. Estimates are based on $2000$ simulation repetitions. $\hat{G}_{Bias}(z)$, $\hat{G}_{MSE}(z)$, $\hat{G}_{Sup}(z)$ and their associated standard errors are rounded to the fourth decimal place, the other statistics are rounded to the second decimal place.}
	\label{tab::appendix_PCO_PCT}
	\endfirsthead
	\endhead
	\toprule
	$n$ & Method & $Z$ & $\hat{G}_{Bias}(z)$ & $\hat{G}_{MSE}(z)$ & $\hat{G}_{Sup}(z)$ & \% OOB & \% NM \\ 
	\midrule
	100 & G-Formula &   0 & -0.0062 [0.0008] & 0.0023 [0.0001] & 0.0915 [0.0008] &  0.00 &  0.00 \\ 
	100 & G-Formula &   1 & -0.0101 [0.0012] & 0.0039 [0.0001] & 0.0997 [0.0008] &  0.00 &  0.00 \\ 
	100 & G-Formula PV &   0 &  0.0031 [0.0008] & 0.0031 [0.0001] & 0.1098 [0.0022] &  0.00 & 12.85 \\ 
	100 & G-Formula PV &   1 & -0.0023 [0.0013] & 0.0052 [0.0001] & 0.1236 [0.0023] &  0.00 & 15.70 \\ 
	100 & IPTW KM &   0 & -0.0081 [0.0009] & 0.0034 [0.0001] & 0.1184 [0.0009] &  0.00 &  0.00 \\ 
	100 & IPTW KM &   1 & -0.0221 [0.0015] & 0.0070 [0.0002] & 0.1366 [0.0011] &  0.00 &  0.00 \\ 
	100 & IPTW Cox &   0 & -0.0155 [0.0008] & 0.0035 [0.0001] & 0.1185 [0.0010] &  0.00 &  0.00 \\ 
	100 & IPTW Cox &   1 & -0.0344 [0.0015] & 0.0074 [0.0002] & 0.1371 [0.0011] &  0.00 &  0.00 \\ 
	100 & IPTW PV &   0 & -0.0080 [0.0009] & 0.0035 [0.0001] & 0.1185 [0.0009] & 13.15 & 76.10 \\ 
	100 & IPTW PV &   1 & -0.0229 [0.0015] & 0.0072 [0.0002] & 0.1380 [0.0011] & 22.05 & 81.45 \\ 
	100 & Matching &   0 &  0.0000 [0.0010] & 0.0048 [0.0001] & 0.1419 [0.0011] &  0.00 &  0.00 \\ 
	100 & Matching &   1 & -0.0124 [0.0016] & 0.0081 [0.0002] & 0.1503 [0.0011] &  0.00 &  0.00 \\ 
	100 & EL &   0 & -0.0091 [0.0009] & 0.0037 [0.0002] & 0.1196 [0.0010] &  0.00 &  0.00 \\ 
	100 & EL &   1 & -0.0202 [0.0014] & 0.0068 [0.0001] & 0.1361 [0.0010] &  0.00 &  0.00 \\ 
	100 & AIPTW &   0 & -0.0036 [0.0008] & 0.0030 [0.0001] & 0.1125 [0.0009] &  3.65 & 77.40 \\ 
	100 & AIPTW &   1 & -0.0096 [0.0014] & 0.0065 [0.0002] & 0.1337 [0.0011] & 28.15 & 83.25 \\ 
	100 & AIPTW PV &   0 & -0.0025 [0.0008] & 0.0032 [0.0001] & 0.1140 [0.0009] & 14.30 & 76.55 \\ 
	100 & AIPTW PV &   1 & -0.0072 [0.0014] & 0.0065 [0.0002] & 0.1355 [0.0012] & 43.35 & 88.75 \\ 
	100 & G-Formula IPTW &   0 & -0.0072 [0.0008] & 0.0024 [0.0001] & 0.0920 [0.0008] &  0.00 &  0.00 \\ 
	100 & G-Formula IPTW &   1 & -0.0132 [0.0013] & 0.0042 [0.0001] & 0.1016 [0.0008] &  0.00 &  0.00 \\ 
	100 & Kaplan-Meier &   0 & -0.0269 [0.0009] & 0.0043 [0.0001] & 0.1254 [0.0010] &  0.00 &  0.00 \\ 
	100 & Kaplan-Meier &   1 &  0.0314 [0.0013] & 0.0061 [0.0001] & 0.1310 [0.0010] &  0.00 &  0.00 \\ 
	300 & G-Formula &   0 & -0.0015 [0.0004] & 0.0007 [0.0000] & 0.0518 [0.0004] &  0.00 &  0.00 \\ 
	300 & G-Formula &   1 & -0.0026 [0.0007] & 0.0013 [0.0000] & 0.0571 [0.0004] &  0.00 &  0.00 \\ 
	300 & G-Formula PV &   0 &  0.0036 [0.0005] & 0.0009 [0.0000] & 0.0579 [0.0009] &  0.00 &  2.95 \\ 
	300 & G-Formula PV &   1 &  0.0019 [0.0008] & 0.0016 [0.0000] & 0.0649 [0.0009] &  0.00 &  5.65 \\ 
	300 & IPTW KM &   0 & -0.0056 [0.0005] & 0.0011 [0.0000] & 0.0672 [0.0005] &  0.00 &  0.00 \\ 
	300 & IPTW KM &   1 & -0.0158 [0.0009] & 0.0024 [0.0001] & 0.0792 [0.0006] &  0.00 &  0.00 \\ 
	300 & IPTW Cox &   0 & -0.0081 [0.0005] & 0.0012 [0.0000] & 0.0674 [0.0005] &  0.00 &  0.00 \\ 
	300 & IPTW Cox &   1 & -0.0205 [0.0009] & 0.0026 [0.0001] & 0.0800 [0.0006] &  0.00 &  0.00 \\ 
	300 & IPTW PV &   0 & -0.0056 [0.0005] & 0.0012 [0.0000] & 0.0674 [0.0005] &  8.55 & 67.10 \\ 
	300 & IPTW PV &   1 & -0.0161 [0.0009] & 0.0025 [0.0001] & 0.0797 [0.0006] & 15.05 & 60.10 \\ 
	300 & Matching &   0 &  0.0067 [0.0006] & 0.0016 [0.0000] & 0.0835 [0.0006] &  0.00 &  0.00 \\ 
	300 & Matching &   1 & -0.0068 [0.0010] & 0.0028 [0.0001] & 0.0878 [0.0006] &  0.00 &  0.00 \\ 
	300 & EL &   0 & -0.0065 [0.0005] & 0.0011 [0.0000] & 0.0673 [0.0005] &  0.00 &  0.00 \\ 
	300 & EL &   1 & -0.0147 [0.0008] & 0.0023 [0.0000] & 0.0783 [0.0006] &  0.00 &  0.00 \\ 
	300 & AIPTW &   0 & -0.0003 [0.0005] & 0.0010 [0.0000] & 0.0637 [0.0005] &  0.65 & 67.40 \\ 
	300 & AIPTW &   1 & -0.0022 [0.0008] & 0.0021 [0.0000] & 0.0760 [0.0005] & 21.80 & 65.65 \\ 
	300 & AIPTW PV &   0 &  0.0003 [0.0005] & 0.0010 [0.0000] & 0.0652 [0.0005] &  7.05 & 63.00 \\ 
	300 & AIPTW PV &   1 & -0.0013 [0.0008] & 0.0021 [0.0000] & 0.0767 [0.0006] & 24.25 & 75.30 \\ 
	300 & G-Formula IPTW &   0 & -0.0018 [0.0004] & 0.0007 [0.0000] & 0.0520 [0.0004] &  0.00 &  0.00 \\ 
	300 & G-Formula IPTW &   1 & -0.0041 [0.0007] & 0.0013 [0.0000] & 0.0577 [0.0004] &  0.00 &  0.00 \\ 
	300 & Kaplan-Meier &   0 & -0.0251 [0.0005] & 0.0020 [0.0000] & 0.0809 [0.0007] &  0.00 &  0.00 \\ 
	300 & Kaplan-Meier &   1 &  0.0381 [0.0008] & 0.0030 [0.0001] & 0.0865 [0.0007] &  0.00 &  0.00 \\ 
	500 & G-Formula &   0 & -0.0011 [0.0003] & 0.0005 [0.0000] & 0.0411 [0.0003] &  0.00 &  0.00 \\ 
	500 & G-Formula &   1 & -0.0010 [0.0006] & 0.0008 [0.0000] & 0.0447 [0.0003] &  0.00 &  0.00 \\ 
	500 & G-Formula PV &   0 &  0.0032 [0.0004] & 0.0006 [0.0000] & 0.0453 [0.0007] &  0.00 &  2.30 \\ 
	500 & G-Formula PV &   1 &  0.0025 [0.0006] & 0.0010 [0.0000] & 0.0507 [0.0007] &  0.00 &  3.95 \\ 
	500 & IPTW KM &   0 & -0.0058 [0.0004] & 0.0007 [0.0000] & 0.0535 [0.0004] &  0.00 &  0.00 \\ 
	500 & IPTW KM &   1 & -0.0151 [0.0007] & 0.0016 [0.0000] & 0.0626 [0.0005] &  0.00 &  0.00 \\ 
	500 & IPTW Cox &   0 & -0.0073 [0.0004] & 0.0007 [0.0000] & 0.0538 [0.0004] &  0.00 &  0.00 \\ 
	500 & IPTW Cox &   1 & -0.0179 [0.0007] & 0.0016 [0.0000] & 0.0633 [0.0005] &  0.00 &  0.00 \\ 
	500 & IPTW PV &   0 & -0.0058 [0.0004] & 0.0007 [0.0000] & 0.0535 [0.0004] &  4.80 & 66.40 \\ 
	500 & IPTW PV &   1 & -0.0152 [0.0007] & 0.0016 [0.0000] & 0.0629 [0.0005] & 10.40 & 43.45 \\ 
	500 & Matching &   0 &  0.0080 [0.0004] & 0.0010 [0.0000] & 0.0660 [0.0005] &  0.00 &  0.00 \\ 
	500 & Matching &   1 & -0.0058 [0.0008] & 0.0017 [0.0000] & 0.0683 [0.0005] &  0.00 &  0.00 \\ 
	500 & EL &   0 & -0.0067 [0.0004] & 0.0007 [0.0000] & 0.0536 [0.0004] &  0.00 &  0.00 \\ 
	500 & EL &   1 & -0.0141 [0.0007] & 0.0015 [0.0000] & 0.0618 [0.0005] &  0.00 &  0.00 \\ 
	500 & AIPTW &   0 & -0.0002 [0.0004] & 0.0006 [0.0000] & 0.0500 [0.0004] &  0.20 & 63.95 \\ 
	500 & AIPTW &   1 & -0.0012 [0.0006] & 0.0013 [0.0000] & 0.0588 [0.0004] & 15.75 & 55.15 \\ 
	500 & AIPTW PV &   0 &  0.0002 [0.0004] & 0.0006 [0.0000] & 0.0511 [0.0004] &  3.60 & 61.50 \\ 
	500 & AIPTW PV &   1 & -0.0007 [0.0006] & 0.0013 [0.0000] & 0.0591 [0.0004] & 15.35 & 62.45 \\ 
	500 & G-Formula IPTW &   0 & -0.0013 [0.0003] & 0.0005 [0.0000] & 0.0412 [0.0003] &  0.00 &  0.00 \\ 
	500 & G-Formula IPTW &   1 & -0.0025 [0.0006] & 0.0008 [0.0000] & 0.0451 [0.0003] &  0.00 &  0.00 \\ 
	500 & Kaplan-Meier &   0 & -0.0251 [0.0004] & 0.0015 [0.0000] & 0.0702 [0.0006] &  0.00 &  0.00 \\ 
	500 & Kaplan-Meier &   1 &  0.0387 [0.0006] & 0.0024 [0.0000] & 0.0741 [0.0006] &  0.00 &  0.00 \\ 
	1000 & G-Formula &   0 & -0.0008 [0.0002] & 0.0003 [0.0000] & 0.0301 [0.0002] &  0.00 &  0.00 \\ 
	1000 & G-Formula &   1 & -0.0005 [0.0004] & 0.0004 [0.0000] & 0.0319 [0.0002] &  0.00 &  0.00 \\ 
	1000 & G-Formula PV &   0 &  0.0033 [0.0003] & 0.0003 [0.0000] & 0.0321 [0.0003] &  0.00 &  0.80 \\ 
	1000 & G-Formula PV &   1 &  0.0014 [0.0004] & 0.0005 [0.0000] & 0.0352 [0.0005] &  0.00 &  1.10 \\ 
	1000 & IPTW KM &   0 & -0.0055 [0.0003] & 0.0004 [0.0000] & 0.0394 [0.0003] &  0.00 &  0.00 \\ 
	1000 & IPTW KM &   1 & -0.0160 [0.0005] & 0.0009 [0.0000] & 0.0461 [0.0004] &  0.00 &  0.00 \\ 
	1000 & IPTW Cox &   0 & -0.0063 [0.0003] & 0.0004 [0.0000] & 0.0395 [0.0003] &  0.00 &  0.00 \\ 
	1000 & IPTW Cox &   1 & -0.0174 [0.0005] & 0.0009 [0.0000] & 0.0467 [0.0004] &  0.00 &  0.00 \\ 
	1000 & IPTW PV &   0 & -0.0056 [0.0003] & 0.0004 [0.0000] & 0.0394 [0.0003] &  1.00 & 61.85 \\ 
	1000 & IPTW PV &   1 & -0.0162 [0.0005] & 0.0009 [0.0000] & 0.0464 [0.0004] &  3.25 & 25.40 \\ 
	1000 & Matching &   0 &  0.0090 [0.0003] & 0.0006 [0.0000] & 0.0485 [0.0004] &  0.00 &  0.00 \\ 
	1000 & Matching &   1 & -0.0068 [0.0005] & 0.0009 [0.0000] & 0.0481 [0.0004] &  0.00 &  0.00 \\ 
	1000 & EL &   0 & -0.0066 [0.0003] & 0.0004 [0.0000] & 0.0395 [0.0003] &  0.00 &  0.00 \\ 
	1000 & EL &   1 & -0.0149 [0.0005] & 0.0009 [0.0000] & 0.0454 [0.0003] &  0.00 &  0.00 \\ 
	1000 & AIPTW &   0 & -0.0001 [0.0003] & 0.0003 [0.0000] & 0.0357 [0.0003] &  0.00 & 50.50 \\ 
	1000 & AIPTW &   1 & -0.0012 [0.0005] & 0.0006 [0.0000] & 0.0418 [0.0003] &  4.30 & 37.45 \\ 
	1000 & AIPTW PV &   0 &  0.0003 [0.0003] & 0.0003 [0.0000] & 0.0367 [0.0003] &  0.50 & 54.30 \\ 
	1000 & AIPTW PV &   1 & -0.0016 [0.0005] & 0.0006 [0.0000] & 0.0423 [0.0003] &  4.90 & 40.05 \\ 
	1000 & G-Formula IPTW &   0 & -0.0008 [0.0002] & 0.0003 [0.0000] & 0.0302 [0.0002] &  0.00 &  0.00 \\ 
	1000 & G-Formula IPTW &   1 & -0.0017 [0.0004] & 0.0004 [0.0000] & 0.0321 [0.0002] &  0.00 &  0.00 \\ 
	1000 & Kaplan-Meier &   0 & -0.0247 [0.0003] & 0.0012 [0.0000] & 0.0598 [0.0004] &  0.00 &  0.00 \\ 
	1000 & Kaplan-Meier &   1 &  0.0374 [0.0004] & 0.0018 [0.0000] & 0.0621 [0.0004] &  0.00 &  0.00 \\
	\bottomrule
\end{longtable}

\normalsize

\newpage

\begin{figure}[!htb]
	\centering
	\includegraphics[width=1\linewidth]{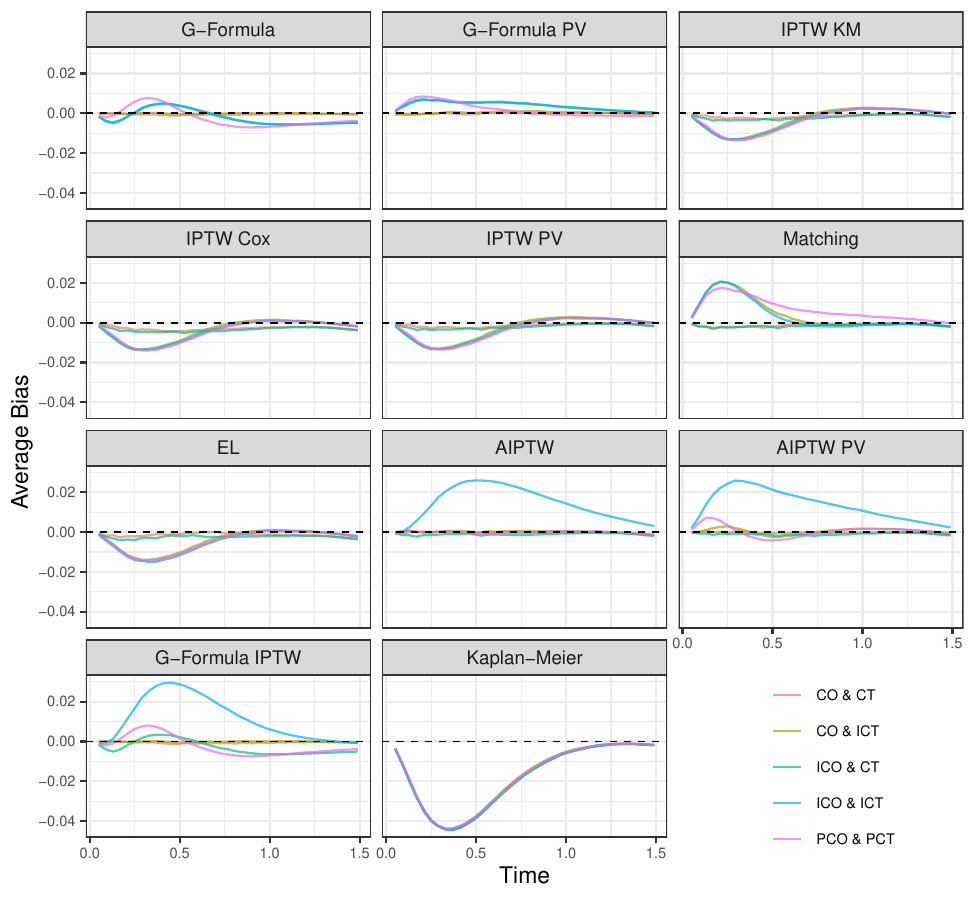}
	\caption{The average bias over time for $Z = 0$ in all simulation scenarios with $n = 1000$. Estimates are based on $2000$ simulation repetitions.}
	\label{fig::bias_over_time_n1000_z0}
\end{figure}

\begin{figure}[!htb]
	\centering
	\includegraphics[width=1\linewidth]{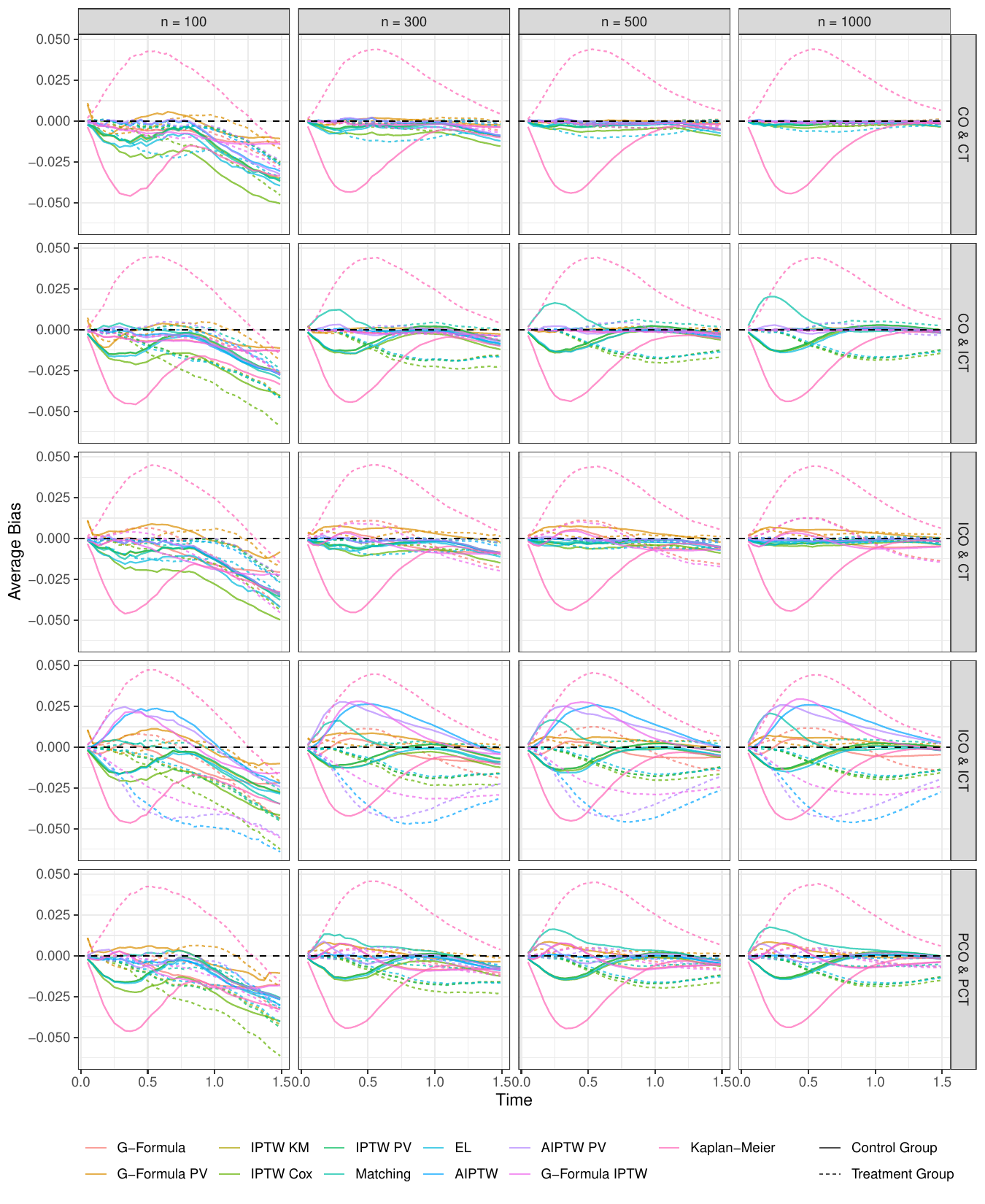}
	\caption{The average bias over time for both groups in all simulation scenarios with varying sample sizes. Estimates are based on $2000$ simulation repetitions.}
	\label{fig::bias_over_time_all}
\end{figure}

\begin{figure}[!htb]
	\centering
	\includegraphics[width=1\linewidth]{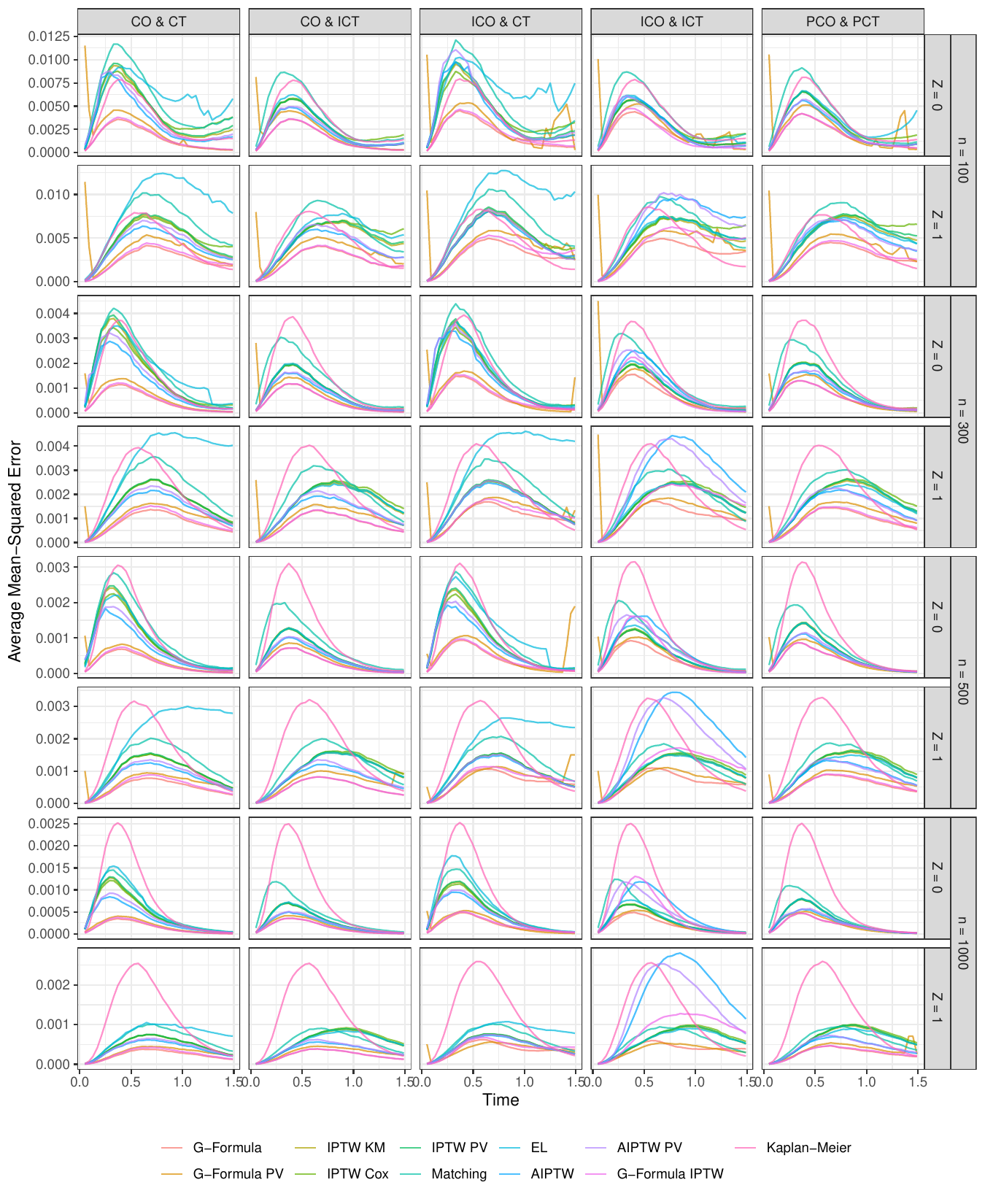}
	\caption{The average mean-squared error over time for both groups in all simulation scenarios with varying sample sizes. Estimates are based on $2000$ simulation repetitions.}
	\label{fig::mse_over_time}
\end{figure}

\begin{figure}[!htb]
	\centering
	\includegraphics[width=1\linewidth]{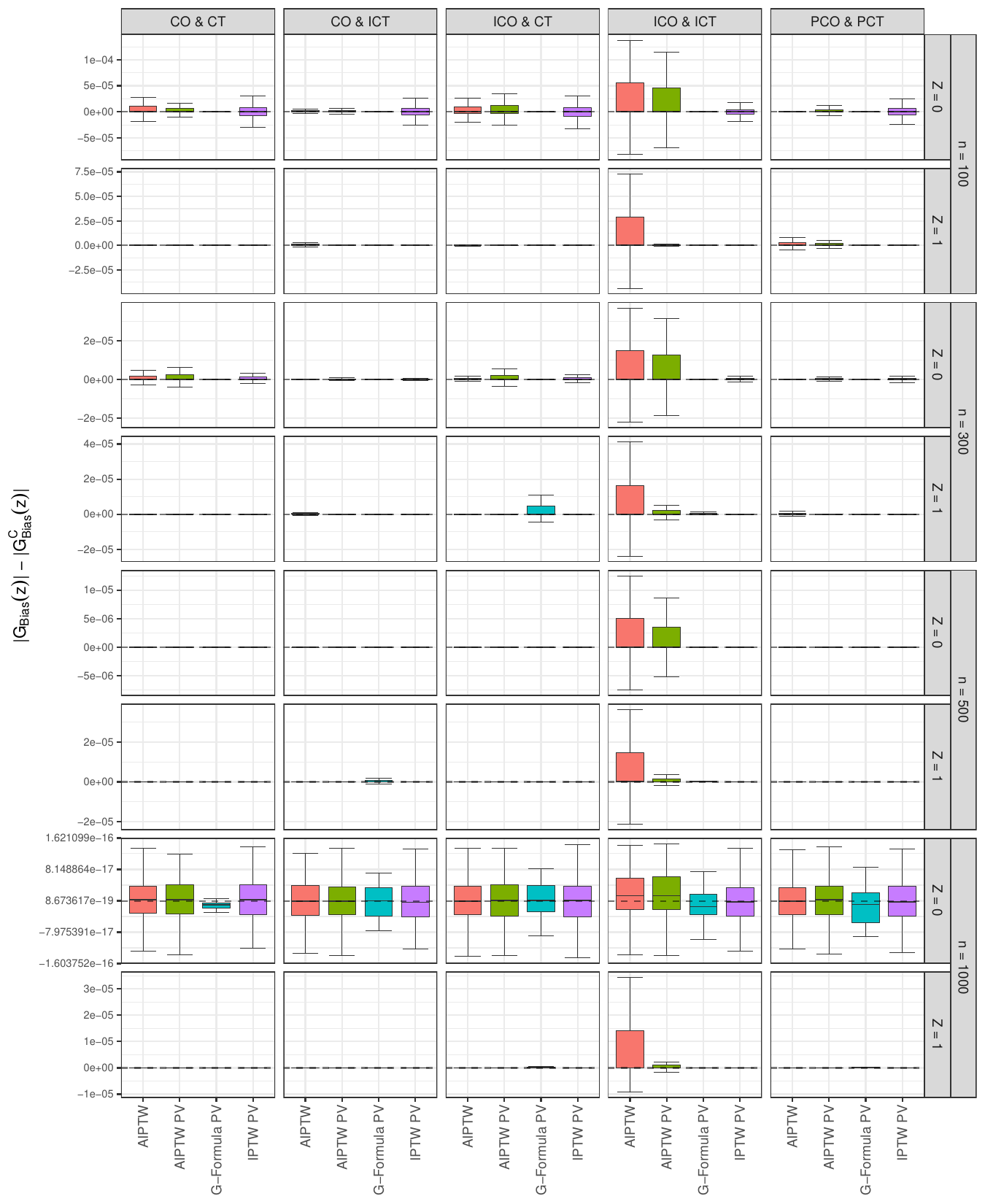}
	\caption{Distributions of $|\hat{\Delta}_{Bias}(z)| - |\hat{\Delta}_{Bias}^C(z)|$ for all methods and both treatment groups in each simulation scenario with varying sample sizes. $\hat{\Delta}_{Bias}^C(z)$ is the standard $\hat{\Delta}_{Bias}(z)$, recalculated after truncating estimates outside of the probability bounds and applying isotonic regression. Only survival curves exhibiting non-monotonicity or out-of-bounds estimates were considered.}
	\label{fig::bias_diff}
\end{figure}

\begin{figure}[!htb]
	\centering
	\includegraphics[width=1\linewidth]{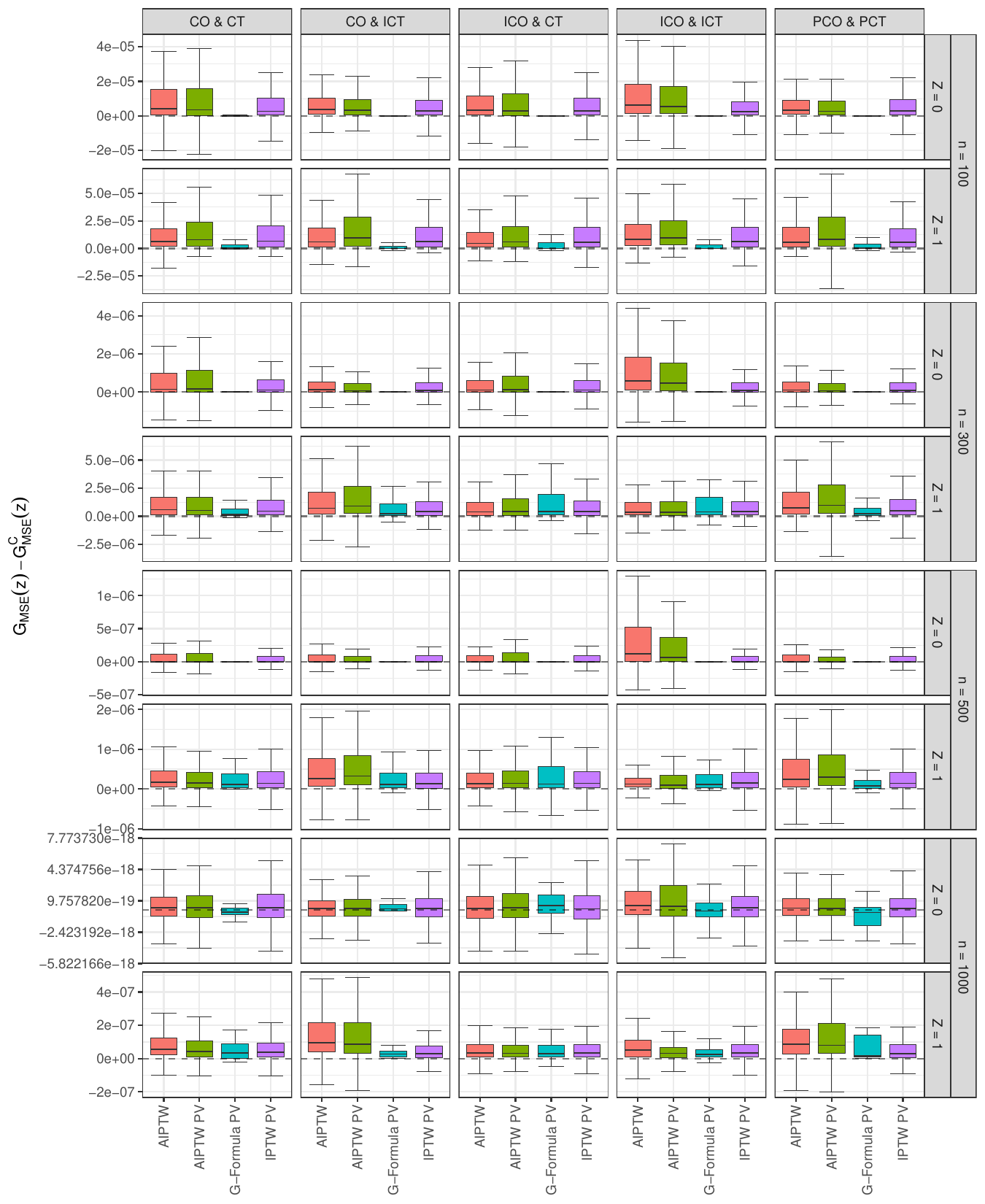}
	\caption{Distributions of $\hat{\Delta}_{MSE}(z) - \hat{\Delta}_{MSE}^C(z)$ for all methods and both treatment groups in each simulation scenario with varying sample sizes. $\hat{\Delta}_{MSE}^C(z)$ is the standard $\hat{\Delta}_{MSE}(z)$, recalculated after truncating estimates outside of the probability bounds and applying isotonic regression. Only survival curves exhibiting non-monotonicity or out-of-bounds estimates were considered.}
	\label{fig::mse_diff}
\end{figure}

\begin{figure}[!htb]
	\centering
	\includegraphics[width=0.8\linewidth]{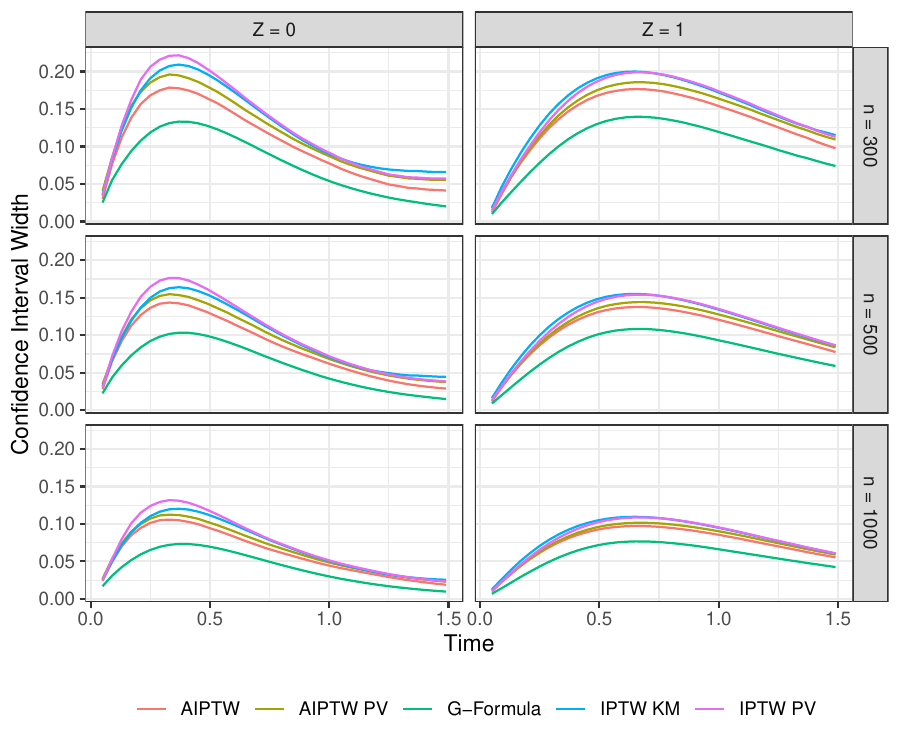}
	\caption{The average approximate 95\% confidence interval width over time for both groups in the CO \& CT simulation scenario with varying sample sizes. Estimates are based on $2000$ simulation repetitions.}
	\label{fig::ci_width}
\end{figure}

\begin{figure}[!htb]
	\centering
	\includegraphics[width=0.8\linewidth]{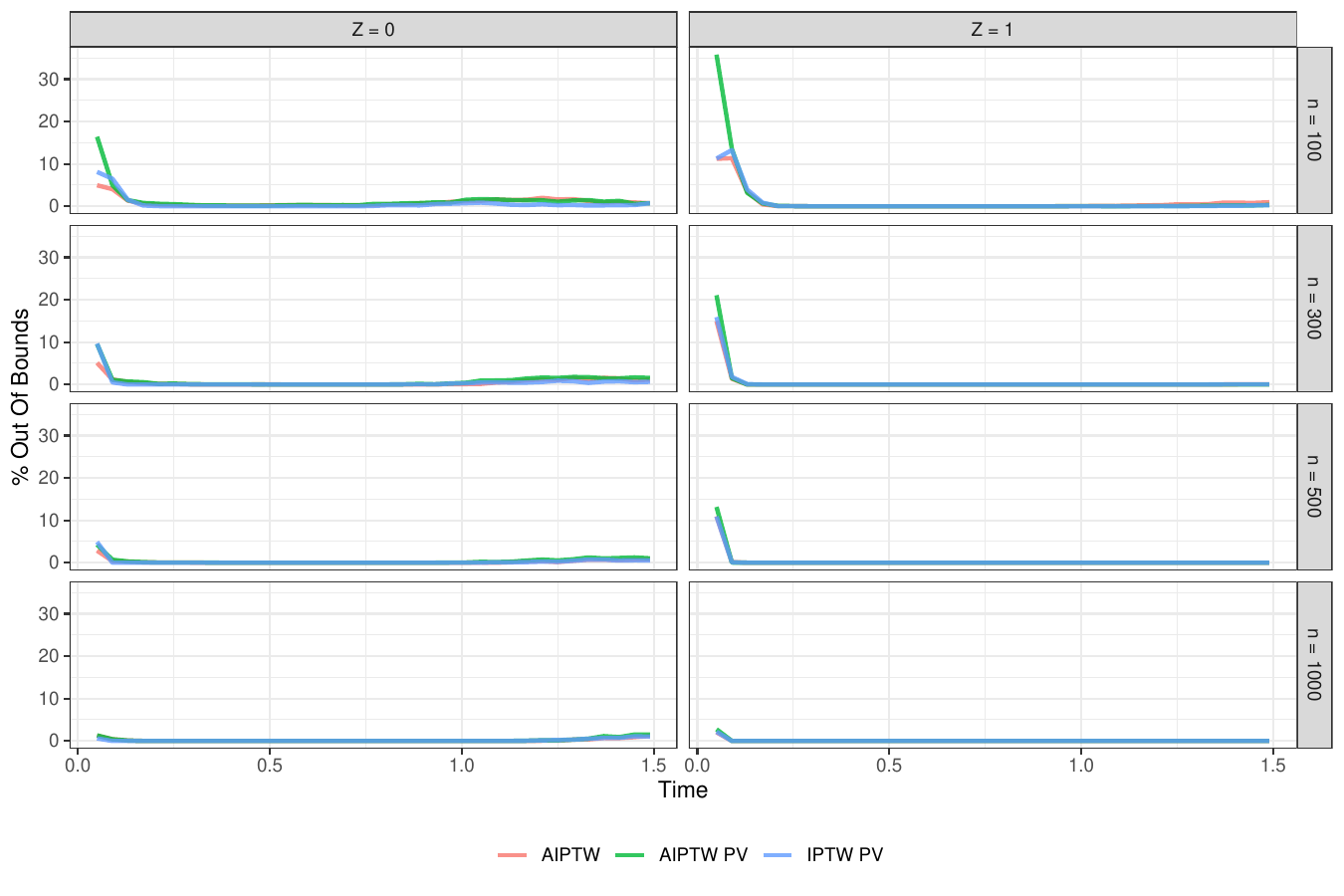}
	\caption{Percentage of estimates falling outside of the 0 and 1 probability bounds over time for both treatment groups and varying sample sizes. All models were correctly specified (CO \& CT Scenario). Based on $2000$ simulation repetitions.}
	\label{fig::oob_over_time}
\end{figure}

\FloatBarrier
\newpage

\section{Simulation with Different Covariate Sets}

To supplement the main simulation scenarios discussed in the main text, we additionally performed a second simulation study in which the set of covariates used in both the treatment-assignment and the outcome models where systematically varied. The simulation follows the same steps as the main simulation study, but with a different underlying causal DAG and causal coefficients. The following causal DAG was used:

\begin{figure}[!htb]
	\centering
	\includegraphics[width=0.8\linewidth]{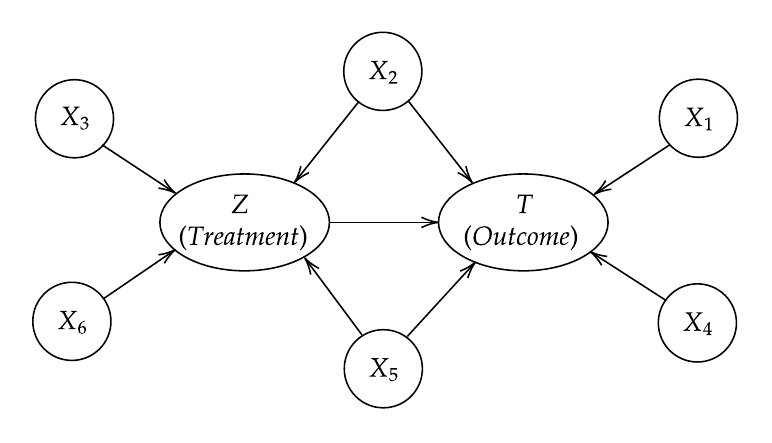}
	\caption{Causal diagram of the data generation mechanism used in the simulation. Adapted from Chatton et al. (2020a).}
	\label{fig::appendix_4_causal_diagram}
\end{figure}

The following structural equations can be used to describe this DAG formally:

\begin{align*} 
	X_1 &\sim Bernoulli(0.5) \\ 
	X_2 &\sim Bernoulli(0.5) \\
	X_3 &\sim Bernoulli(0.5) \\
	X_4 &\sim N(0, 1) \\
	X_5 &\sim N(0, 1) \\
	X_6 &\sim N(0, 1) \\
	Z &\sim  Bernoulli\left(\frac{1}{1 + \exp(-(-0.5 + \log(2) \cdot X_2 + \log(1.5) \cdot X_3 + \log(1.5) \cdot X_5 + \log(2) \cdot X_6))}\right) \\
	T &\sim  \left(-\frac{\log\left(Unif(0, 1)\right)}{\exp(\log(1.8) \cdot X_1 + \log(1.3) \cdot X_2 + \log(1.8) \cdot X_4 + \log(1.3) \cdot X_5 - 1 \cdot Z)}\right)^{0.5} \\
\end{align*}

\newpage

We considered the following six scenarios:

\begin{enumerate}
	\item \textbf{Correct Outcome Mechanism \& Correct Treatment Assignment} (CO \& CT): Using $X_1, X_2, X_4, X_5$ and $Z$ as independent variables to model the outcome mechanism and $X_2,X_5$ as independent variables to model the treatment assignment mechanism.
	\item \textbf{Correct Outcome Mechanism \& Incorrect Treatment Assignment} (CO \& ICT):  Using $X_1, X_2, X_4, X_5$ and $Z$ as independent variables to model the outcome mechanism and $X_2$ as independent variables to model the treatment assignment mechanism.
	\item \textbf{Incorrect Outcome Mechanism \& Correct Treatment Assignment} (ICO \& CT):  Using $X_1, X_2$ and $Z$ as independent variables to model the outcome mechanism and $X_2, X_5$ as independent variables to model the treatment assignment mechanism.
	\item \textbf{Incorrect Outcome Mechanism \& Incorrect Treatment Assignment} (ICO \& ICT):  Using $X_1, X_2$ and $Z$ as independent variables to model the outcome mechanism and $X_2$ as independent variables to model the treatment assignment mechanism.
	\item \textbf{Correct Outcome Mechanism \& Correct Treatment Assignment \& Treatment Assignment Predictors} (CO \& CT \& TP): Using $X_1, X_2, X_3, X_4, X_5, X_6$ and $Z$ as independent variables to model the outcome mechanism and $X_2, X_3, X_5, X_6$ as independent variables to model the treatment assignment mechanism.
	\item \textbf{Correct Outcome Mechanism \& Correct Treatment Assignment \& Outcome Predictors} (CO \& CT \& OP): Using $X_1, X_2, X_4, X_5$ and $Z$ as independent variables to model the outcome mechanism and $X_1, X_2, X_4, X_5$ as independent variables to model the treatment assignment mechanism.
\end{enumerate}

As was mentioned in the main text, the EL method does not use any models to create the estimates. It therefore only received the raw design matrix in all scenarios. To keep the comparisons fair, it received the union of all variables that were used in both the outcome model and the treatment model for the other methods. For example, in the CO \& CT scenario, the EL method received $X_1, X_2, X_4, X_5$ and in the ICO \& CT it only received $X_1, X_2, X_5$.
\par
The results include the same tables as in appendix~\ref{chap::appendix_add_results} and some similar plots.

\newpage

\footnotesize

\begin{longtable}{rlrlllll}
	\caption{Estimates for all performance criteria in the \textbf{CO \& CT} scenario. The values in parentheses are the Monte-Carlo standard errors of the associated estimates. Estimates are based on $2000$ simulation repetitions. $\hat{G}_{Bias}(z)$, $\hat{G}_{MSE}(z)$, $\hat{G}_{Sup}(z)$ and their associated standard errors are rounded to the fourth decimal place, the other statistics are rounded to the second decimal place.}
	\label{tab::appendix3_CO_CT}
	\endfirsthead
	\endhead
	\toprule
	$n$ & Method & $Z$ & $\hat{G}_{Bias}(z)$ & $\hat{G}_{MSE}(z)$ & $\hat{G}_{Sup}(z)$ & \% OOB & \% NM \\ 
	\midrule
	100 & G-Formula &   0 & -0.0041 [0.0009] & 0.0029 [0.0001] & 0.0979 [0.0008] &  0.00 &  0.00 \\ 
	100 & G-Formula &   1 & -0.0130 [0.0014] & 0.0045 [0.0001] & 0.1023 [0.0009] &  0.00 &  0.00 \\ 
	100 & G-Formula PV &   0 &  0.0031 [0.0010] & 0.0044 [0.0001] & 0.1320 [0.0031] &  0.00 & 43.55 \\ 
	100 & G-Formula PV &   1 & -0.0056 [0.0014] & 0.0061 [0.0002] & 0.1370 [0.0032] &  0.00 & 45.65 \\ 
	100 & IPTW KM &   0 & -0.0022 [0.0010] & 0.0047 [0.0001] & 0.1311 [0.0011] &  0.00 &  0.00 \\ 
	100 & IPTW KM &   1 & -0.0065 [0.0015] & 0.0063 [0.0001] & 0.1280 [0.0010] &  0.00 &  0.00 \\ 
	100 & IPTW Cox &   0 & -0.0119 [0.0010] & 0.0047 [0.0001] & 0.1297 [0.0011] &  0.00 &  0.00 \\
\bottomrule
\end{longtable}

\newpage

\begin{longtable}{rlrlllll}
	\caption{Estimates for all performance criteria in the \textbf{CO \& ICT} scenario. The values in parentheses are the Monte-Carlo standard errors of the associated estimates. Estimates are based on $2000$ simulation repetitions. $\hat{G}_{Bias}(z)$, $\hat{G}_{MSE}(z)$, $\hat{G}_{Sup}(z)$ and their associated standard errors are rounded to the fourth decimal place, the other statistics are rounded to the second decimal place.}
	\label{tab::appendix3_CO_ICT}
	\endfirsthead
	\endhead
	\toprule
	$n$ & Method & $Z$ & $\hat{G}_{Bias}(z)$ & $\hat{G}_{MSE}(z)$ & $\hat{G}_{Sup}(z)$ & \% OOB & \% NM \\ 
	\midrule
	100 & G-Formula &   0 & -0.0045 [0.0009] & 0.0030 [0.0001] & 0.0987 [0.0009] &  0.00 &  0.00 \\ 
	100 & G-Formula &   1 & -0.0128 [0.0013] & 0.0044 [0.0001] & 0.1020 [0.0008] &  0.00 &  0.00 \\ 
	100 & G-Formula PV &   0 &  0.0027 [0.0010] & 0.0045 [0.0001] & 0.1337 [0.0033] &  0.00 & 42.15 \\ 
	100 & G-Formula PV &   1 & -0.0037 [0.0014] & 0.0062 [0.0002] & 0.1390 [0.0033] &  0.00 & 43.95 \\ 
	100 & IPTW KM &   0 & -0.0102 [0.0010] & 0.0046 [0.0001] & 0.1292 [0.0011] &  0.00 &  0.00 \\ 
	100 & IPTW KM &   1 &  0.0043 [0.0014] & 0.0061 [0.0001] & 0.1262 [0.0009] &  0.00 &  0.00 \\ 
	100 & IPTW Cox &   0 & -0.0193 [0.0010] & 0.0047 [0.0001] & 0.1291 [0.0011] &  0.00 &  0.00 \\
\bottomrule
\end{longtable}

\newpage

\begin{longtable}{rlrlllll}
	\caption{Estimates for all performance criteria in the \textbf{ICO \& CT} scenario. The values in parentheses are the Monte-Carlo standard errors of the associated estimates. Estimates are based on $2000$ simulation repetitions. $\hat{G}_{Bias}(z)$, $\hat{G}_{MSE}(z)$, $\hat{G}_{Sup}(z)$ and their associated standard errors are rounded to the fourth decimal place, the other statistics are rounded to the second decimal place.}
	\label{tab::appendix3_ICO_CT}
	\endfirsthead
	\endhead
	\toprule
	$n$ & Method & $Z$ & $\hat{G}_{Bias}(z)$ & $\hat{G}_{MSE}(z)$ & $\hat{G}_{Sup}(z)$ & \% OOB & \% NM \\ 
	\midrule
	100 & G-Formula &   0 & -0.0145 [0.0010] & 0.0037 [0.0001] & 0.1055 [0.0010] &  0.00 &  0.00 \\ 
	100 & G-Formula &   1 &  0.0004 [0.0015] & 0.0050 [0.0001] & 0.1071 [0.0009] &  0.00 &  0.00 \\ 
	100 & G-Formula PV &   0 & -0.0051 [0.0011] & 0.0047 [0.0001] & 0.1300 [0.0031] &  0.00 & 43.70 \\ 
	100 & G-Formula PV &   1 &  0.0096 [0.0015] & 0.0062 [0.0002] & 0.1342 [0.0031] &  0.00 & 45.50 \\ 
	100 & IPTW KM &   0 & -0.0019 [0.0011] & 0.0049 [0.0001] & 0.1331 [0.0011] &  0.00 &  0.00 \\ 
	100 & IPTW KM &   1 & -0.0027 [0.0015] & 0.0067 [0.0001] & 0.1313 [0.0010] &  0.00 &  0.00 \\ 
	100 & IPTW Cox &   0 & -0.0115 [0.0011] & 0.0049 [0.0001] & 0.1317 [0.0011] &  0.00 &  0.00 \\
\bottomrule
\end{longtable}

\newpage

\begin{longtable}{rlrlllll}
	\caption{Estimates for all performance criteria in the \textbf{ICO \& ICT} scenario. The values in parentheses are the Monte-Carlo standard errors of the associated estimates. Estimates are based on $2000$ simulation repetitions. $\hat{G}_{Bias}(z)$, $\hat{G}_{MSE}(z)$, $\hat{G}_{Sup}(z)$ and their associated standard errors are rounded to the fourth decimal place, the other statistics are rounded to the second decimal place.}
	\label{tab::appendix3_ICO_ICT}
	\endfirsthead
	\endhead
	\toprule
	$n$ & Method & $Z$ & $\hat{G}_{Bias}(z)$ & $\hat{G}_{MSE}(z)$ & $\hat{G}_{Sup}(z)$ & \% OOB & \% NM \\ 
	\midrule
	100 & G-Formula &   0 & -0.0153 [0.0010] & 0.0035 [0.0001] & 0.1035 [0.0009] &  0.00 &  0.00 \\ 
	100 & G-Formula &   1 & -0.0021 [0.0014] & 0.0049 [0.0001] & 0.1063 [0.0009] &  0.00 &  0.00 \\ 
	100 & G-Formula PV &   0 & -0.0059 [0.0010] & 0.0044 [0.0001] & 0.1283 [0.0032] &  0.00 & 42.50 \\ 
	100 & G-Formula PV &   1 &  0.0076 [0.0015] & 0.0061 [0.0002] & 0.1340 [0.0031] &  0.00 & 43.90 \\ 
	100 & IPTW KM &   0 & -0.0100 [0.0010] & 0.0046 [0.0001] & 0.1297 [0.0010] &  0.00 &  0.00 \\ 
	100 & IPTW KM &   1 &  0.0040 [0.0015] & 0.0062 [0.0001] & 0.1272 [0.0010] &  0.00 &  0.00 \\ 
	100 & IPTW Cox &   0 & -0.0191 [0.0010] & 0.0048 [0.0001] & 0.1298 [0.0010] &  0.00 &  0.00 \\
\bottomrule
\end{longtable}

\newpage

\begin{longtable}{rlrlllll}
	\caption{Estimates for all performance criteria in the \textbf{CO \& CT \& TP} scenario. The values in parentheses are the Monte-Carlo standard errors of the associated estimates. Estimates are based on $2000$ simulation repetitions. $\hat{G}_{Bias}(z)$, $\hat{G}_{MSE}(z)$, $\hat{G}_{Sup}(z)$ and their associated standard errors are rounded to the fourth decimal place, the other statistics are rounded to the second decimal place.}
	\label{tab::appendix3_CO_CT_TP}
	\endfirsthead
	\endhead
	\toprule
	$n$ & Method & $Z$ & $\hat{G}_{Bias}(z)$ & $\hat{G}_{MSE}(z)$ & $\hat{G}_{Sup}(z)$ & \% OOB & \% NM \\ 
	\midrule
	100 & G-Formula &   0 & -0.0040 [0.0010] & 0.0031 [0.0001] & 0.0996 [0.0009] &  0.00 &  0.00 \\ 
	100 & G-Formula &   1 & -0.0129 [0.0014] & 0.0048 [0.0001] & 0.1052 [0.0009] &  0.00 &  0.00 \\ 
	100 & G-Formula PV &   0 &  0.0026 [0.0011] & 0.0049 [0.0001] & 0.1376 [0.0032] &  0.00 & 44.85 \\ 
	100 & G-Formula PV &   1 & -0.0029 [0.0015] & 0.0068 [0.0002] & 0.1429 [0.0033] &  0.00 & 47.35 \\ 
	100 & IPTW KM &   0 & -0.0025 [0.0012] & 0.0060 [0.0001] & 0.1473 [0.0012] &  0.00 &  0.00 \\ 
	100 & IPTW KM &   1 & -0.0038 [0.0016] & 0.0076 [0.0002] & 0.1390 [0.0011] &  0.00 &  0.00 \\ 
	100 & IPTW Cox &   0 & -0.0137 [0.0012] & 0.0059 [0.0001] & 0.1444 [0.0012] &  0.00 &  0.00 \\
\bottomrule
\end{longtable}

\newpage

\begin{longtable}{rlrlllll}
	\caption{Estimates for all performance criteria in the \textbf{CO \& CT \& OP} scenario. The values in parentheses are the Monte-Carlo standard errors of the associated estimates. Estimates are based on $2000$ simulation repetitions. $\hat{G}_{Bias}(z)$, $\hat{G}_{MSE}(z)$, $\hat{G}_{Sup}()$ and their associated standard errors are rounded to the fourth decimal place, the other statistics are rounded to the second decimal place.}
	\label{tab::appendix3_CO_CT_OP}
	\endfirsthead
	\endhead
	\toprule
	$n$ & Method & $Z$ & $\hat{G}_{Bias}(z)$ & $\hat{G}_{MSE}(z)$ & $\hat{G}_{Sup}(z)$ & \% OOB & \% NM \\ 
	\midrule
	100 & G-Formula &   0 & -0.0035 [0.0009] & 0.0029 [0.0001] & 0.0981 [0.0008] &  0.00 &  0.00 \\ 
	100 & G-Formula &   1 & -0.0117 [0.0014] & 0.0045 [0.0001] & 0.1038 [0.0008] &  0.00 &  0.00 \\ 
	100 & G-Formula PV &   0 &  0.0028 [0.0010] & 0.0044 [0.0001] & 0.1293 [0.0030] &  0.00 & 42.80 \\ 
	100 & G-Formula PV &   1 & -0.0031 [0.0014] & 0.0061 [0.0002] & 0.1355 [0.0030] &  0.00 & 45.50 \\ 
	100 & IPTW KM &   0 & -0.0009 [0.0010] & 0.0047 [0.0001] & 0.1318 [0.0010] &  0.00 &  0.00 \\ 
	100 & IPTW KM &   1 & -0.0040 [0.0015] & 0.0066 [0.0001] & 0.1304 [0.0010] &  0.00 &  0.00 \\ 
	100 & IPTW Cox &   0 & -0.0108 [0.0010] & 0.0046 [0.0001] & 0.1302 [0.0010] &  0.00 &  0.00 \\
\bottomrule
\end{longtable}

\begin{figure}[!htb]
	\centering
	\includegraphics[width=1\linewidth]{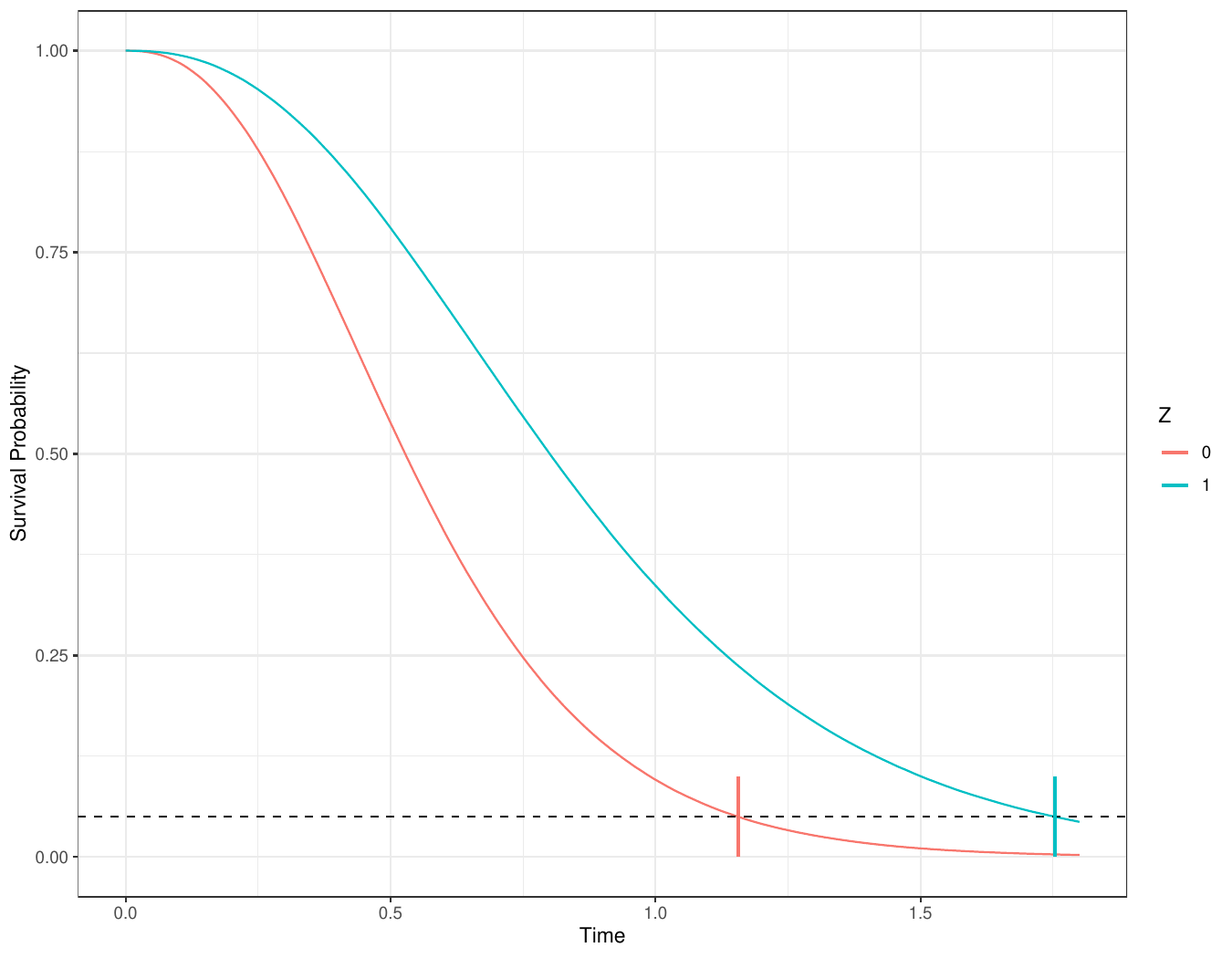}
	\caption{The true survival curves in the simulation study for both treatment groups ($Z$). The vertical lines indicate the 95\% quantile of the survival times.}
	\label{fig::appendix_4_true_surv}
\end{figure}

\begin{figure}[!htb]
	\centering
	\includegraphics[width=1\linewidth]{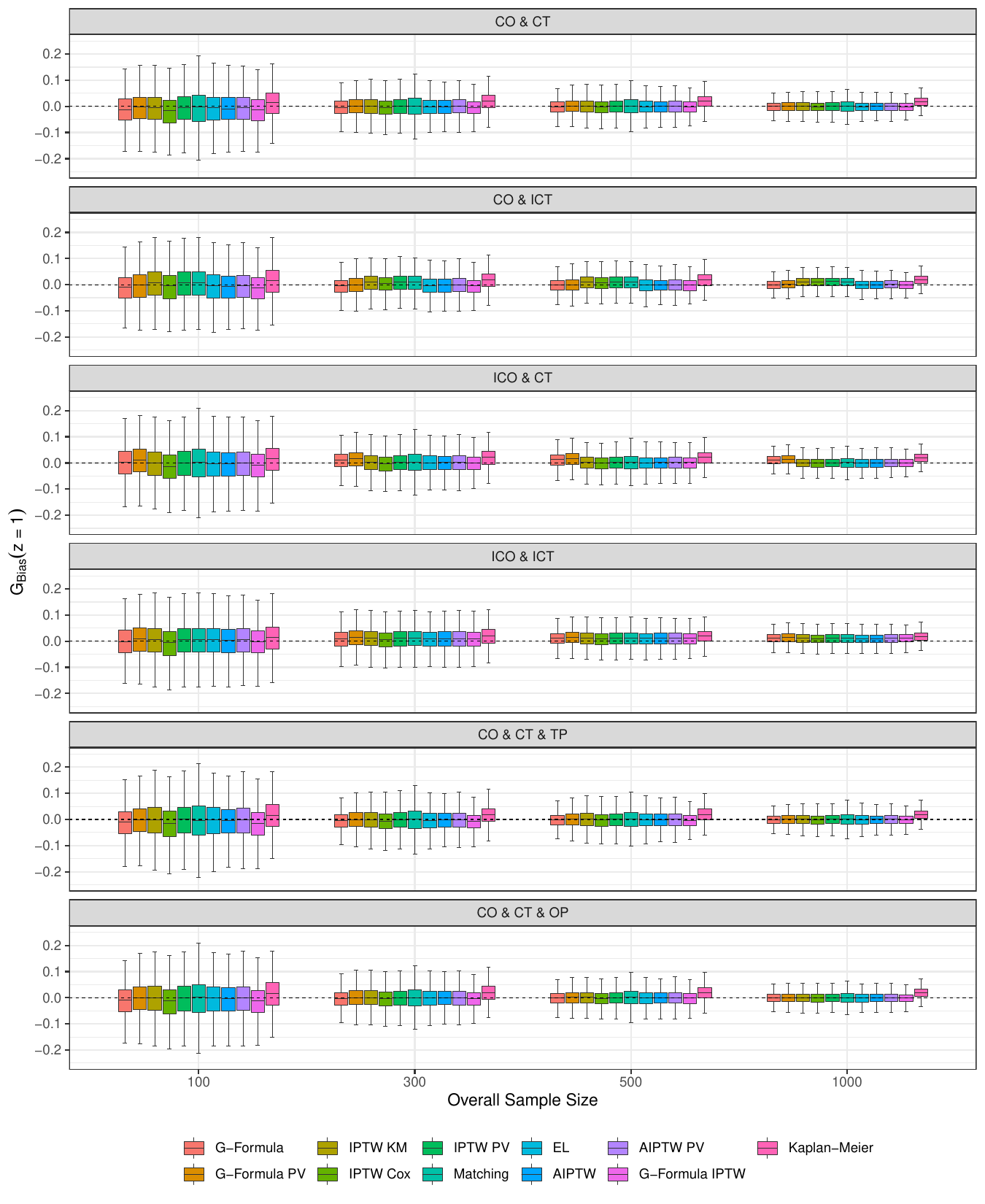}
	\caption{Distributions of $\hat{\Delta}_{Bias}(z = 0)$ (control group) for all methods in each simulation scenario with varying sample sizes. Outliers are not shown. Estimates are based on $2000$ simulation repetitions.}
	\label{fig::appendix_4_bias_over_scenarios_treatment}
\end{figure}

\begin{figure}[!htb]
	\centering
	\includegraphics[width=1\linewidth]{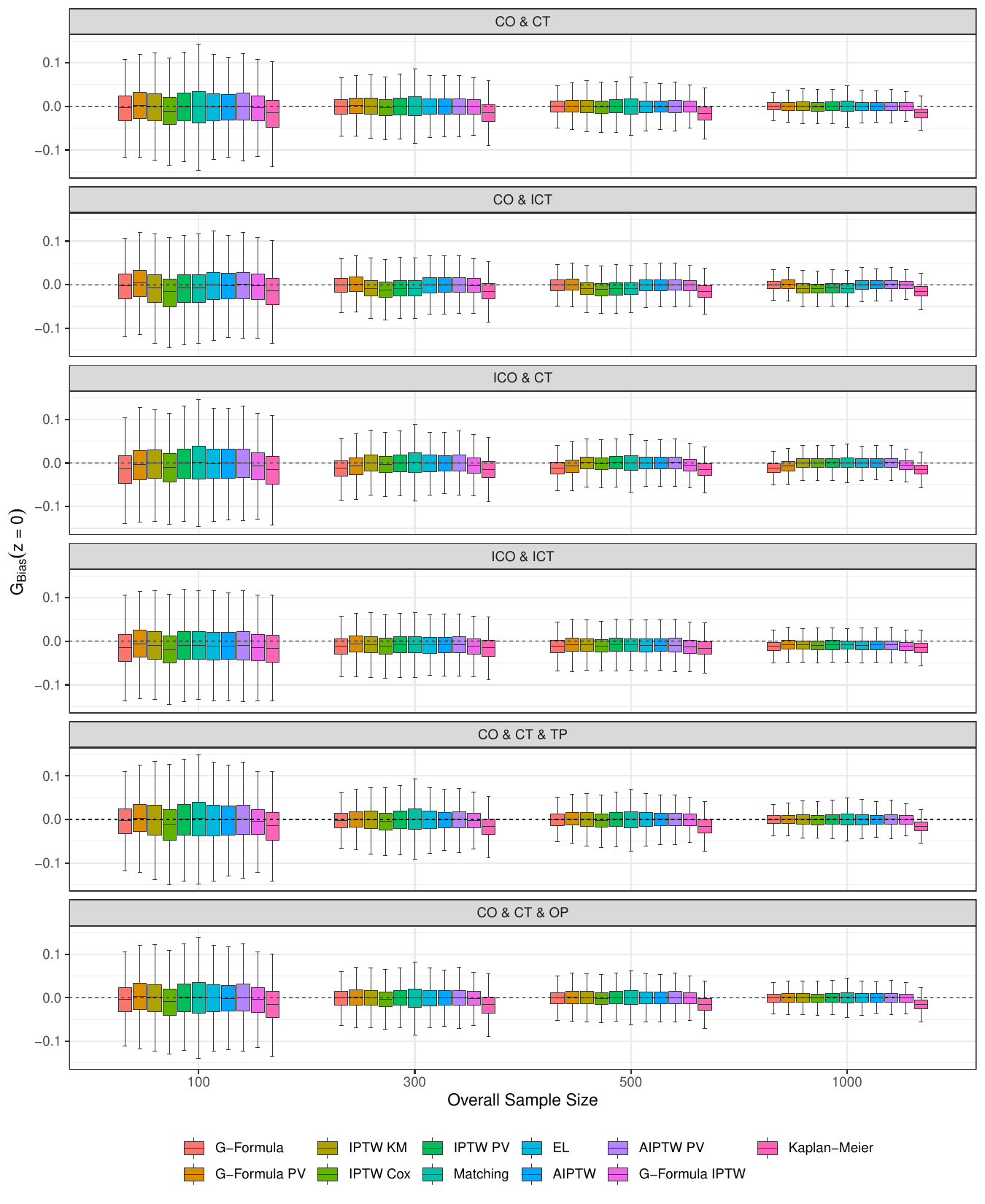}
	\caption{Distributions of $\hat{\Delta}_{Bias}(z = 0)$ (control group) for all methods in each simulation scenario with varying sample sizes. Outliers are not shown. Estimates are based on $2000$ simulation repetitions.}
	\label{fig::appendix_4_bias_over_scenarios_control}
\end{figure}

\begin{figure}[!htb]
	\centering
	\includegraphics[width=1\linewidth]{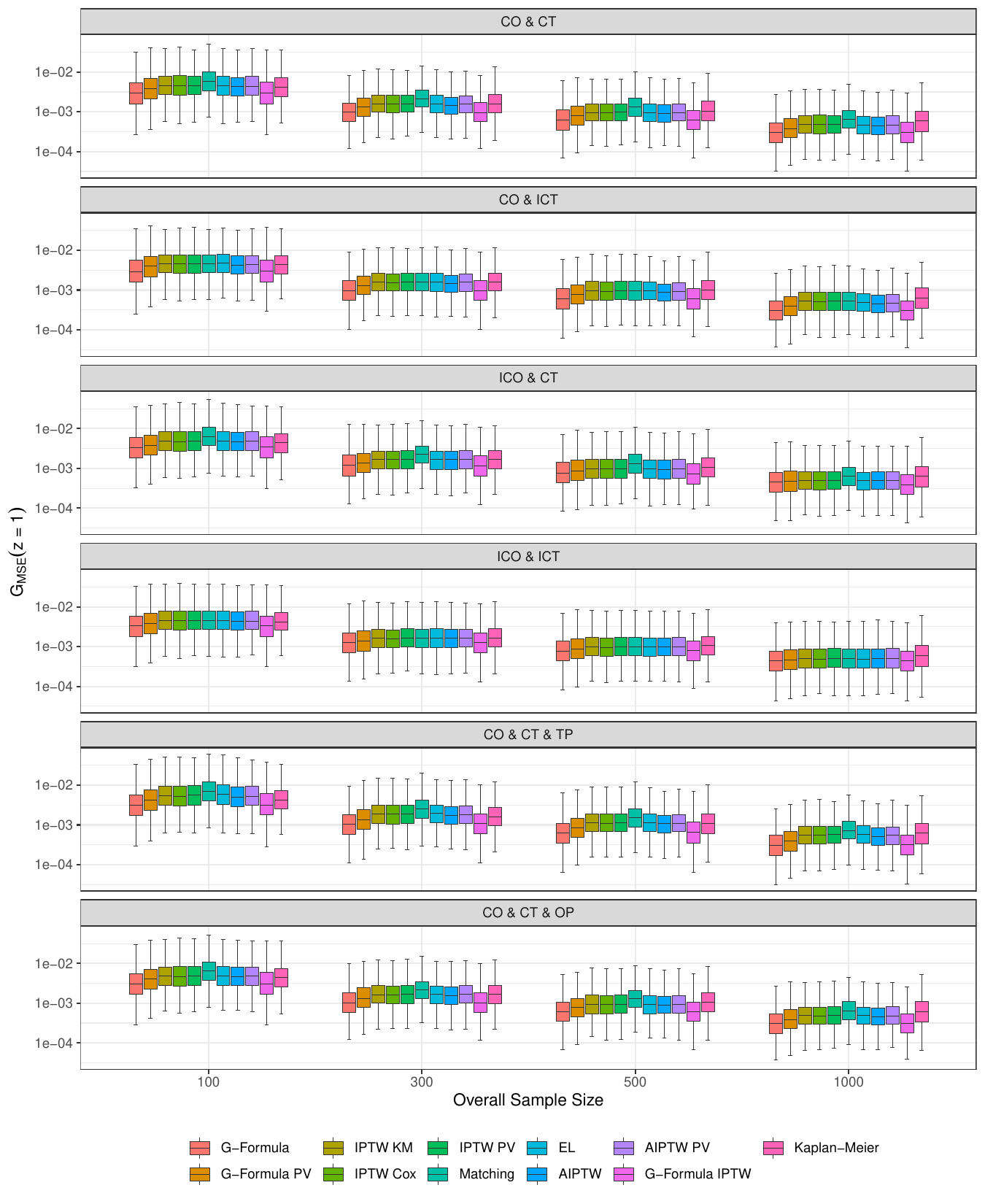}
	\caption{Distributions of $\hat{\Delta}_{MSE}(z = 0)$ (control group) on the log-scale for all methods in each simulation scenario with varying sample sizes. Outliers are not shown. Estimates are based on $2000$ simulation repetitions.}
	\label{fig::appendix_4_MSE_over_scenarios_treatment}
\end{figure}

\begin{figure}[!htb]
	\centering
	\includegraphics[width=1\linewidth]{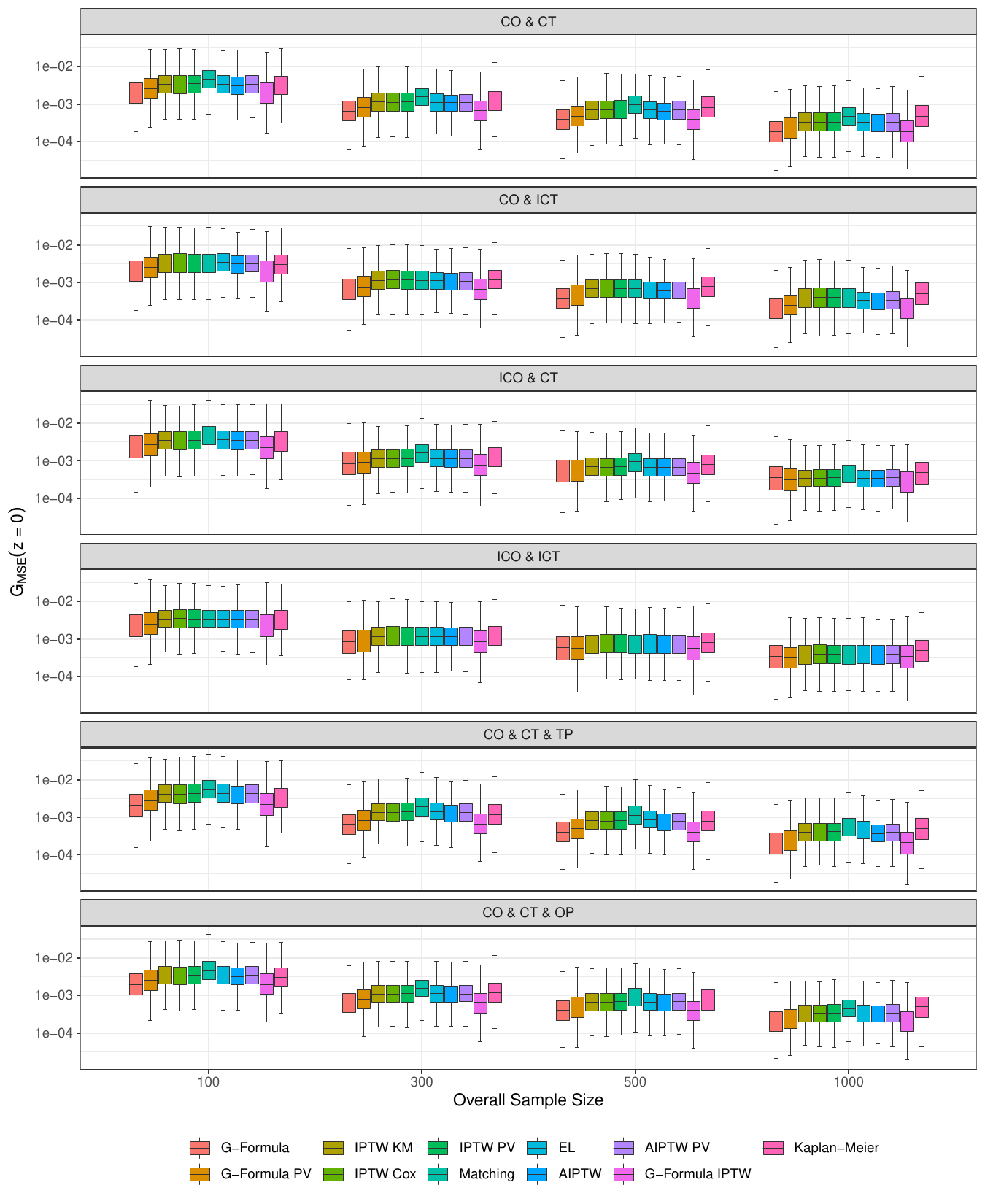}
	\caption{Distributions of $\hat{\Delta}_{MSE}(z = 0)$ (control group) on the log-scale for all methods in each simulation scenario with varying sample sizes. Outliers are not shown. Estimates are based on $2000$ simulation repetitions.}
	\label{fig::appendix_4_MSE_over_scenarios_control}
\end{figure}

\begin{figure}[!htb]
	\centering
	\includegraphics[width=1\linewidth]{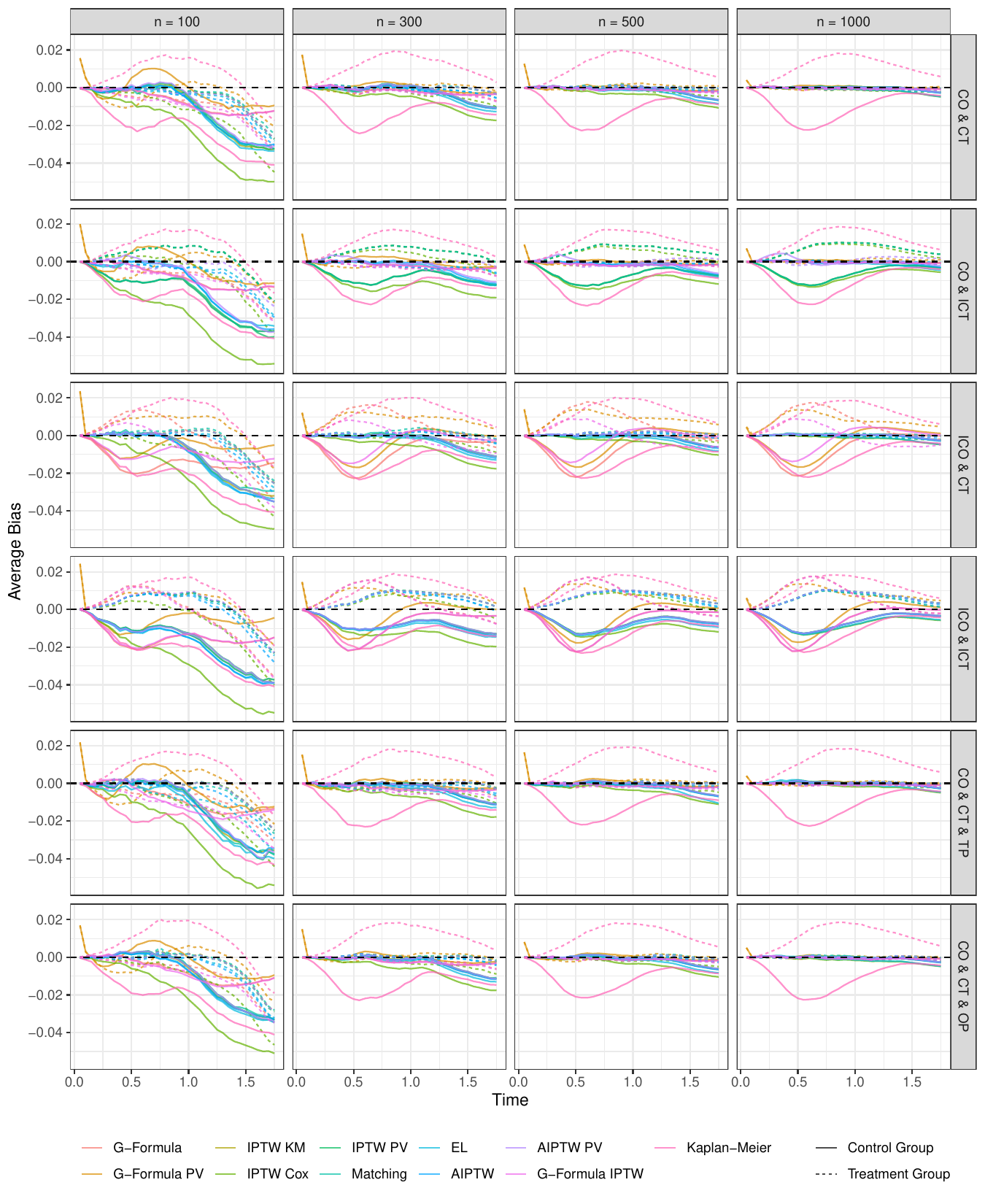}
	\caption{The average bias over time for both groups in all simulation scenarios with varying sample sizes. Estimates are based on $2000$ simulation repetitions.}
	\label{fig::appendix_4_bias_over_time}
\end{figure}

\begin{figure}[!htb]
	\centering
	\includegraphics[width=1\linewidth]{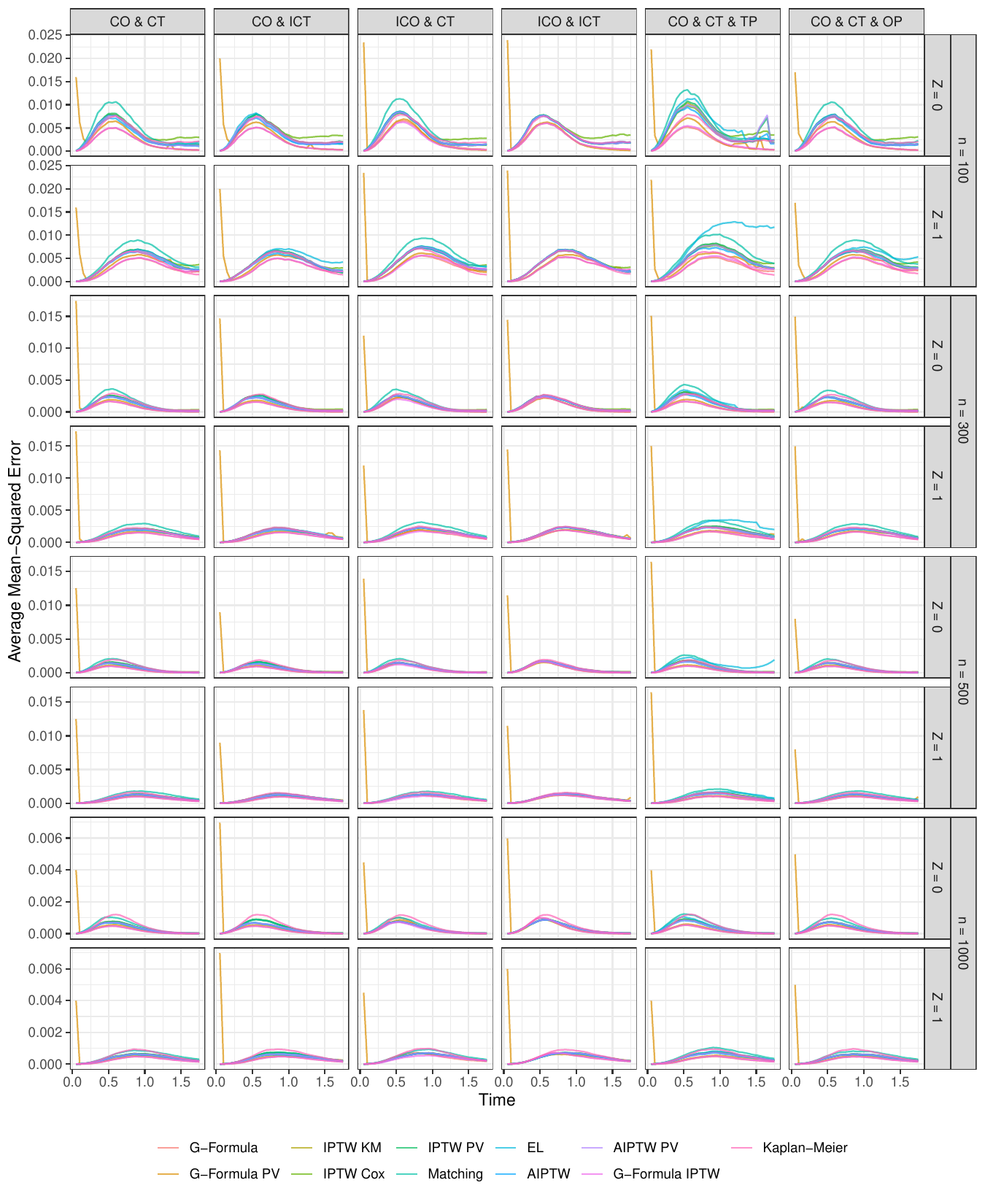}
	\caption{The average mean-squared error over time for both groups in all simulation scenarios with varying sample sizes. Estimates are based on $2000$ simulation repetitions.}
	\label{fig::appendix_4_mse_over_time}
\end{figure}

\begin{figure}[!htb]
	\centering
	\includegraphics[width=0.8\linewidth]{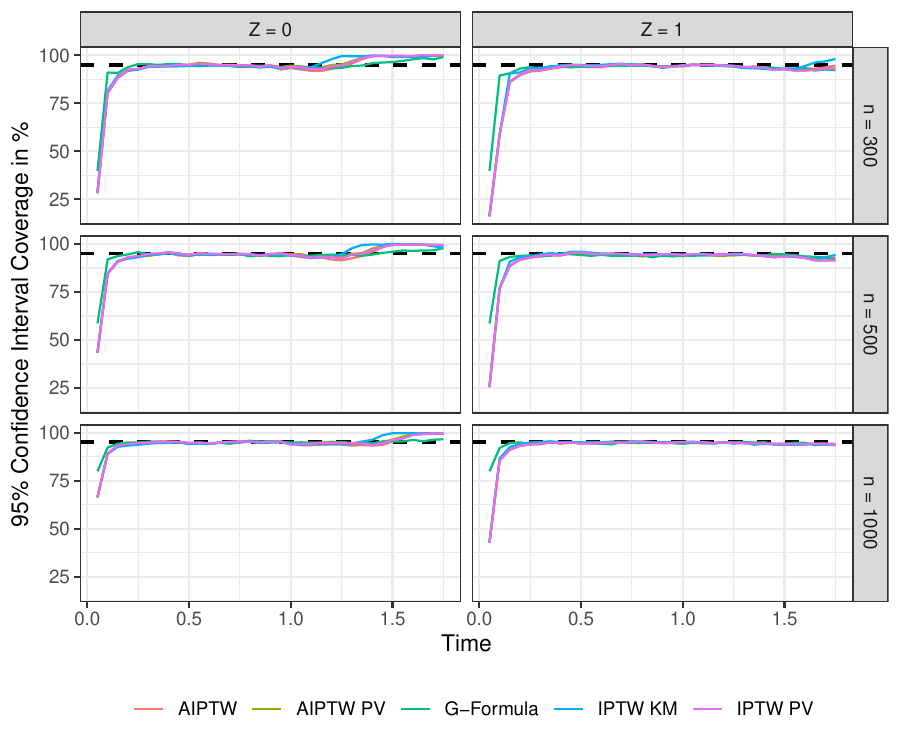}
	\caption{The coverage of approximate 95\% confidence interval width over time for both groups in the CO \& CT simulation scenario with varying sample sizes. Estimates are based on $2000$ simulation repetitions.}
	\label{fig::appendix_4_ci_coverage}
\end{figure}

\begin{figure}[!htb]
	\centering
	\includegraphics[width=0.8\linewidth]{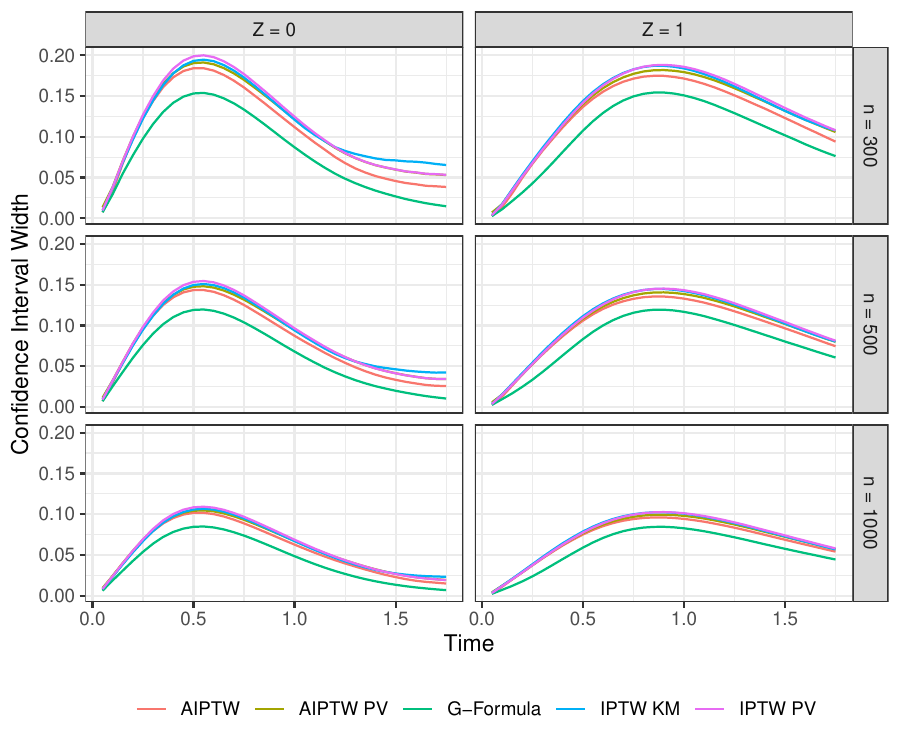}
	\caption{The average approximate 95\% confidence interval width over time for both groups in the CO \& CT simulation scenario with varying sample sizes. Estimates are based on $2000$ simulation repetitions.}
	\label{fig::appendix_4_ci_width}
\end{figure}

\end{document}